\pdftrailerid{}
\PassOptionsToPackage{unicode}{hyperref}
\PassOptionsToPackage{hyphens}{url}
\PassOptionsToPackage{dvipsnames,svgnames,x11names}{xcolor}
\documentclass[12pt]{article}
\usepackage[letterpaper,margin=1in]{geometry}

\usepackage{amsmath,amssymb,amsthm,mathtools}
\usepackage{iftex}
\ifPDFTeX
  \usepackage[T1]{fontenc}
  \usepackage[utf8]{inputenc}
  \usepackage{textcomp}
\else
  \usepackage{unicode-math}
  \defaultfontfeatures{Scale=MatchLowercase}
  \defaultfontfeatures[\rmfamily]{Ligatures=TeX,Scale=1}
\fi
\usepackage{lmodern}
\IfFileExists{upquote.sty}{\usepackage{upquote}}{}
\IfFileExists{microtype.sty}{
  \usepackage[]{microtype}
  \UseMicrotypeSet[protrusion]{basicmath}
}{}
\makeatletter
\@ifundefined{KOMAClassName}{
  \IfFileExists{parskip.sty}{
    \usepackage{parskip}
  }{
    \setlength{\parindent}{0pt}
    \setlength{\parskip}{6pt plus 2pt minus 1pt}}
}{
  \KOMAoptions{parskip=half}}
\makeatother
\usepackage{xcolor}
\makeatletter
\ifx\paragraph\undefined\else
  \let\oldparagraph\paragraph
  \renewcommand{\paragraph}{\@ifstar\xxxParagraphStar\xxxParagraphNoStar}
  \newcommand{\xxxParagraphStar}[1]{\oldparagraph*{#1}\mbox{}}
  \newcommand{\xxxParagraphNoStar}[1]{\oldparagraph{#1}\mbox{}}
\fi
\ifx\subparagraph\undefined\else
  \let\oldsubparagraph\subparagraph
  \renewcommand{\subparagraph}{\@ifstar\xxxSubParagraphStar\xxxSubParagraphNoStar}
  \newcommand{\xxxSubParagraphStar}[1]{\oldsubparagraph*{#1}\mbox{}}
  \newcommand{\xxxSubParagraphNoStar}[1]{\oldsubparagraph{#1}\mbox{}}
\fi
\makeatother

\usepackage{longtable,booktabs,array}
\usepackage{calc}
\usepackage{etoolbox}
\makeatletter
\patchcmd\longtable{\par}{\if@noskipsec\mbox{}\fi\par}{}{}
\makeatother
\IfFileExists{footnotehyper.sty}{\usepackage{footnotehyper}}{\usepackage{footnote}}
\makesavenoteenv{longtable}
\usepackage{graphicx}
\makeatletter
\def\maxwidth{\ifdim\Gin@nat@width>\linewidth\linewidth\else\Gin@nat@width\fi}
\def\maxheight{\ifdim\Gin@nat@height>\textheight\textheight\else\Gin@nat@height\fi}
\makeatother
\setkeys{Gin}{width=\maxwidth,height=\maxheight,keepaspectratio}
\makeatletter
\def\fps@figure{htbp}
\makeatother
\graphicspath{{figures/}}

\makeatletter
\@ifpackageloaded{caption}{}{\usepackage{caption}}
\AtBeginDocument{

  \renewcommand*\tablename{Table}}
\@ifpackageloaded{float}{}{\usepackage{float}}
\floatstyle{ruled}
\@ifundefined{c@chapter}{\newfloat{codelisting}{h}{lop}}{\newfloat{codelisting}{h}{lop}[chapter]}
\floatname{codelisting}{Listing}

\makeatother
\makeatletter
\@ifpackageloaded{subcaption}{}{\usepackage{subcaption}}
\makeatother
\ifLuaTeX
  \usepackage{selnolig}
\fi
\usepackage{enumitem}
\usepackage[]{natbib}
\usepackage{bookmark}
\IfFileExists{xurl.sty}{\usepackage{xurl}}{}
\theoremstyle{plain}
\newtheorem{theorem}{Theorem}
\newtheorem{lemma}{Lemma}

\newtheorem{proposition}{Proposition}
\theoremstyle{definition}
\newtheorem{assumption}{Assumption}

\theoremstyle{remark}
\newtheorem{remark}{Remark}

\setlist[enumerate,1]{label=(\alph*)}
\setlist[enumerate,2]{label=(\roman*)}

\newcommand{\eff}{\mathrm{eff}}
\newcommand{\trans}{\mathsf T}
\newcommand{\E}{\mathbb E}
\newcommand{\Prob}{\mathbb P}
\newcommand{\Var}{\operatorname{Var}}

\newcommand{\AVar}{\operatorname{AVar}}

\newcommand{\union}{\mathrm{NP}\cup\mathrm{P}}
\newcommand{\Pp}{\mathrm{P}}
\newcommand{\NP}{\mathrm{NP}}
\newcommand{\RA}{\mathrm{sub}}
\DeclareMathOperator{\expit}{expit}

\def\spacingset#1{\renewcommand{\baselinestretch}{#1}\small\normalsize}
\hypersetup{
  pdftitle={Semiparametric Efficient Data Integration Using the Dual-Frame Sampling Framework},
  pdfauthor={Kosuke Morikawa and Jae Kwang Kim},
  pdfkeywords={informative sampling; non-probability samples; efficient influence function; debiased machine learning; survey sampling},
  colorlinks=true,
  linkcolor={blue},
  filecolor={Maroon},
  citecolor={Blue},
  urlcolor={Blue},
  pdfsubject={Semiparametric data integration and survey sampling},
  pdfcreator={LaTeX}}

\AtBeginEnvironment{thebibliography}{%
  \hypersetup{citecolor=black,linkcolor=black,urlcolor=black}%
  \color{black}%
}

\begin{document}
\spacingset{1}
\begin{center}
{\LARGE\bf Semiparametric Efficient Data Integration Using the Dual-Frame Sampling Framework}\\[1.1em]
{\large Kosuke Morikawa$^{1,*}$ and Jae Kwang Kim$^{1}$}\\[0.55em]
$^{1}$Department of Statistics, Iowa State University, Ames, Iowa 50011, USA\\
\end{center}

\vspace{0.6em}

\begin{abstract}
Integrating probability and non-probability samples is increasingly important, yet unknown non-probability inclusion mechanisms complicate identification and efficient estimation. We develop semiparametric theory for dual-frame data integration and propose two complementary estimators. The first models the conditional inclusion probability parametrically and, under correct specification, attains the semiparametric efficiency bound. The probability sample serves as an outcome-observed validation sample, identifying the conditional inclusion parameters without instrumental variables even under informative selection. We derive the efficient influence function while leaving the conditional distribution of the design inclusion probability unrestricted. The resulting estimating equations satisfy exact augmentation invariance, which supports both same-sample sieve estimation and pooled cross-fitting under weak nuisance-rate conditions. The second estimator, motivated by a two-stage sampling approximation, does not require a non-probability inclusion model; although not fully efficient, it is efficient within a restricted augmentation class, is robust to misspecification of the non-probability inclusion model, and its oracle version weakly dominates optimally augmented probability-sample-only inference. Simulations and a public-data repeated-sampling study show efficiency gains under correct specification and stable performance under misspecification and weak identification. Across the main simulations, the efficient procedures substantially reduce RMSE relative to probability-sample-only estimation, while the restricted estimator remains stable under inclusion-model misspecification and weak identification.
\end{abstract}


\vspace{0.8em}
\spacingset{1.15}

\section{Introduction}
\label{sec:1}

The integration of probability and non-probability samples has become a central challenge in modern survey statistics. Probability samples support design-based inference through known inclusion probabilities, but they are often expensive and limited in size. Non-probability samples---including administrative records, opt-in web panels, and operational data sources---can be much larger, yet their unknown non-probability inclusion mechanisms may produce substantial selection bias. Combining these sources can improve precision only if the additional information is incorporated without obscuring identification or invalidating uncertainty quantification \citep{bethlehem2010selection,  wu2022,  paddock2025}.

A natural framework for such integration is dual-frame sampling, traditionally developed for two probability-based frames with partial overlap. Foundational contributions include \citet{hartleyMultipleFrameSurveys1962,hartleyMultipleFrameMethodology1974}, \citet{fullerEstimatorsSamplesSelected1972}, and \citet{skinnerEstimationDualFrame1996}. Later work developed calibration, empirical-likelihood, and semiparametric procedures for overlapping frames and merged samples \citep{lohrEstimationMultipleFrameSurveys2006,raoPseudoEmpiricalLikelihood2010,  kimDataIntegrationCombining2021, saegusaSemiparametricInferenceMerged2022}. Recent empirical comparisons in official-statistics settings also show that undercoverage and measurement error can materially alter the relative performance of non-probability integration methods \citep{angEmpiricalComparisonMethods2025}. When one source is non-probability based, outcome-dependent inclusion introduces an additional identification problem resembling missing-not-at-random sampling \citep{morikawaSemiparametricOptimalEstimation2021a,miaoIdentificationSemiparametricEfficiency2024a}. We use the dual-frame structure to make explicit which information identifies this non-probability inclusion mechanism and which information is used only to improve efficiency.

We study two complementary semiparametric models. In the first, the probability and non-probability samples are conditionally independent surveys from the same population. The probability sample observes the outcome and the non-probability inclusion indicator and therefore acts as a validation sample for the finite-dimensional non-probability inclusion model \citep{robinsEstimationRegressionCoefficients1994,breslowWeightedLikelihood2007,saegusaWeightedLikelihood2013}. Starting from the observed-data semiparametric model, we characterize the relevant tangent and augmentation spaces and derive the joint efficient influence function (EIF) for the target parameter and the conditional inclusion parameter, while leaving the conditional law of the design inclusion probability unrestricted \citep{bickelEfficientAdaptiveEstimation1993,tsiatisSemiparametricTheoryMissing2006}. In the second model, the non-probability source is conceptualized as being selected first and the probability sample is drawn from the remaining units. A restricted semiparametric projection in this two-stage model yields an estimator that does not require specifying a non-probability inclusion model. This two-stage construction is only a conceptual device: the resulting estimating equation remains design-unbiased under the original independent-survey design even when the data were not literally generated by this two-stage scheme; see Proposition~\ref{prop:ra-centering}.

Our first contribution is a global identification result for the finite-dimensional non-probability inclusion model. By leveraging the information available from the probability sample, the conditional inclusion probability can be identified without relying on instrumental variables, distributional restrictions, or shadow variables, as required by alternative identification strategies \citep{wangInstrumentalVariableApproach2014,miaoIdentifiabilityNormalNormal2016,miaoIdentificationSemiparametricEfficiency2024a}. Under a weighted full-rank condition, the resulting population map has a unique root. The same score provides a stable preliminary estimator and empirical diagnostics for weak identification based on its singular values.

Our second contribution is the derivation of the EIFs for the target parameter $\theta$ and the finite-dimensional conditional inclusion parameter $\phi$ \citep{bickelEfficientAdaptiveEstimation1993,tsiatisSemiparametricTheoryMissing2006}. The estimating equations associated with these EIFs satisfy \emph{exact augmentation invariance}: at the true target and conditional inclusion parameters, the population moments are unchanged for every admissible choice of the augmentation functions. This property is stronger than local Neyman orthogonality, which requires only that the first derivative of the population moment vanish under local nuisance perturbations. Exact invariance separates validity from variance optimization and permits both a same-sample locally profiled sieve implementation, where a sieve is a sequence of finite-dimensional approximation spaces \citep{chenLargeSampleSieve2007}, and a pooled cross-fitted implementation that learns auxiliary regressions outside each evaluation fold \citep{chernozhukov2018double}. Here ``profiled'' means that the augmentation functions are refitted whenever the finite-dimensional parameter changes; both implementations solve the resulting joint estimating equation locally.

Our third contribution is a sub-efficient estimator from the two-stage model. The term ``sub-efficient'' refers to the restriction of the augmentation space relative to the full dual-frame model; it does not mean inferior performance to probability-sample-only inference. The resulting sub-efficient estimating function remains exactly centered under the independent-survey experiment for every working augmentation. Its oracle version is optimal within the restricted class and weakly dominates the optimally augmented probability-sample-only estimator under the same probability sampling design. Combining the two results yields an explicit ordering from the full efficiency bound through the restricted estimator and the probability-sample-only procedures.

Section~\ref{sec:2} introduces the setup. Section~\ref{sec:3} studies identification and preliminary estimation. Sections~\ref{sec:4} and~\ref{sec:5} derive the efficient and sub-efficient procedures. Sections~\ref{sec:6} and~\ref{sec:7} report the empirical studies, and Section~\ref{sec:discussion} concludes. Complete proofs, numerical algorithms, and full empirical results are in the Supplementary Material.

\section{Basic Setup}
\label{sec:2}

\subsection{Population and Sampling Structure}

Consider a finite population of $N$ units indexed by $i=1,\ldots,N$. For each unit, let $Y$ denote the outcome of interest and let $X$ be a vector of frame covariates. The complete data are $L_i=(X_i^{\trans},Y_i)^{\trans}$. The finite-population target $\theta_N$ solves
\begin{equation*}
\sum_{i=1}^N U(\theta_N;L_i)=0,
\end{equation*}
where $U(\theta;L)$ is a known estimating function. For the population mean, $U(\theta;L)=\theta-Y$.

For the semiparametric calculations, the realized population is embedded in an independent and identically distributed (i.i.d.) superpopulation model. The corresponding parameter $\theta_0$ is defined by $
E\{U(\theta_0;L)\}=0.
$

The semiparametric theory below is formulated for the superpopulation target $\theta_0$ under the independent Poisson sampling experiment specified below, and the efficient influence function is derived for that experiment. The main Monte Carlo coverage calculations therefore use $\theta_0$. The realized finite-population target $\theta_N$ is retained as a secondary error criterion, and the public-data application is treated as a separate fixed-population repeated-sampling experiment.

Throughout, $a^{\trans}$ denotes transpose, $\|a\|$ is the Euclidean norm, $a^{\otimes2}=aa^{\trans}$, and $A^{-\trans}=(A^{-1})^{\trans}$. For symmetric matrices, $A\preceq B$ means that $B-A$ is positive semidefinite (the Loewner order), and $\lambda_{\min}(A)$ is the smallest eigenvalue. All expectations and variances are under the true law unless stated otherwise. We use $\AVar(\widehat\theta)$ for the limiting variance of $\sqrt N(\widehat\theta-\theta_0)$, $o_p(1)$ for convergence to zero in probability, and $O_p(1)$ for boundedness in probability. Finally, $\expit(t)=\{1+\exp(-t)\}^{-1}$ and $\operatorname{logit}(p)=\log\{p/(1-p)\}$.

Two samples are drawn. Let $\delta_{\mathrm{NP}}=1$ indicate inclusion in the non-probability source and let $\delta_{\mathrm P}=1$ indicate inclusion in the probability sample. The main sampling law is
\begin{align}
\delta_{\mathrm{NP}}\mid(L,\pi_{\mathrm P})
&\sim \operatorname{Bernoulli}\{\pi_{\mathrm{NP}}(L;\phi_0)\},
\nonumber\\
\delta_{\mathrm P}\mid(L,\pi_{\mathrm P})
&\sim \operatorname{Bernoulli}(\pi_{\mathrm P}),
\qquad
\delta_{\mathrm{NP}}\perp\delta_{\mathrm P}\mid(L,\pi_{\mathrm P}).
\label{eq:sampling-law}
\end{align}
For each unit in the probability sample, $\pi_{\mathrm P}$ is the known design inclusion probability, whereas the conditional inclusion probability for the non-probability source is modeled by the finite-dimensional family $\pi_{\mathrm{NP}}(L;\phi)$. Throughout, $\pi_{\mathrm{NP}}(L;\phi)$ denotes the conditional inclusion probability under parameter $\phi$, $\{\pi_{\mathrm{NP}}(\cdot;\phi):\phi\in\Phi\}$ is the non-probability inclusion model, and $\pi_{\mathrm{NP}}(L;\phi_0)$ is the true non-probability inclusion mechanism. The positivity condition used below means that the relevant inclusion probabilities and conditional-moment denominators are bounded away from zero on the support used for inference.

\begin{assumption}[Frame and sample inclusion information]
\label{ass:observed-frame}
A population frame supplies $X$ for every population unit and records both sample inclusion indicators $(\delta_{\mathrm{NP}},\delta_{\mathrm P})$ for each population unit. The outcome $Y$ is observed whenever $\delta_{\mathrm{NP}}+\delta_{\mathrm P}>0$, and $\pi_{\mathrm P}$ is observed whenever $\delta_{\mathrm P}=1$.
\end{assumption}

Assumption~\ref{ass:observed-frame} is the observed-data experiment for the main theory. In particular, the non-probability inclusion indicator is observed for probability-sample units, which makes the probability sample a validation sample for the non-probability inclusion model. Section~\ref{sec:6} includes a sensitivity experiment that coarsens this overlap information while retaining one record per population unit; that experiment lies outside the identification theorem.

\subsection{Observed Data Structure}

Under Assumption~\ref{ass:observed-frame}, the unit-level observed data can be written as
$
O=(X,\delta_{\mathrm{NP}},\delta_{\mathrm P},\allowbreak
\delta_{\mathrm{NP}\cup\mathrm P}Y,\allowbreak
\delta_{\mathrm P}\pi_{\mathrm P})$, and 
$
\delta_{\mathrm{NP}\cup\mathrm P}
=\delta_{\mathrm{NP}}+\delta_{\mathrm P}
-\delta_{\mathrm{NP}}\delta_{\mathrm P}.
$
A key feature is that $\pi_{\mathrm P}$ is observed only in the probability sample. Consequently, the conditional law $f(\pi_{\mathrm P}\mid L)$ is an unknown nuisance component of the general semiparametric model. Section~\ref{sec:4} shows that the conditional moments required by the efficient projection are nevertheless identified from probability-sample records.

\begin{table}[htbp]
\centering
\caption{Response patterns and observed variables under the main experiment.}
\label{tab:response}
\begin{tabular}{ccccc}
\toprule
$\delta_{\mathrm{NP}}$ & $\delta_{\mathrm P}$ & $X$ & $Y$ & $\pi_{\mathrm P}$ \\
\midrule
1 & 1 & observed & observed & observed \\
1 & 0 & observed & observed & missing \\
0 & 1 & observed & observed & observed \\
0 & 0 & observed & missing & missing \\
\bottomrule
\end{tabular}
\end{table}

\begin{remark}[Coarsened overlap information]
\label{rem:coarsened-overlap}
The identification theorem requires the non-probability inclusion indicator to be observed for probability-sample units. When overlap information is missing or coarsened, the validation score can be misclassified and Theorem~\ref{thm:identification} no longer applies. The coarsened-overlap sensitivity experiment in Section~\ref{sec:6} retains one record per population unit; no exact record-linkage-free or duplicate-invariance claim is made.
\end{remark}

\subsection{Baseline Estimators}

Using only the probability sample, one may solve
\begin{equation}
\sum_{i=1}^N\frac{\delta_{\mathrm P,i}}{\pi_{\mathrm P,i}}
U(\theta;L_i)=0,
\label{eq:ipw_prob}
\end{equation}
which gives the survey-weighted estimating-equation estimator $\widehat\theta_{\mathrm P}$ \citep{horvitzGeneralizationSampling1952, binderVariancesAsymptoticallyNormal1983}. This estimator is valid under the probability sampling design but discards all outcomes observed only in the non-probability source.

If the non-probability inclusion model is available, a non-probability inverse-probability estimator solves
\begin{equation}
\sum_{i=1}^N
\frac{\delta_{\mathrm{NP},i}}
{\pi_{\mathrm{NP}}(L_i;\widehat\phi)}
U(\theta;L_i)=0.
\label{eq:ipw_nonprob}
\end{equation}
Here $\widehat\phi$ denotes a consistent estimator of the true conditional inclusion parameter $\phi_0$. Section~\ref{sec:3} constructs the preliminary estimator $\widetilde\phi$ from the outcome-observed probability validation sample in~\eqref{eq:validation-preliminary}; Section~\ref{sec:4} subsequently refines it using the estimating equations associated with the EIF\@. Its validity depends on correct specification and identification of the non-probability inclusion model.

Calibration equations based only on frame covariates have also been proposed \citep{kottUsingCalibrationWeighting2010}. Such equations can be useful when only $X$ is available. However, it is well known that models allowing the conditional inclusion probability to depend on the outcome are generally not identifiable from such information alone; identification typically requires an instrumental-variable-type assumption, often formulated in terms of a shadow variable \citep{wangInstrumentalVariableApproach2014,miaoIdentifiabilityNormalNormal2016,miaoIdentificationSemiparametricEfficiency2024a}. In contrast, the validation score used here directly exploits the outcome observed in the probability sample. The next section establishes identification using this validation information.

\section[Identification and Preliminary Estimation of the Non-Probability Inclusion Model]{Identification and Preliminary Estimation of the Non-Probability Inclusion Model}
\label{sec:3}

The conditional inclusion parameter must be identified before the full semiparametric information can be used. The probability sample plays a special role: under Assumption~\ref{ass:observed-frame}, it observes both $L$ and the non-probability inclusion indicator. Conditional independence in~\eqref{eq:sampling-law} therefore turns the probability sample into an outcome-observed validation sample for $\pi_{\mathrm{NP}}(L;\phi)$. For a differentiable non-probability inclusion model, write $\dot\pi_{\mathrm{NP}}(L;\phi)=\partial\pi_{\mathrm{NP}}(L;\phi)/\partial\phi$ for its column gradient with respect to $\phi$.

The observed-data likelihood score for $\phi$ contains the four response-pattern contributions
\begin{align}
S_{\phi}^{\mathrm{obs}}(O;\phi)
={}&\delta_{\mathrm{NP}}\delta_{\mathrm P}
\frac{\dot\pi_{\mathrm{NP}}}{\pi_{\mathrm{NP}}}
-(1-\delta_{\mathrm{NP}})\delta_{\mathrm P}
\frac{\dot\pi_{\mathrm{NP}}}{1-\pi_{\mathrm{NP}}}
\nonumber\\
&+\delta_{\mathrm{NP}}(1-\delta_{\mathrm P})
\frac{\dot\pi_{\mathrm{NP}}}{\pi_{\mathrm{NP}}}
-(1-\delta_{\mathrm{NP}})(1-\delta_{\mathrm P})m_{00}(X;\phi),
\label{eq:score_phi}
\end{align}
where
\begin{equation}
m_{00}(X;\phi)
=E\left\{
\frac{\dot\pi_{\mathrm{NP}}(L;\phi)}
{1-\pi_{\mathrm{NP}}(L;\phi)}
\,\middle|\,
\delta_{\mathrm{NP}}=\delta_{\mathrm P}=0,X
\right\}.
\label{eq:m00-identification}
\end{equation}
Direct evaluation of~\eqref{eq:score_phi} therefore requires an unknown conditional distribution among units observed in neither source. This difficulty concerns direct likelihood-score calculation; it does not imply that the finite-dimensional non-probability inclusion model is unidentified. The probability validation sample provides a simpler unbiased score that avoids~\eqref{eq:m00-identification}.

\subsection{Validation-Sample Score}

For a differentiable non-probability inclusion model, define
\begin{equation}
S_{\mathrm{val}}(O;\phi)
=\delta_{\mathrm P}
\{\delta_{\mathrm{NP}}-\pi_{\mathrm{NP}}(L;\phi)\}
\frac{\dot\pi_{\mathrm{NP}}(L;\phi)}
{\pi_{\mathrm{NP}}(L;\phi)
\{1-\pi_{\mathrm{NP}}(L;\phi)\}}.
\label{eq:validation-score}
\end{equation}
All quantities in~\eqref{eq:validation-score} are observed when $\delta_{\mathrm P}=1$. Moreover, by~\eqref{eq:sampling-law},
\begin{equation*}
E\{S_{\mathrm{val}}(O;\phi_0)\mid L,\pi_{\mathrm P}\}=0.
\end{equation*}

We state the identification result for the logistic non-probability inclusion model
\begin{equation}
\pi_{\mathrm{NP}}(L;\phi)
=\operatorname{expit}\{\phi^{\trans}V(L)\},
\label{eq:logistic-selection}
\end{equation}
where $V(L)$ contains an intercept and the analyst-specified variables in the non-probability inclusion model. Let
$
\bar\pi_{\mathrm P}(L)=E(\pi_{\mathrm P}\mid L).
$
The population validation moment is
\begin{equation*}
G_{\mathrm{val}}(\phi)
=E\{S_{\mathrm{val}}(O;\phi)\}
=E\left[
\bar\pi_{\mathrm P}(L)
\{\pi_{\mathrm{NP}}(L;\phi_0)
-\pi_{\mathrm{NP}}(L;\phi)\}V(L)
\right].
\end{equation*}

\begin{assumption}[Validation information]
\label{ass:validation-information}
The parameter space $\Phi$ is compact and convex, $\phi_0$ is an interior point, and for some $c_{\mathrm{val}}>0$,
\begin{equation*}
\inf_{\phi\in\Phi}
\lambda_{\min}
E\left[
\bar\pi_{\mathrm P}(L)
\pi_{\mathrm{NP}}(L;\phi)
\{1-\pi_{\mathrm{NP}}(L;\phi)\}
V(L)V(L)^{\trans}
\right]
\ge c_{\mathrm{val}}.
\end{equation*}
\end{assumption}

Assumption~\ref{ass:validation-information} is a weighted full-rank condition. It permits arbitrary conditional variation in $\pi_{\mathrm P}\mid L$ and requires only that the validation sample provide positive information in every direction of the non-probability inclusion model.

\begin{theorem}[Global identification from the probability validation sample]
\label{thm:identification}
Under~\eqref{eq:sampling-law}, Assumption~\ref{ass:observed-frame}, the logistic model~\eqref{eq:logistic-selection}, and Assumption~\ref{ass:validation-information}, $G_{\mathrm{val}}(\phi)=0$ has the unique solution $\phi_0$.
\end{theorem}

Theorem~\ref{thm:identification} separates model identification from efficiency. The validation score identifies $\phi_0$ using only probability-sample records. Section~\ref{sec:4} then applies semiparametric projection to the full-data target estimating function and the observed-data likelihood score for $\phi$ to construct the EIFs for $\theta$ and $\phi$ using all response patterns.

The outcome-observed probability sample plays a role analogous to a validation sample in incomplete-data inference \citep{robinsEstimationRegressionCoefficients1994} and to a phase-two sample in semiparametric two-phase designs \citep{breslowWeightedLikelihood2007,saegusaWeightedLikelihood2013}. Here, however, the validation information is used specifically to identify a non-probability inclusion model that may depend on the outcome, and the two data sources need not form a nested two-phase sample.

\subsection{Preliminary Estimation and Weak-Identification Diagnostics}

Let $\widetilde\phi$ solve the sample validation equation
\begin{equation}
\frac{1}{N}\sum_{i=1}^N S_{\mathrm{val}}(O_i;\widetilde\phi)=0.
\label{eq:validation-preliminary}
\end{equation}
For the logistic model, this is ordinary logistic regression of $\delta_{\mathrm{NP}}$ on $V(L)$ within the probability sample. Under standard uniform-law and differentiability conditions, Theorem~\ref{thm:identification} implies consistency, and
\begin{equation*}
\sqrt N(\widetilde\phi-\phi_0)
=-J_{\mathrm{val},0}^{-1}
\frac1{\sqrt N}\sum_{i=1}^N
S_{\mathrm{val}}(O_i;\phi_0)+o_p(1),
\end{equation*}
where
\begin{equation}
J_{\mathrm{val},0}
=-E\left[
\bar\pi_{\mathrm P}(L)
\pi_{\mathrm{NP}}(L;\phi_0)
\{1-\pi_{\mathrm{NP}}(L;\phi_0)\}
V(L)V(L)^{\trans}
\right].
\label{eq:validation-jacobian}
\end{equation}
The estimator $\widetilde\phi$ initializes the local profiled EIF-based estimating-equation algorithm in Section~\ref{sec:4}.

The empirical analogue of~\eqref{eq:validation-jacobian} also provides a direct weak-identification diagnostic. We report its smallest singular value and its condition number, defined as the ratio of the largest to the smallest singular value. A small singular value or a large condition number indicates that some linear combination of the components of the conditional inclusion parameter is weakly informed by the probability validation sample, even though the population model remains formally identified.

\section{Semiparametric Efficient Estimation under Two Independent Surveys}
\label{sec:4}

We now derive the efficient influence function from the observed-data semiparametric model and then construct feasible estimators based on it. The tangent-space representation and projection normal equations are given in the Supplementary Material.

\subsection{Semiparametric model and projection argument}

Under~\eqref{eq:sampling-law}, the complete-data density factors as
\begin{align*}
&f(\delta_{\mathrm{NP}},\delta_{\mathrm P},\pi_{\mathrm P},L;
\phi,\eta_1,\eta_2) \nonumber\\
&\quad=
\pi_{\mathrm{NP}}(L;\phi)^{\delta_{\mathrm{NP}}}
\{1-\pi_{\mathrm{NP}}(L;\phi)\}^{1-\delta_{\mathrm{NP}}}
\pi_{\mathrm P}^{\delta_{\mathrm P}}(1-\pi_{\mathrm P})^{1-\delta_{\mathrm P}}
 f(\pi_{\mathrm P}\mid L;\eta_1)f(L;\eta_2),
\end{align*}
where $\eta_1$ and $\eta_2$ index the unrestricted nuisance laws $f(\pi_{\mathrm P}\mid L)$ and $f(L)$, respectively. Let $\Lambda_2$ denote the conditional-mean-zero augmentation space induced by the observed-data patterns. Its formal definition, pattern representation, and projection equations are given in Supplementary Section S.2.

For the Z-estimand defined by $E\{U(\theta;L)\}=0$, begin with the inverse-probability-weighted estimating function $\delta_{\mathrm P}U(\theta;L)/\pi_{\mathrm P}$. Theorem S1 in Supplementary Section S.2 makes explicit the EIF construction used implicitly in the proof of Lemma 4.1 of \citet{morikawaSemiparametricOptimalEstimation2021a} and extends it to the present setting. For a Z-estimand, after a nonsingular constant transformation, the efficient estimating equations for $\theta$ and $\phi$ are obtained by projecting a conditionally unbiased estimating function for $\theta$ onto $\Lambda_2^\perp$ and the observed-data likelihood score for $\phi$ onto $\Lambda_2$, respectively. The corresponding EIFs are obtained by normalizing these projected equations by their joint Jacobian.

\subsection{Efficient influence function}

Write
$
q_\phi(L,\pi_{\mathrm P})
=\pi_{\mathrm{NP}}(L;\phi)+\pi_{\mathrm P}
-\pi_{\mathrm{NP}}(L;\phi)\pi_{\mathrm P}$, where $ O_{\mathrm P}=(1-\pi_{\mathrm P})/\pi_{\mathrm P}$.
Here $q_\phi(L,\pi_{\mathrm P})$ is the conditional probability of appearing in at least one source. Within this section, suppressed arguments in $\pi_{\mathrm{NP}}$, $\dot\pi_{\mathrm{NP}}$, and $q_\phi$ are evaluated at the displayed $(L,\phi,\pi_{\mathrm P})$. A \emph{working function} means any admissible square-integrable candidate used in place of an augmentation function when constructing the EIFs for $\theta$ and $\phi$. For a working function $k_2(L)$,
\begin{align*}
R_\phi\{O;k_2\}
={}&1-\frac{\delta_{\mathrm{NP}}\delta_{\mathrm P}}
{\pi_{\mathrm{NP}}(L;\phi)\pi_{\mathrm P}}
-\delta_{\mathrm{NP}}\left(1-\frac{\delta_{\mathrm P}}{\pi_{\mathrm P}}\right)\{1+k_2(L)\}\nonumber\\
&-\delta_{\mathrm P}\left\{1-\frac{\delta_{\mathrm{NP}}}{\pi_{\mathrm{NP}}(L;\phi)}\right\}
\left[1+O_{\mathrm P}\{1-\pi_{\mathrm{NP}}(L;\phi)-\pi_{\mathrm{NP}}(L;\phi)k_2(L)\}\right].
\end{align*}
 Recall $m_{00}(X;\phi)$ from Section~\ref{sec:3}. The semiparametric projection yields the following augmentation functions used to construct the EIFs for $\theta$ and $\phi$:
\begin{align*}
k_1^*(L)
&=\frac{U(\theta;L)E(O_{\mathrm P}\mid L)}
{E\{O_{\mathrm P}q_\phi(L,\pi_{\mathrm P})\mid L\}},\\
k_2^*(L)
&=\frac{E[O_{\mathrm P}\{1-q_\phi(L,\pi_{\mathrm P})\}\mid L]}
{E\{O_{\mathrm P}q_\phi(L,\pi_{\mathrm P})\mid L\}},\\
\eta^*(X)
&=\frac{E[\{1-q_\phi\}\{U(\theta;L)/\pi_{\mathrm P}-k_1^*\pi_{\mathrm{NP}}O_{\mathrm P}\}\mid X]}
{E[\{1-q_\phi\}\{1+O_{\mathrm P}[1-\pi_{\mathrm{NP}}(1+k_2^*)]\}\mid X]},\\
\kappa_1^*(L)
&=-\frac{E[\{1-\pi_{\mathrm P}\}\dot\pi_{\mathrm{NP}}/\pi_{\mathrm{NP}}\mid L]}
{E\{O_{\mathrm P}q_\phi(L,\pi_{\mathrm P})\mid L\}},\\
\kappa^*(X)
&=\frac{E[\{1-q_\phi\}\{m_{00}(X;\phi)-\kappa_1^*\pi_{\mathrm{NP}}O_{\mathrm P}\}\mid X]}
{E[\{1-q_\phi\}\{1+O_{\mathrm P}[1-\pi_{\mathrm{NP}}(1+k_2^*)]\}\mid X]}.
\end{align*}
 Feasible estimation of these augmentation functions is described in the next subsection, with further details given in Supplementary Section~S.3. Let
$
\zeta^*=(k_1^*,k_2^*,\eta^*,\kappa_1^*,\kappa^*)
$
denote the five augmentation functions used to construct the EIFs for $\theta$ and $\phi$.

\begin{theorem}[Efficient influence functions for the target and conditional inclusion parameters]
\label{thm:general-efficient-score}
Under the conditions in the Supplementary Material, define
\begin{align}
S_{\mathrm{eff},\theta}
={}&\frac{\delta_{\mathrm P}}{\pi_{\mathrm P}}U(\theta;L)
+\left\{\delta_{\mathrm P}(\delta_{\mathrm{NP}}-\pi_{\mathrm{NP}})O_{\mathrm P}
+\delta_{\mathrm{NP}}\left(1-\frac{\delta_{\mathrm P}}{\pi_{\mathrm P}}\right)\right\}k_1^*(L)
+\eta^*(X)R_\phi\{O;k_2^*\},
\label{eq:general-eff-theta}
\end{align}
and
\begin{align}
S_{\mathrm{eff},\phi}
={}&\delta_{\mathrm P}(\delta_{\mathrm{NP}}-\pi_{\mathrm{NP}})
\left\{\frac{\dot\pi_{\mathrm{NP}}}{\pi_{\mathrm{NP}}(1-\pi_{\mathrm{NP}})}-O_{\mathrm P}\kappa_1^*(L)\right\}\nonumber\\
&-\delta_{\mathrm{NP}}\left(1-\frac{\delta_{\mathrm P}}{\pi_{\mathrm P}}\right)\kappa_1^*(L)
-\kappa^*(X)R_\phi\{O;k_2^*\}.
\label{eq:general-eff-phi}
\end{align}
Let $\xi=(\theta^{\trans},\phi^{\trans})^{\trans}$ and $\xi_0=(\theta_0^{\trans},\phi_0^{\trans})^{\trans}$. Define the joint estimating function
\[
S_{\mathrm{eff}}(O;\xi)
=\{S_{\mathrm{eff},\theta}(O;\xi)^{\trans},S_{\mathrm{eff},\phi}(O;\xi)^{\trans}\}^{\trans}.
\]
If $J_0=E\{\partial S_{\mathrm{eff}}(O;\xi_0)/\partial\xi^{\trans}\}$ is nonsingular, then
\begin{equation}
\varphi_{\xi,\mathrm{eff}}(O)
=-J_0^{-1}S_{\mathrm{eff}}(O;\xi_0)
\label{eq:eif}
\end{equation}
is the joint EIF for $(\theta,\phi)$. Writing
$
\varphi_{\xi,\mathrm{eff}}
=\{\varphi_{\theta,\mathrm{eff}}^{\trans},\varphi_{\phi,\mathrm{eff}}^{\trans}\}^{\trans},
$
$\varphi_{\theta,\mathrm{eff}}$ and $\varphi_{\phi,\mathrm{eff}}$ are the EIFs for $\theta$ and $\phi$, respectively.
\end{theorem}

The target-score component in Theorem~\ref{thm:general-efficient-score} has a limited robustness property for probability-sample records. Setting $\delta_{\mathrm P}=1$ in~\eqref{eq:general-eff-theta} gives
\[
S_{\mathrm{eff},\theta}\big|_{\delta_{\mathrm P}=1}
=\frac{U(\theta;L)}{\pi_{\mathrm P}}
-\pi_{\mathrm{NP}}(L;\phi)O_{\mathrm P}k_1^*(L)
-\eta^*(X)O_{\mathrm P}
\left[1-\pi_{\mathrm{NP}}(L;\phi)\{1+k_2^*(L)\}\right],
\]
which does not involve $\delta_{\mathrm{NP}}$. Thus, for fixed $\phi$ and augmentation functions, the contribution of a probability-sample unit to the $\theta$ estimating equation is the same whether that unit is classified as probability-only or as belonging to both samples. This algebraic property does not remove the need to observe $\delta_{\mathrm{NP}}$ in the validation equation or in $S_{\mathrm{eff},\phi}$, and it does not cover duplicated unlinked records. Supplementary Section S.2 gives the simplification.

For any admissible collection $\zeta=(k_1,k_2,\eta,\kappa_1,\kappa)$, let $S_{\theta}(O;\xi,\zeta)$ and $S_{\phi}(O;\xi,\zeta)$ denote the two expressions in~\eqref{eq:general-eff-theta} and~\eqref{eq:general-eff-phi}, respectively, after replacing $\zeta^*$ by $\zeta$, and write
\[
S(O;\xi,\zeta)
=\{S_{\theta}(O;\xi,\zeta)^{\trans},
S_{\phi}(O;\xi,\zeta)^{\trans}\}^{\trans}.
\]

\begin{proposition}[Exact augmentation invariance]
\label{prop:exact-invariance}
For every admissible square-integrable $\zeta=(k_1,k_2,\eta,\kappa_1,\kappa)$, at $(\theta_0,\phi_0)$,
\begin{align*}
E\{S_{\theta}\mid L,\pi_{\mathrm P}\}&=U(\theta_0;L),
&
E\{S_{\phi}\mid L,\pi_{\mathrm P}\}&=0.
\end{align*}
Hence the population moments are constant over the admissible augmentation class.
\end{proposition}

Exact augmentation invariance has two distinct consequences. First, at $(\theta_0,\phi_0)$, the population moment is unchanged by any admissible working augmentation, although the influence function and variance may change; under identification, the true parameter therefore remains a root. Second, the nuisance-estimation drift is identically zero, so one does not need the conventional product-rate condition that products of nuisance-estimation errors vanish faster than $N^{-1/2}$. Same-sample fitting still requires a stochastic-equicontinuity condition for evaluating the joint estimating function at a data-dependent fit, whereas cross-fitting replaces this requirement by convergence of the joint estimating function evaluated on held-out folds. The result does not protect against misspecification of $\pi_{\mathrm{NP}}(L;\phi)$, incorrect probability weights, misclassified sample inclusion indicators, or failure of conditional independence.

\begin{proposition}[Efficiency gain from the non-probability source]
\label{prop:efficient-dominates-p}
Let $V_{\theta,\mathrm{eff}}$ be the target block of the variance of~\eqref{eq:eif}. For any regular asymptotically linear estimator based only on the probability-sample subexperiment, with asymptotic variance $V_{\theta,\mathrm P}$,
\begin{equation*}
V_{\theta,\mathrm{eff}}\preceq V_{\theta,\mathrm P}.
\end{equation*}
\end{proposition}

\subsection{Feasible implementation of the efficient procedure}

We estimate the two bounded $L$-level reparameterizations by weighted fractional-logit sieves, recover $k_2$ from its bounded reparameterization, and reconstruct $k_1$ and $\kappa_1$ from the corresponding shape identities. The remaining real-valued $X$-conditional functions, $\eta$ and $\kappa$, are estimated by weighted linear sieves. This construction avoids separately fitting a numerator and denominator and then taking their ratio. The exact reparameterizations, pseudo-outcomes, weights, shape restrictions, and objective functions are given in Supplementary Section~S.3.

In the simulations, where the frame vector is $(X,Z)^{\trans}$ with scalar continuous $X$ and binary $Z$, the $L$-conditional regressions use training-sample-standardized natural cubic splines in the two continuous variables $(X,Y)$, including their tensor interaction and a main effect for $Z$; the $X$-conditional regressions use a $Z$-varying natural spline in $X$. If $n_{\mathrm P,\mathrm{tr}}$ is the relevant probability-sample training size, the prespecified primary rule takes $s=2$ and sets the number of one-dimensional spline columns to
\[
K_{\mathrm{req}}(n_{\mathrm P,\mathrm{tr}},d)
=\max\left\{2,\left\lceil n_{\mathrm P,\mathrm{tr}}^{1/(2s+d)}\right\rceil\right\},
\]
with $d=2$ for the $L$-conditional regressions and $d=1$ for the $X$-conditional regressions, subject to deterministic total-term caps. A richer prespecified sensitivity rule multiplies the rate by $1.5$ and uses a larger minimum and cap; neither rule is selected using residual-based fit criteria or Monte Carlo performance. The complete basis matrices, caps, and realized dimensions are reported in Supplementary Sections S.3--S.4. More generally, balancing the usual series variance and approximation errors gives $J_N\asymp n_{\mathrm P,\mathrm{tr}}^{d/(2s+d)}$ for an $s$-smooth function of $d$ continuous variables \citep{chenLargeSampleSieve2007}. Exact augmentation invariance removes the conventional $N^{-1/4}$ product-rate requirement, but convergence of the fitted joint estimating function remains necessary.

Partition the unit indices into $K$ fixed folds $I_1,\ldots,I_K$. For a candidate $\xi$, let $\widehat\zeta_{\mathrm{sieve}}(\xi)$ denote the augmentation functions fitted on the full sample, and let $\widehat\zeta^{[-k]}(\xi)$ denote the functions fitted after excluding fold $I_k$. We use DML as an abbreviation for \emph{double/debiased machine learning}. The pooled cross-fitted estimating function below follows the DML2 construction of \citet{chernozhukov2018double}, which pools held-out estimating-function contributions before estimating the parameter. For tables and figures, we call the full-sample implementation \emph{Eff-sieve} and the pooled cross-fitted implementation \emph{Eff-DML}; the prefix ``Eff'' indicates that both use the joint efficient estimating function. Define
\begin{align*}
\widehat\Psi_{\mathrm{sieve}}(\xi)
&=\frac{1}{N}\sum_{i=1}^N S\{O_i;\xi,\widehat\zeta_{\mathrm{sieve}}(\xi)\},
&
\widehat\Psi_{\mathrm{DML}}(\xi)
&=\frac1N\sum_{k=1}^K\sum_{i\in I_k}
S\{O_i;\xi,\widehat\zeta^{[-k]}(\xi)\}.
\end{align*}
The two estimators are the local approximate roots
\begin{equation}
\widehat\Psi_m(\widehat\xi_m)=o_p(N^{-1/2}),
\qquad m\in\{\mathrm{sieve},\mathrm{DML}\},
\label{eq:profiled-roots}
\end{equation}
initialized by the globally identified validation estimator.

Cross-fitting avoids the same-sample stochastic-equicontinuity requirement and accommodates highly adaptive learners such as random forests, boosted regression trees, and neural networks \citep{chernozhukov2018double}. However, each nuisance learner is then trained on only a fraction of the probability sample. Because the $L$-level augmentation regressions are learned exclusively from probability-sample records, this loss of training data can impair finite-sample nuisance estimation when the validation sample is small. Eff-sieve instead uses all available probability-sample records; under controlled sieve complexity and the same-sample stochastic-equicontinuity condition in Theorem~\ref{thm:profiled-efficient}, it remains first-order valid without sample splitting. We therefore regard Eff-sieve as a natural implementation for this setting and Eff-DML as a complementary cross-fitted implementation.

For Eff-DML, the fold-specific augmentation functions are learned outside $I_k$ and evaluated on $I_k$, so every unit contributes once to the pooled estimating function. In both implementations, the augmentation functions are refitted whenever the candidate $(\theta,\phi)$ changes. The robust local solvers and convergence diagnostics are described in Supplementary Section S.3. For numerical root finding, the profiled Jacobian includes the effect of refitting the augmentation functions at nearby candidate values; the final sandwich variance instead uses the partial Jacobian described in Supplementary Section~S.3.

\begin{theorem}[Locally profiled efficient estimators]
\label{thm:profiled-efficient}
Suppose $\widehat\xi_m$ is consistent and satisfies~\eqref{eq:profiled-roots}. Let $S_m^{\dagger}$ denote the $L^2$ limit of the fitted joint estimating function under the true observed-data law, and let $J_m$ denote the corresponding nonsingular limiting profiled Jacobian.

For Eff-sieve, define
$
\widehat S_{\mathrm{sieve},0}(O)
=
S\{O;\xi_0,\widehat\zeta_{\mathrm{sieve}}(\xi_0)\},
$
where the subscript $0$ denotes evaluation at the true finite-dimensional parameter $\xi_0$, and assume the componentwise same-sample stochastic-equicontinuity condition
\begin{equation}
\frac1{\sqrt N}\sum_{i=1}^N
\left\{
\widehat S_{\mathrm{sieve},0}(O_i)
-
S_{\mathrm{sieve}}^{\dagger}(O_i)
\right\}
=o_p(1).
\label{eq:sieve-random-index}
\end{equation}
For Eff-DML, assume that the number of folds $K$ is fixed, that the out-of-fold fitted joint estimating functions stabilize in $L^2$ under the true observed-data law, and that their $(2+c)$ moments are uniformly bounded for some $c>0$. Then
\[
\sqrt N(\widehat\xi_m-\xi_0)
=
-J_m^{-1}\frac1{\sqrt N}
\sum_{i=1}^N S_m^{\dagger}(O_i)
+o_p(1).
\]
If the fitted joint estimating function converges to $S_{\mathrm{eff}}$, the corresponding estimator attains the semiparametric efficiency bound. Exact augmentation invariance removes the conventional $N^{-1/4}$ product-rate requirement for the augmentation functions.
\end{theorem}

Condition~\eqref{eq:sieve-random-index} controls the reuse of the same observations to fit and evaluate the sieve augmentation functions and is not implied by $L^2$ convergence alone. A primitive sufficient route for Assumption S2 in the Supplementary Material, detailed in Supplementary Section~S.3, is that the local score-difference class be pointwise measurable, have uniform entropy of order $J_N\log(A/\epsilon)$ for a fixed constant $A>1$, and admit either a uniformly bounded envelope or a uniformly sub-Weibull envelope \citep{vanDerVaartWellnerLocalMaximal2011}. Combined with the usual series $L^2$ rate and $J_N\asymp n_{\mathrm P,\mathrm{tr}}^{d/(2s+d)}$, these conditions reduce, up to logarithmic factors, to $s>d/2$. Both prespecified simulation rules use $s=2$, with $d=2$ and $d=1$ for the $L$- and $X$-conditional regressions, respectively. Cross-fitting avoids this same-sample entropy condition because each fitted function is evaluated only on observations outside its training fold.

\section{Restricted Projection under Two-Stage Sampling}
\label{sec:5}

The second construction begins with the original two-stage dual-frame model: the non-probability source is selected first and the probability sample is drawn from the remaining units. Although this model supplies the derivation, the samples need not have been collected literally in this order. This model motivates a restricted projection that avoids specifying a non-probability inclusion model. We call the resulting procedure \emph{sub-efficient} because it is optimal in a restricted augmentation space rather than in the full space of Section~\ref{sec:4}.

\subsection{Two-stage model and restricted projection}

The complete-data density is
\begin{align*}
f_{\mathrm{2st}}(\delta_{\mathrm{NP}},\delta_{\mathrm P},\pi_{\mathrm P},L)
={}&\pi_{\mathrm{NP}}(L)^{\delta_{\mathrm{NP}}}
\{1-\pi_{\mathrm{NP}}(L)\}^{1-\delta_{\mathrm{NP}}}
\left\{\pi_{\mathrm P}^{\delta_{\mathrm P}}(1-\pi_{\mathrm P})^{1-\delta_{\mathrm P}}\right\}^{1-\delta_{\mathrm{NP}}}
 f(\pi_{\mathrm P}\mid L)f(L).
\end{align*}
A generic square-integrable observed-data function is
\[
g(O)
=
\delta_{\mathrm{NP}}g_1(L)
+
(1-\delta_{\mathrm{NP}})\delta_{\mathrm P}
g_2(L,\pi_{\mathrm P})
+
(1-\delta_{\mathrm{NP}})(1-\delta_{\mathrm P})g_3(X),
\]
where $g_1$, $g_2$, and $g_3$ are arbitrary measurable functions for,
respectively, non-probability-sample units, probability-only units, and units
in neither source.
Imposing the conditional-mean-zero restriction and solving for the probability-only component $g_2$ yields the space $\Lambda_{2,\mathrm{2st}}$, where the subscript ``2st'' indicates the two-stage model:
\begin{align*}
\Lambda_{2,\mathrm{2st}}
=\Bigg\{&g_1(L)\left[
\left(1-\frac{\delta_{\mathrm{NP}}}{\pi_{\mathrm{NP}}}\right)
-(1-\delta_{\mathrm{NP}})\left(1-\frac{\delta_{\mathrm P}}{\pi_{\mathrm P}}\right)
\right]+g_3(X)\left(1-\frac{\delta_{\mathrm{NP}}}{\pi_{\mathrm{NP}}}\right):
\\[-2pt]
&\quad g_1,g_3\text{ square integrable}\Bigg\}.
\end{align*}
The oracle unrestricted projection onto $\Lambda_{2,\mathrm{2st}}^\perp$ generally involves the unknown non-probability inclusion mechanism. We therefore restrict the general representation by setting $g_1(L)=-g(X)$ and $g_3(X)=g(X)$. This restriction yields the space
\begin{equation*}
\widetilde\Lambda_{2,\mathrm{2st}}
=
\left\{
(1-\delta_{\mathrm{NP}})
\left(1-\frac{\delta_{\mathrm P}}{\pi_{\mathrm P}}\right)
g(X)
:
g\text{ square integrable}
\right\}.
\end{equation*}
The restricted projection has a deliberate robustness interpretation. The full augmentation space uses the non-probability inclusion model and can attain the efficiency bound when that model is credible. The subspace $\widetilde\Lambda_{2,\mathrm{2st}}$ removes this dependence and retains only an $X$-based correction for the portion of the population not observed in the non-probability source. Thus the restricted procedure trades some efficiency under the full model for protection against misspecification of the non-probability inclusion model. Proposition~\ref{prop:ra-centering} below shows that the estimating function remains exactly centered under the independent-survey experiment used in Section~\ref{sec:4}, not only under the conceptual two-stage design.

\begin{figure}[htbp]
\centering
\includegraphics[width=0.68\linewidth]{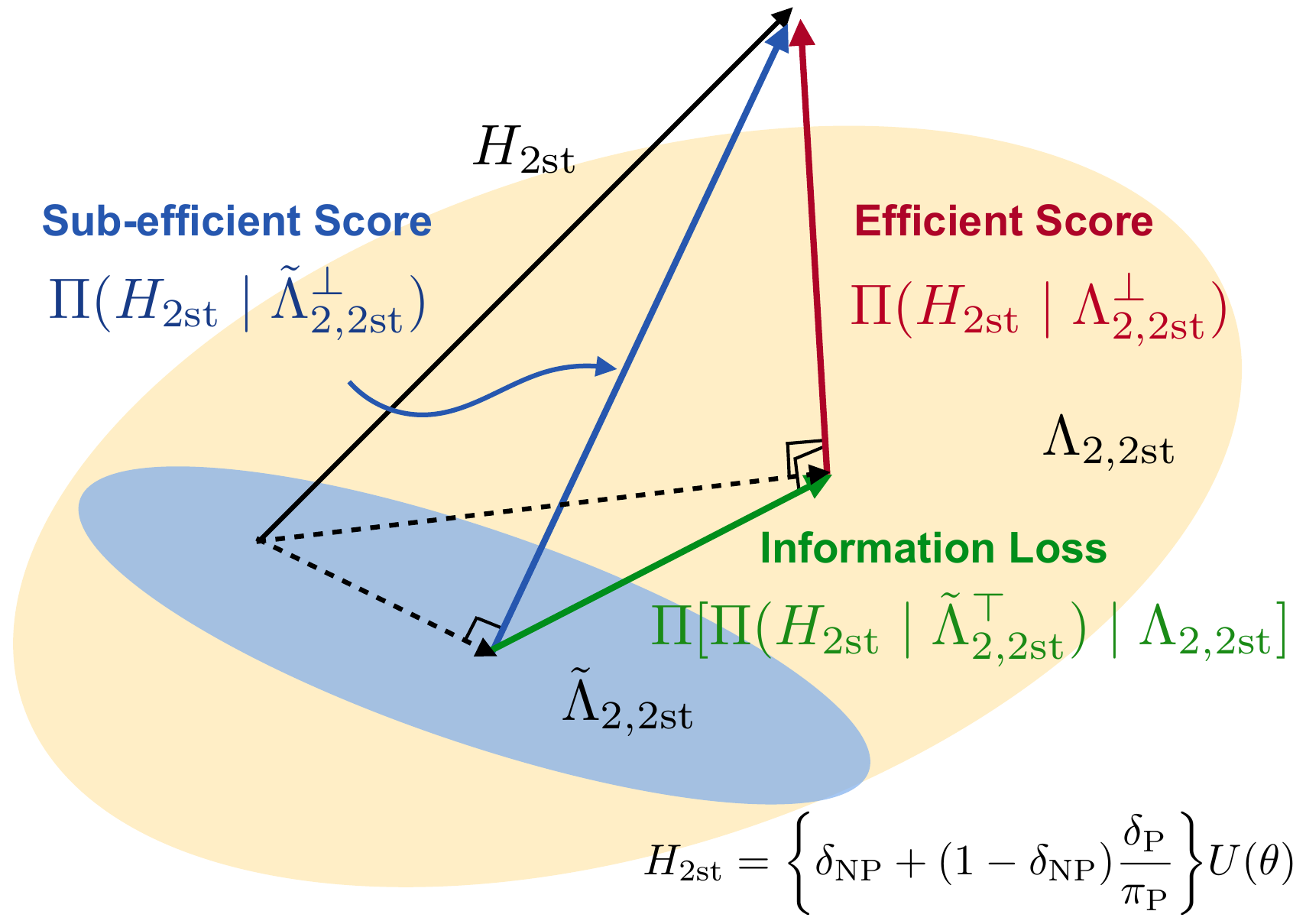}
\caption[Projection geometry in the conceptual two-stage model.]{Projection geometry in the conceptual two-stage model. The oracle full two-stage efficient score and the sub-efficient score are the projections of $H_{\mathrm{2st}}(O;\theta)$ onto $\Lambda_{2,\mathrm{2st}}^{\perp}$ and $\widetilde\Lambda_{2,\mathrm{2st}}^{\perp}$, respectively; the difference is the component of the sub-efficient score lying in the full augmentation space.}
\label{fig:geometry}
\end{figure}

Under the conceptual two-stage model, both projections start from the unbiased observed-data representation
$H_{\mathrm{2st}}(O;\theta)=\delta_{\mathrm{NP}}U(\theta;L)+(1-\delta_{\mathrm{NP}})\delta_{\mathrm P}U(\theta;L)/\pi_{\mathrm P}$. As in the derivation of the efficient score in Section \ref{sec:4}, projecting $H_{\mathrm{2st}}$ onto the orthogonal complement of the restricted augmentation space yields the efficient score within this restricted space. We refer to this score as the \emph{sub-efficient score}. A geometric illustration is provided in Figure~\ref{fig:geometry}.

\subsection{Sub-efficient estimating function and efficiency ordering}

\begin{theorem}[Sub-efficient estimating function]
\label{thm:subefficient-score}
The restricted class is
\begin{align}
\widetilde S_{\mathrm{eff}}(O;\theta,\gamma)
={}&\delta_{\mathrm{NP}}U(\theta;L)
+(1-\delta_{\mathrm{NP}})\left\{\frac{\delta_{\mathrm P}}{\pi_{\mathrm P}}U(\theta;L)
+\left(1-\frac{\delta_{\mathrm P}}{\pi_{\mathrm P}}\right)\gamma(X)\right\}.
\label{eq:subeff-score}
\end{align}
The optimal $\gamma$-function within the restricted class is
\begin{equation*}
\gamma^*(X)
=\frac{E[\{1-\pi_{\mathrm{NP}}(L)\}O_{\mathrm P}U(\theta_0;L)\mid X]}
{E[\{1-\pi_{\mathrm{NP}}(L)\}O_{\mathrm P}\mid X]}.
\end{equation*}
\end{theorem}

For the population mean, let $U(\theta;L)=\theta-Y$ and write $\gamma_{\theta}(X)=\theta-m(X)$. If $k(i)$ denotes the fold containing unit $i$, then solving~\eqref{eq:subeff-score} with a cross-fitted working regression $\widehat m^{[-k(i)]}$ gives
\begin{equation*}
\widehat\theta_{\mathrm{sub}}
=\frac1N\sum_{i=1}^N\left[
\delta_{\mathrm{NP},i}Y_i
+(1-\delta_{\mathrm{NP},i})\left\{
\frac{\delta_{\mathrm P,i}}{\pi_{\mathrm P,i}}Y_i
+\left(1-\frac{\delta_{\mathrm P,i}}{\pi_{\mathrm P,i}}\right)
\widehat m^{[-k(i)]}(X_i)
\right\}
\right].
\end{equation*}
If both inclusion mechanisms are noninformative given $X$, then
$\gamma^*(X)=E\{U(\theta_0;L)\mid X\}=\theta_0-E(Y\mid X)$, so $m^*(X)=E(Y\mid X)$, which is the augmentation function used by \citet{kimDataIntegrationCombining2021}.

The restricted projection yielding $\gamma^*$ is given in Supplementary Section S.2, and its observable weighted-regression construction is given in Supplementary Section S.3.
We use the label \emph{Eff-sub} for estimators based on~\eqref{eq:subeff-score}.

\begin{proposition}[Exact centering under independent surveys]
\label{prop:ra-centering}
Under~\eqref{eq:sampling-law}, for every square-integrable $\gamma$,
\begin{equation*}
E\{\widetilde S_{\mathrm{eff}}(O;\theta_0,\gamma)\mid L,\pi_{\mathrm P}\}=U(\theta_0;L).
\end{equation*}
Thus the working augmentation affects variance but not consistency.
\end{proposition}

Let $\gamma_{\mathrm P}^*(X)=E\{O_{\mathrm P}U(\theta_0;L)\mid X\}/E(O_{\mathrm P}\mid X)$. Here $\widehat\theta_{\mathrm P,\gamma_{\mathrm P}^*}$ denotes the optimally augmented probability-sample-only estimator and $\widehat\theta_{\mathrm P,0}$ denotes its unaugmented counterpart.

\begin{proposition}[Efficiency ordering]
\label{prop:ra-dominance}
The full efficient, oracle sub-efficient, optimally augmented probability-sample-only, and unaugmented probability-sample-only estimators satisfy
\[
V_{\theta,\mathrm{eff}}
\preceq\AVar(\widehat\theta_{\mathrm{sub},\gamma^*})
\preceq\AVar(\widehat\theta_{\mathrm P,\gamma_{\mathrm P}^*})
\preceq\AVar(\widehat\theta_{\mathrm P,0}).
\]
\end{proposition}

The common-augmentation variance identity and the proof of this ordering are given in Supplementary Section S.2. A feasible implementation of Eff-sub, including the weighted-regression estimator of $\gamma$ and its cross-fitted asymptotic expansion, is given in Supplementary Section S.3.

\section{Simulation Study}
\label{sec:6}

We evaluate the estimators under the general experiment in Section~\ref{sec:4}. Each setting uses $N=10{,}000$, 500 replications, and two cross-fitting folds ($K=2$). The primary target is the superpopulation mean $\theta_0=0$; finite-population errors are reported in the Supplementary Material. The design inclusion probability $\pi_{\mathrm P}$ has residual variation conditional on $L$ in all main scenarios.

Let $X\sim N(0,1)$, $Z\sim\operatorname{Bernoulli}(0.5)$, and let the outcome error be independent of $(X,Z)$. We label the three outcome mechanisms O1--O3:
\begin{align*}
\mathrm{O1}:\quad &Y=-\exp(-2)+\cos(2X)+0.5X+\epsilon,
&&\epsilon\sim N(0,0.25),\\
\mathrm{O2}:\quad &Y=0.8X+\epsilon,
&&\epsilon\sim N(0,0.10^2),\\
\mathrm{O3}:\quad &Y=0.2+0.8X-0.4Z+\epsilon,
&&\epsilon\sim N(0,0.25).
\end{align*}
We label the two non-probability inclusion mechanisms M1 and M2, where ``M'' stands for mechanism:
\begin{align*}
\mathrm{M1}:\quad
&\operatorname{logit}\{\pi_{\mathrm{NP}}(L)\}
=-2.15-0.5X-0.75Y,\\
\mathrm{M2}:\quad
&\operatorname{logit}\{\pi_{\mathrm{NP}}(L)\}
=-1.5-0.3\cos(2Y)-0.1(Y-1)^2.
\end{align*}
For the probability source, define
\begin{equation*}
\bar\pi_{\mathrm P}(X)
=0.005+\expit\{-3-0.25(X-2)^2\},
\end{equation*}
and generate a bounded beta perturbation of this conditional mean as specified in Supplementary Section S.4. The two sample inclusion indicators are then generated independently conditional on $(L,\pi_{\mathrm P})$.

\subsection{Design and methods}

We combine the outcome and non-probability inclusion mechanisms into four scenarios, labeled S1--S4, summarized in Table~\ref{tab:scenarios}. Exact data-generating equations, the distribution of $\pi_{\mathrm P}\mid L$, and all tuning parameters are in Supplementary Section S.4.

\begin{table}[H]
\centering
\caption{Main simulation scenarios.}
\label{tab:scenarios}
\begin{tabular}{@{}>{\raggedright\arraybackslash}p{0.10\linewidth}>{\raggedright\arraybackslash}p{0.10\linewidth}>{\raggedright\arraybackslash}p{0.23\linewidth}>{\raggedright\arraybackslash}p{0.44\linewidth}@{}}
\toprule
Scenario & Outcome & \shortstack[l]{Non-probability\\ inclusion mechanism} & Main feature\\

\midrule
S1 & O1 & M1 & Correct informative non-probability inclusion model\\
S2 & O2 & M1 & Near-collinearity and weak validation information\\
S3 & O3 & M2 & Misspecified fitted logistic non-probability inclusion model\\
S4 & O1 & M1 & Coarsened overlap information with one record per population unit\\
\bottomrule
\end{tabular}
\end{table}

P and NP denote the estimators in~\eqref{eq:ipw_prob} and~\eqref{eq:ipw_nonprob}, respectively; P-AUG is the augmented probability-sample estimator. CLW is the pseudo-likelihood weighting benchmark of \citet{chenDoublyRobustInference2020}, and CLW+P is the minimum-variance combination of CLW and P. Eff-sub, Eff-sieve, and Eff-DML are defined in Sections~\ref{sec:4}--\ref{sec:5}. Oracle-Eff uses independently trained augmentation functions and serves only as an efficiency diagnostic. Additional diagnostic methods reported only in the Supplementary Material are defined in Supplementary Table~S1.

The primary spline rule is used for the main table and figure. A richer rule, prespecified before examining the new results and evaluated with the same Monte Carlo seeds, is reported as a sensitivity analysis in the Supplementary Material; the rule is not selected post hoc by comparing RMSEs.

We report bias, empirical standard deviation (SD), root mean squared error (RMSE), mean estimated standard error (SE), and coverage relative to $\theta_0$. Failure rates, Monte Carlo standard errors, finite-population errors, conditional inclusion parameter estimates, and mechanism-equation and validation-Jacobian diagnostics are in the Supplementary Material.

The expected probability-sample size is approximately 250 in the main scenarios, while the non-probability source is substantially larger. This imbalance makes the validation regression and the estimation of conditional moments involving $\pi_{\mathrm P}$ the main finite-sample bottlenecks.

\subsection{Main results}

\begin{figure}[!htbp]
\centering
\includegraphics[width=0.62\linewidth]{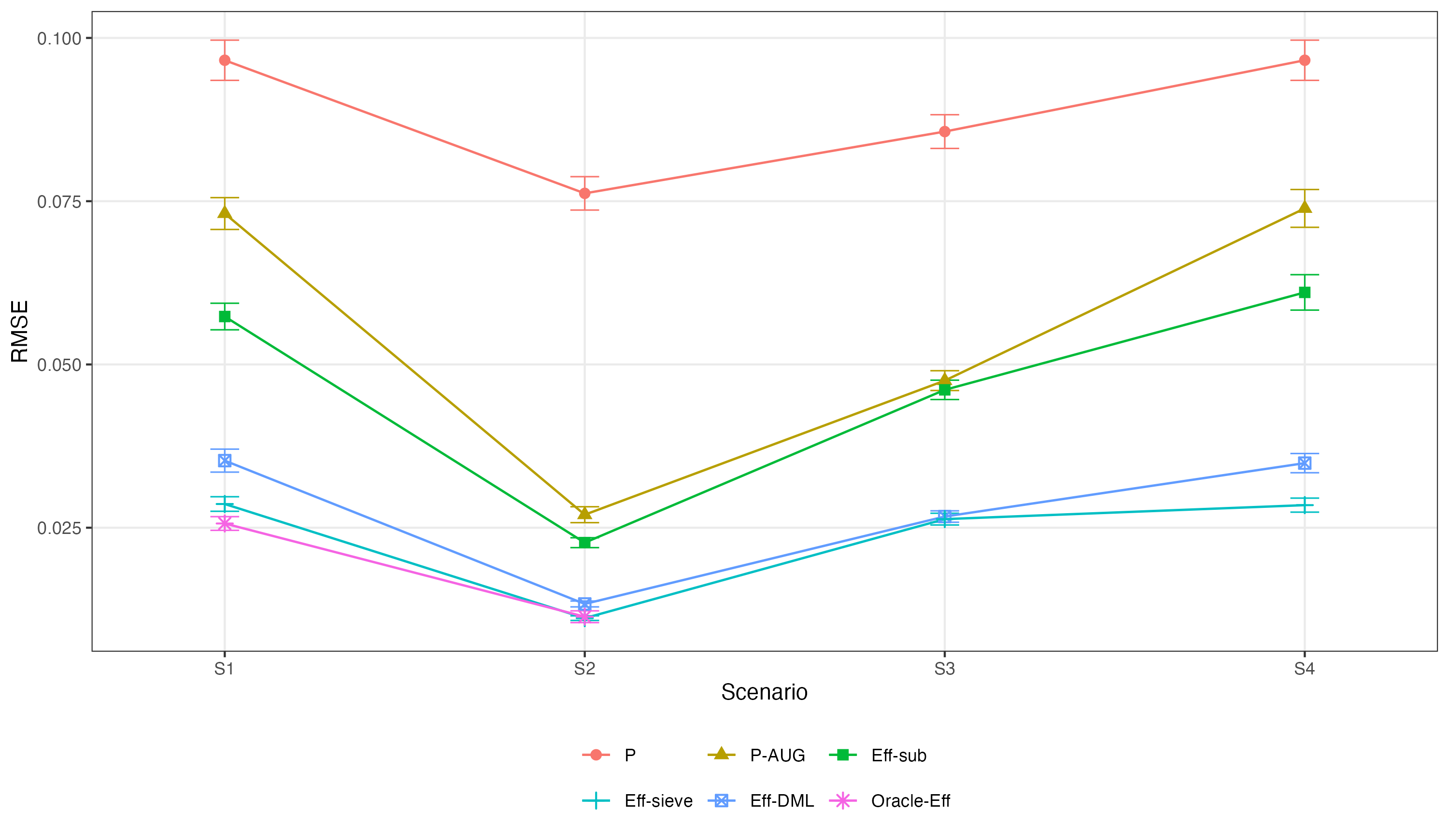}
\caption[RMSE for the primary estimators in Scenarios S1--S4. Error bars show Monte Carlo standard errors. Oracle-Eff is reported only when the non-probability inclusion model is correctly specified.]{RMSE for the primary estimators in Scenarios S1--S4. Error bars show Monte Carlo standard errors. Oracle-Eff is reported only when the non-probability inclusion model is correctly specified.}
\label{fig:sim-updated}
\end{figure}

Table~\ref{tab:simulation-performance} and Figure~\ref{fig:sim-updated} summarize the primary methods. In S1, where the fitted non-probability inclusion model is correct, Eff-sieve and Eff-DML reduce RMSE from $0.097$ for P to $0.029$ and $0.035$, while Eff-sub reduces it to $0.057$. The two efficient implementations have small bias and coverage of $0.932$ and $0.966$; their RMSEs are close to the independently trained Oracle-Eff benchmark of $0.026$.

S2 separates weak identification of a component of the conditional inclusion parameter from estimation of the target. The validation regressors are nearly collinear, and the outcome component of the conditional inclusion parameter is imprecise, yet the target RMSE is $0.011$ for Eff-sieve and $0.013$ for Eff-DML, compared with $0.076$ for P, because $X$ nearly determines $Y$. P-AUG and Eff-sub have RMSEs of $0.027$ and $0.023$. A continuous weak-identification sequence, including singular-value and condition-number diagnostics, is reported in the Supplementary Material.

In S3 the fitted logistic non-probability inclusion model is deliberately misspecified. Eff-sieve and Eff-DML have small negative biases of $-0.006$ and RMSEs of $0.026$ and $0.027$, reductions of approximately 69\% relative to P. This is empirical robustness in this particular mechanism, not a consequence of exact augmentation invariance, which does not protect against misspecification of $\pi_{\mathrm{NP}}(L;\phi)$. S4 coarsens overlap information and lies outside the identification theorem; Eff-sieve and Eff-DML have RMSEs of $0.028$ and $0.035$, but these results are interpreted only as sensitivity evidence. Under the richer prespecified spline rule, RMSEs decrease further in S1 and S4 and change by at most $0.002$ in S2 and S3, leaving the qualitative conclusions unchanged.

Pure propensity benchmarks are much less stable when selection depends on the outcome. NP has substantially larger variance, and CLW produces extreme weights under its covariate-only ignorability model. CLW and CLW+P are evaluated in every scenario, but their complete results are reported in the Supplementary Material and omitted from Table~\ref{tab:simulation-performance} and Figure~\ref{fig:sim-updated} because the extreme CLW values would compress the scale of the primary comparisons.

\begin{longtable}{@{}llrrrrr@{}}
\caption{Selected Monte Carlo results in Scenarios S1--S4. Each scenario uses 500 replications. SD denotes empirical standard deviation, RMSE denotes root mean squared error, SE denotes estimated standard error, and Cov.\ denotes coverage. Full results and Monte Carlo standard errors are in the Supplementary Material.}
\label{tab:simulation-performance}\\
\toprule
Scenario & Method & Bias & SD & RMSE & SE & Cov. \\
\midrule
\endfirsthead
\multicolumn{7}{c}{Table \thetable\ continued}\\
\toprule
Scenario & Method & Bias & SD & RMSE & SE & Cov. \\
\midrule
\endhead
\midrule
\multicolumn{7}{r}{Continued on next page}\\
\endfoot
\bottomrule
\endlastfoot
S1 & P          &  0.006 & 0.096 & 0.097 & 0.096 & 0.920 \\
S1 & P-AUG      &  0.006 & 0.073 & 0.073 & 0.068 & 0.906 \\
S1 & Eff-sub    &  0.003 & 0.057 & 0.057 & 0.054 & 0.936 \\
S1 & Eff-sieve  &  0.001 & 0.029 & 0.029 & 0.026 & 0.932 \\
S1 & Eff-DML    &  0.005 & 0.035 & 0.035 & 0.042 & 0.966 \\
S1 & Oracle-Eff &  0.006 & 0.025 & 0.026 & 0.021 & 0.924 \\
\addlinespace
S2 & P          & -0.003 & 0.076 & 0.076 & 0.075 & 0.946 \\
S2 & P-AUG      &  0.000 & 0.027 & 0.027 & 0.025 & 0.916 \\
S2 & Eff-sub    &  0.000 & 0.023 & 0.023 & 0.021 & 0.926 \\
S2 & Eff-sieve  &  0.000 & 0.011 & 0.011 & 0.011 & 0.938 \\
S2 & Eff-DML    &  0.001 & 0.013 & 0.013 & 0.018 & 0.962 \\
S2 & Oracle-Eff &  0.000 & 0.011 & 0.011 & 0.010 & 0.942 \\
\addlinespace
S3 & P          &  0.000 & 0.086 & 0.086 & 0.085 & 0.940 \\
S3 & P-AUG      &  0.003 & 0.047 & 0.048 & 0.048 & 0.936 \\
S3 & Eff-sub    &  0.003 & 0.046 & 0.046 & 0.045 & 0.932 \\
S3 & Eff-sieve  & -0.006 & 0.026 & 0.026 & 0.024 & 0.930 \\
S3 & Eff-DML    & -0.006 & 0.026 & 0.027 & 0.029 & 0.956 \\
\addlinespace
S4 & P          &  0.006 & 0.096 & 0.097 & 0.096 & 0.920 \\
S4 & P-AUG      &  0.006 & 0.074 & 0.074 & 0.069 & 0.918 \\
S4 & Eff-sub    & -0.010 & 0.060 & 0.061 & 0.062 & 0.934 \\
S4 & Eff-sieve  & -0.004 & 0.028 & 0.028 & 0.027 & 0.936 \\
S4 & Eff-DML    &  0.000 & 0.035 & 0.035 & 0.048 & 0.944 \\
\end{longtable}

\subsection{Heterogeneity in the Probability Sampling Design}

Holding the S1 outcome and non-probability inclusion mechanisms fixed, we increase the conditional variation of $\pi_{\mathrm P}\mid L$ while preserving its conditional mean. As the concentration parameter of the beta distribution for $\pi_{\mathrm P}\mid L$ decreases from $100$ to $5$, the RMSE of P rises from $0.099$ to $0.133$, whereas the corresponding RMSEs rise from $0.028$ to $0.039$ for Eff-sieve, from $0.035$ to $0.042$ for Eff-DML, and from $0.055$ to $0.074$ for Eff-sub. Thus the gains persist under substantial heterogeneity in the probability sampling design. Eff-sieve has one diagnosed local-profile failure among 500 replications at the most heterogeneous setting after the default start and ten prespecified reinitializations; the corresponding summary uses the 499 finite fits. The rich-rule sensitivity analysis has no failures and yields the same qualitative conclusion. Complete results are reported in the Supplementary Material.

\section{A Real Data Finite-Population Experiment}
\label{sec:7}

We use the public Culture and Community in a Time of Crisis (CCTC) data as a data-based finite-population experiment \citep{benoitbryanMulrow2020}. The cleaned Frame~1 records, which define the full target frame, form a fixed population of $N=113{,}549$. The target is the finite-population proportion for \texttt{q7\_22}, a binary indicator of attendance at a classical music event. Repeated probability and non-probability samples are analyzed, so the exercise evaluates repeated-sampling performance rather than presenting a single substantive estimate.

In each replication, the probability sample contains $1{,}000$ units with known inclusion probabilities and the non-probability source contains $4{,}000$ units drawn from an undercoverage frame through a complex mechanism unknown to the analyst. The analysis retains the frame covariates and both sample inclusion indicators for all population records.

For the non-probability inclusion model we set
\[
V(L)=\{1,Y,\mathrm{age},\mathrm{METRO},\mathrm{REGION9},\mathrm{race}\}^{\mathsf T},
\qquad \operatorname{logit}\{\pi_{\mathrm{NP}}(L;\phi)\}=\phi^{\mathsf T}V(L).
\]
Here \texttt{METRO} is the metropolitan-area indicator, \texttt{REGION9} is the nine-category region indicator, and \texttt{race} is represented by indicators. After preprocessing, $X$ contains 112 fully observed frame predictors. The $L$-conditional augmentations use $(X,Y)$, whereas the $X$-conditional augmentations use $X$ and can be evaluated for all frame units. Because a tensor-product nonparametric basis is not practical with 112 frame predictors, the CCTC augmentation regressions use an additive linear basis with ridge coefficient $10^{-2}$; this application-specific regularization is distinct from the growing-spline rules used in the simulations.

We report Eff-sieve, Eff-DML, Eff-sub, and the benchmarks P, P-AUG, NP, CLW, CLW+P, and NP-mean, where NP-mean is the unweighted non-probability-sample mean. Complete preprocessing, solver, and weight diagnostics are in the Supplementary Material.

Correct specification of the low-dimensional $\pi_{\mathrm{NP}}(L;\phi)$ model is required for Eff-sieve, Eff-DML, and NP, whereas misspecification of the richer working augmentation models does not shift the population root. Because the archived probability samples may depart from independent Poisson sampling and the non-probability source arises from an unknown undercoverage mechanism, the CCTC experiment is interpreted as a robustness study rather than a direct verification of the efficiency bound. We therefore report the Eff-sub procedure alongside the full efficient procedures.

\subsection{Results}

The fixed population proportion is $0.4359$. Table~\ref{tab:cctc-results} shows three patterns. First, P-AUG and Eff-sub are stable and nearly unbiased. P has RMSE $0.0179$, whereas P-AUG and Eff-sub have RMSEs $0.0151$ and $0.0152$, respectively, an improvement of approximately 15\%. Their coverage probabilities are $0.974$ and $0.964$; the unaugmented P interval is conservative.

Second, methods relying on the working non-probability inclusion model are sensitive in this experiment. Eff-sieve and Eff-DML have biases of $-0.045$ and $-0.048$ and RMSEs of $0.051$ and $0.055$. Their similar centering indicates that the discrepancy is not primarily caused by cross-fitting. NP is considerably more unstable, with 150 of the 500 estimates outside $[0,1]$. These findings are consistent with the unknown undercoverage mechanism, for which the low-dimensional non-probability inclusion model and positivity assumptions are less credible than in the simulations. Exact augmentation invariance cannot repair this form of misspecification.

Third, the stabilized low-dimensional CLW implementation is numerically well behaved but remains biased. CLW and CLW+P have biases of $0.021$ and $0.019$, and NP-mean has small sampling variation but bias $0.017$. Thus precision alone does not remove selection bias.

\begin{table}[H]
\centering
\caption{Performance in the CCTC finite-population experiment over 500 replications. SD denotes empirical standard deviation, RMSE denotes root mean squared error, and SE denotes estimated standard error. ``Outside $[0,1]$'' counts estimates outside the parameter space.}
\label{tab:cctc-results}
\begin{tabular}{@{}lrrrrrr@{}}
\toprule
Method & Bias & SD & RMSE & Mean SE & Coverage & \shortstack{Outside\\$[0,1]$} \\
\midrule
P & -0.001 & 0.018 & 0.018 & 0.025 & 0.986 & 0 \\
P-AUG & -0.001 & 0.015 & 0.015 & 0.016 & 0.974 & 0 \\
Eff-sub & -0.001 & 0.015 & 0.015 & 0.016 & 0.964 & 0 \\
Eff-sieve & -0.045 & 0.023 & 0.051 & 0.009 & 0.116 & 0 \\
Eff-DML & -0.048 & 0.028 & 0.055 & 0.014 & 0.252 & 0 \\
CLW & 0.021 & 0.008 & 0.022 & 0.008 & 0.284 & 0 \\
CLW+P & 0.019 & 0.008 & 0.020 & 0.008 & 0.342 & 0 \\
NP & 2.187 & 11.005 & 11.209 & 0.779 & 0.718 & 150 \\
NP-mean & 0.017 & 0.008 & 0.018 & 0.008 & 0.466 & 0 \\
\bottomrule
\end{tabular}
\end{table}

\section{Discussion}
\label{sec:discussion}

The proposed estimators are subject to several limitations. The finite-dimensional non-probability inclusion model must be correctly specified for the full efficient estimator, the probability weights must be valid, and the two sample inclusion indicators must be available for each population unit for the validation argument. Exact augmentation invariance does not repair violations of these assumptions.

Two extensions are especially relevant. First, the finite-dimensional model $\pi_{\mathrm{NP}}(L;\phi)$ could be replaced by a semiparametric inclusion-probability model with an unspecified covariate component, in the spirit of \citet{shettyRobustEstimationSemiparametric2025}; developing identification and efficiency theory for such a model under dual-frame validation would relax reliance on a fully parametric inclusion model. Second, it would be useful to accommodate the common data-integration setting in which the probability sample provides $X$ and design weights but not $Y$ \citep{chenDoublyRobustInference2020,kimMassImputation2021}. Because the present validation argument uses $Y$ observed in the probability sample, this extension would require alternative identifying restrictions or additional outcome information. Cost-aware allocation between the probability and non-probability sources is another practically important extension.

\bibliography{references}

\clearpage
\phantomsection
\begin{center}
{\LARGE\bf Supplementary Material}\\[0.7em]
{\large Semiparametric Efficient Data Integration Using the Dual-Frame Sampling Framework}
\end{center}
\vspace{0.7em}

% Reset and prefix all supplementary counters while keeping hyperlinks unique.
\setcounter{section}{0}
\setcounter{subsection}{0}
\setcounter{equation}{0}
\setcounter{table}{0}
\setcounter{figure}{0}
\setcounter{theorem}{0}
\setcounter{lemma}{0}
\setcounter{corollary}{0}
\setcounter{proposition}{0}
\setcounter{assumption}{0}
\setcounter{definition}{0}
\setcounter{remark}{0}
\setcounter{example}{0}
\renewcommand{\thesection}{S\arabic{section}}
\renewcommand{\thesubsection}{S\arabic{section}.\arabic{subsection}}
\numberwithin{equation}{section}
\renewcommand{\thetable}{S\arabic{table}}
\renewcommand{\thefigure}{S\arabic{figure}}
\renewcommand{\thetheorem}{S\arabic{theorem}}
\renewcommand{\theproposition}{S\arabic{proposition}}
\renewcommand{\thelemma}{S\arabic{lemma}}
\renewcommand{\thecorollary}{S\arabic{corollary}}
\renewcommand{\theassumption}{S\arabic{assumption}}
\renewcommand{\thedefinition}{S\arabic{definition}}
\renewcommand{\theremark}{S\arabic{remark}}
\renewcommand{\theexample}{S\arabic{example}}
\makeatletter
\renewcommand{\theHsection}{supp.\arabic{section}}
\renewcommand{\theHsubsection}{supp.\arabic{section}.\arabic{subsection}}
\renewcommand{\theHequation}{supp.\arabic{section}.\arabic{equation}}
\renewcommand{\theHtable}{supp.\arabic{table}}
\renewcommand{\theHfigure}{supp.\arabic{figure}}
\@ifundefined{theHtheorem}{}{\renewcommand{\theHtheorem}{supp.\arabic{theorem}}}
\@ifundefined{theHlemma}{}{\renewcommand{\theHlemma}{supp.\arabic{lemma}}}
\@ifundefined{theHcorollary}{}{\renewcommand{\theHcorollary}{supp.\arabic{corollary}}}
\@ifundefined{theHproposition}{}{\renewcommand{\theHproposition}{supp.\arabic{proposition}}}
\@ifundefined{theHassumption}{}{\renewcommand{\theHassumption}{supp.\arabic{assumption}}}
\@ifundefined{theHdefinition}{}{\renewcommand{\theHdefinition}{supp.\arabic{definition}}}
\@ifundefined{theHremark}{}{\renewcommand{\theHremark}{supp.\arabic{remark}}}
\@ifundefined{theHexample}{}{\renewcommand{\theHexample}{supp.\arabic{example}}}
\makeatother
\spacingset{1.15}

\section{Notation, regularity conditions, and validation-sample identification}
\label{supp:regularity}

\subsection{Notation and regularity conditions}

This supplement uses the notation of the main paper. In particular, $L=(X^{\trans},Y)^{\trans}$ denotes the complete study variables, where $X$ is the vector of frame covariates and $Y$ is the outcome. The symbols P and NP denote the probability and non-probability sources, respectively. Let $P_0$ denote the true observed-data law and write $P_0f=\E(f)$. A subscript $0$ on a parameter-dependent quantity denotes evaluation at the true finite-dimensional parameter values; in particular, $U_0=U(\theta_0;L)$. For vectors, $a^{\otimes2}=aa^{\trans}$; for symmetric matrices, $A\preceq B$ means that $B-A$ is positive semidefinite. The empirical average and empirical process are
\[
\mathbb P_N f=N^{-1}\sum_{i=1}^N f(O_i),
\qquad
\mathbb G_N=\sqrt N(\mathbb P_N-P_0),
\]
and $o_{\Prob}(1)$ denotes convergence to zero in probability.

Let
\[
\xi=(\theta^{\trans},\phi^{\trans})^{\trans},
\qquad
\xi_0=(\theta_0^{\trans},\phi_0^{\trans})^{\trans},
\qquad
\zeta=(k_1,k_2,\eta,\kappa_1,\kappa).
\]
Here $\zeta$ denotes a generic collection of the five augmentation functions in Section 4 of the main paper, and $S(O;\xi,\zeta)$ denotes the corresponding stacked estimating function after the optimal augmentation functions are replaced by $\zeta$. We write $\widehat\zeta_{\mathrm{sieve}}(\xi)$ for the full-sample fitted augmentation map and $\widehat\zeta^{[-k]}(\xi)$ for the map fitted without fold $k$.

Write
\[
q_\phi(L,\pi_{\Pp})
=\pi_{\NP}(L;\phi)+\pi_{\Pp}
-\pi_{\NP}(L;\phi)\pi_{\Pp},
\qquad
O_{\Pp}=\frac{1-\pi_{\Pp}}{\pi_{\Pp}},
\]
and, at the truth,
\[
\pi_0(L)=\pi_{\NP}(L;\phi_0),
\qquad
q_0(L,\pi_{\Pp})=q_{\phi_0}(L,\pi_{\Pp}).
\]
The conditional law of $\pi_{\Pp}\mid L$ is unrestricted throughout the main model. Because the design inclusion probability $\pi_{\Pp}$ is observed only in the probability sample, the full-data likelihood uses the pair $(L,\pi_{\Pp})$. Accordingly, full-data conditional expectations below condition on this pair, while the target estimating function remains $U(\theta;L)$. The unit-level observed data are
\[
O=(X,\delta_{\NP},\delta_{\Pp},
\delta_{\union}Y,\delta_{\Pp}\pi_{\Pp}),
\qquad
\delta_{\union}
=\delta_{\NP}+\delta_{\Pp}-\delta_{\NP}\delta_{\Pp}.
\]
All expectations are under the true joint law.

Let $\mathcal H_O=L^2_0(P_0)$ be the Hilbert space of scalar mean-zero, square-integrable observed-data functions with inner product $\langle a,b\rangle=\E(ab)$. For a closed linear subspace $\mathcal A\subseteq\mathcal H_O$, write $\Pi(h\mid\mathcal A)$ for the orthogonal projection of $h$ onto $\mathcal A$ and $\mathcal A^\perp$ for its orthogonal complement. For a vector $h=(h_1,\ldots,h_d)^{\trans}$ whose components belong to $\mathcal H_O$, define $\lVert h\rVert_{P_0,2}^2=\sum_{j=1}^d P_0\{h_j^2\}$; vector projections are taken componentwise. Every $h\in\mathcal H_O$ can be written in the pattern form
\begin{align}
h(O)
={}&
\delta_{\NP}\delta_{\Pp}h_{11}(L,\pi_{\Pp})
+(1-\delta_{\NP})\delta_{\Pp}h_{01}(L,\pi_{\Pp})
\nonumber\\
&+
\delta_{\NP}(1-\delta_{\Pp})h_{10}(L)
-(1-\delta_{\NP})(1-\delta_{\Pp})h_{00}(X).
\label{eq:s-pattern}
\end{align}
The conditional-mean-zero augmentation space for the independent-survey model is
\begin{equation*}
\Lambda_2
=
\{h\in\mathcal H_O:\E(h\mid L,\pi_{\Pp})=0\}.
\end{equation*}
It is the orthogonal complement of the observed-data tangent space generated by the unrestricted law of $(L,\pi_{\Pp})$. Section~\ref{supp:efficiency} uses this space to derive the efficient and restricted projections.

We use the following regularity conditions in addition to the sampling, positivity, and identification conditions in the main paper.

\begin{assumption}[Moment, positivity, and Jacobian conditions]
\label{ass:s-moments}
For some $c>0$, the two projected components used to construct the efficient influence functions (EIFs) for $\theta$ and $\phi$ have finite $(2+c)$ moments. The conditional-moment denominators defining $k_1^*$, $k_2^*$, $\eta^*$, $\kappa_1^*$, and $\kappa^*$ are bounded away from zero. The joint Jacobians appearing in the main paper are nonsingular, and their empirical analogues are uniformly nonsingular with probability tending to one in a neighborhood of the truth.
\end{assumption}

\begin{assumption}[Same-sample sieve regularity]
\label{ass:s-sieve}
For the full-sample sieve estimator, there is a deterministic limit
$\zeta^\dagger=(k_1^\dagger,k_2^\dagger,
\eta^\dagger,\kappa_1^\dagger,\kappa^\dagger)$ such that the fitted joint estimating function converges in $L^2(P_0)$ to $S^\dagger(O)=S(O;\xi_0,\zeta^\dagger)$ and
\[
\mathbb G_N\left[
S\{O;\xi_0,\widehat\zeta_{\mathrm{sieve}}(\xi_0)\}
-S^\dagger(O)
\right]=o_{\Prob}(1),
\]
componentwise. The empirical Jacobian converges uniformly to the corresponding nonsingular limit. In addition, the squared score-difference class obeys a uniform law of large numbers over the same local neighborhood. Thus, for the consistent local root considered below,
\[
\mathbb P_N\left\|
S\{O;\widehat\xi_{\mathrm{sieve}},
\widehat\zeta_{\mathrm{sieve}}(\widehat\xi_{\mathrm{sieve}})\}
-S(O;\widehat\xi_{\mathrm{sieve}},\zeta^\dagger)
\right\|^2=o_{\Prob}(1).
\]
\end{assumption}

Assumption~\ref{ass:s-sieve} is the empirical-process form of the same-sample stochastic-equicontinuity condition stated without empirical-process notation in the main paper. Primitive sufficient conditions are given in Section~\ref{supp:sieve-tuning}. Under the entropy and bounded- or sub-Weibull-envelope conditions stated there, the balanced sieve choice $J_N\asymp n_{\Pp,\mathrm{tr}}^{d/(2s+d)}$, with $n_{\Pp,\mathrm{tr}}\asymp N$, verifies both the stochastic-equicontinuity and empirical second-moment requirements, up to logarithmic factors, when $s>d/2$. Assumption~\ref{ass:s-sieve} is not imposed on the cross-fitted estimator.

\begin{assumption}[Cross-fitted learner regularity]
\label{ass:s-crossfit}
For a fixed number of folds with $|I_k|/N\to\rho_k\in(0,1)$, there is a deterministic limit $\zeta^\dagger$ such that, for every fold $k$,
\[
\left\|
S\{\cdot;\xi_0,\widehat\zeta^{[-k]}(\xi_0)\}
-S(\cdot;\xi_0,\zeta^\dagger)
\right\|_{P_0,2}=o_{\Prob}(1).
\]
The components of the fitted estimating functions have uniformly bounded $(2+c)$ moments for some $c>0$, and the pooled empirical Jacobian converges to the nonsingular Jacobian of $S(\cdot;\xi_0,\zeta^\dagger)$.
\end{assumption}

Neither Assumption~\ref{ass:s-sieve} nor Assumption~\ref{ass:s-crossfit} imposes a product-rate condition on the augmentation estimators. Exact augmentation invariance removes the population drift. Same-sample estimation still requires empirical-process control, whereas cross-fitting instead requires the fitted joint estimating function to converge in probability in $L^2(P_0)$.

\subsection{Validation identification}
\label{supp:identification}

Let $V(L)$ be the regressor vector in the logistic non-probability
inclusion model and let $\Phi$ be its compact convex parameter space. Write
\begin{align*}
\pi_0(L)&=\pi_{\NP}(L;\phi_0),
&
\pi_\phi(L)&=\expit\{\phi^{\trans}V(L)\},\\
\bar\pi_{\Pp}(L)&=\E(\pi_{\Pp}\mid L),
&
\dot\pi_\phi(L)&=\frac{\partial\pi_\phi(L)}{\partial\phi}.
\end{align*}
For the logistic model, the validation score in the main paper reduces to
\begin{equation*}
S_{\mathrm{val}}(O;\phi)
=\delta_{\Pp}\{\delta_{\NP}-\pi_\phi(L)\}V(L).
\end{equation*}

\begin{lemma}[Population validation map]
\label{lem:s-validation-map}
Under the conditional-independence sampling law,
\begin{equation}
G_{\mathrm{val}}(\phi)
:=\E\{S_{\mathrm{val}}(O;\phi)\}
=\E\left[
\bar\pi_{\Pp}(L)\{\pi_0(L)-\pi_\phi(L)\}V(L)
\right].
\label{eq:s-validation-map}
\end{equation}
\end{lemma}

\begin{proof}
Conditional on $(L,\pi_{\Pp})$, the two sample inclusion indicators are independent and
\[
\E(\delta_{\Pp}\mid L,\pi_{\Pp})=\pi_{\Pp},
\qquad
\E(\delta_{\NP}\mid L,\pi_{\Pp})=\pi_0(L).
\]
Therefore,
\begin{align*}
\E\{S_{\mathrm{val}}(O;\phi)\mid L,\pi_{\Pp}\}
&=\pi_{\Pp}\{\pi_0(L)-\pi_\phi(L)\}V(L).
\end{align*}
Taking expectation over $\pi_{\Pp}\mid L$ gives~\eqref{eq:s-validation-map}.
\end{proof}

\begin{lemma}[Uniformly negative validation Jacobian]
\label{lem:s-validation-jacobian}
The Jacobian of~\eqref{eq:s-validation-map} is
\begin{equation}
\partial_{\phi^{\trans}}G_{\mathrm{val}}(\phi)
=-\E\left[
\bar\pi_{\Pp}(L)
\pi_\phi(L)\{1-\pi_\phi(L)\}
V(L)V(L)^{\trans}
\right].
\label{eq:s-validation-jacobian}
\end{equation}
Under the validation-information condition in the main paper,
\[
a^{\trans}\partial_{\phi^{\trans}}G_{\mathrm{val}}(\phi)a
\leq-c_{\mathrm{val}}\|a\|^2
\]
uniformly over $\phi\in\Phi$ and $a\in\mathbb R^{\dim(\phi)}$, where
$c_{\mathrm{val}}>0$ is the constant in the validation-information condition.
\end{lemma}

\begin{proof}
For the logistic non-probability inclusion model,
\[
\dot\pi_\phi(L)
=\pi_\phi(L)\{1-\pi_\phi(L)\}V(L).
\]
Differentiating~\eqref{eq:s-validation-map} gives~\eqref{eq:s-validation-jacobian}; the quadratic-form bound is exactly the validation-information condition in the main paper.
\end{proof}

\begin{proof}[Proof of the validation-identification theorem in the main paper]
Since $G_{\mathrm{val}}(\phi_0)=0$ and $\Phi$ is convex, the segment
$\phi_0+t(\phi-\phi_0)$ remains in $\Phi$ for $0\leq t\leq1$. Hence
\begin{align*}
(\phi-\phi_0)^{\trans}G_{\mathrm{val}}(\phi)
&=\int_0^1
(\phi-\phi_0)^{\trans}
\partial_{\phi^{\trans}}G_{\mathrm{val}}
\{\phi_0+t(\phi-\phi_0)\}
(\phi-\phi_0)\,dt\\
&\leq-c_{\mathrm{val}}\|\phi-\phi_0\|^2.
\end{align*}
If $G_{\mathrm{val}}(\phi)=0$, the left side is zero, so the inequality implies $\phi=\phi_0$.
\end{proof}

\subsection{Preliminary estimator and local information}

Let
\[
\widehat G_{\mathrm{val},N}(\phi)
=\mathbb P_N S_{\mathrm{val}}(O;\phi).
\]
Under a uniform law of large numbers over the compact parameter set, every approximate root $\widetilde\phi$ satisfying
$\|\widehat G_{\mathrm{val},N}(\widetilde\phi)\|=o_{\Prob}(1)$ is consistent. Indeed, if a subsequence remained at least $\epsilon$ away from $\phi_0$, strong monotonicity would imply
\[
\|G_{\mathrm{val}}(\widetilde\phi)\|
\geq c_{\mathrm{val}}\epsilon,
\]
contradicting uniform convergence and the approximate-root condition.

If, in addition, $\widetilde\phi$ is an exact root or satisfies
$\|\widehat G_{\mathrm{val},N}(\widetilde\phi)\|=o_{\Prob}(N^{-1/2})$, a Taylor expansion around $\phi_0$ yields
\[
\sqrt N(\widetilde\phi-\phi_0)
=-J_{\mathrm{val},0}^{-1}
\frac1{\sqrt N}\sum_{i=1}^N
S_{\mathrm{val}}(O_i;\phi_0)+o_{\Prob}(1),
\]
where
\[
J_{\mathrm{val},0}
=-\E\left[
\bar\pi_{\Pp}(L)\pi_0(L)\{1-\pi_0(L)\}
V(L)V(L)^{\trans}
\right].
\]
The sample analogue is
\[
\widehat J_{\mathrm{val}}(\phi)
=-\mathbb P_N\left[
\delta_{\Pp}\pi_\phi(L)\{1-\pi_\phi(L)\}
V(L)V(L)^{\trans}
\right].
\]
Its smallest singular value is a local conditioning diagnostic. A small value warns that some linear combination of the components of the conditional inclusion parameter is weakly informed by the validation sample. It is not a test of model specification: a well-conditioned Jacobian does not imply that the logistic non-probability inclusion model is correct.

The validation equation also supplies the globally identified preliminary estimate used to initialize the local profiled procedures. Starting from $\widetilde\phi$ and the accompanying preliminary target estimate, the implementation refits the augmentation functions and takes damped local Newton updates of the joint estimating function. Under the conditions in Sections~\ref{supp:regularity} and~\ref{supp:dml}, any approximate root that converges locally in the shrinking neighborhood selected by this initializer has the expansion stated in the main paper. No global root, or assumption about remote roots, is required.

\section{Projection arguments for Sections 4 and 5}
\label{supp:efficiency}

\subsection{The Morikawa--Kim reduction for Z-estimands}
\label{supp:projection-principle}

Let $p$ and $q$ be the dimensions of $\theta$ and $\phi$, respectively, and order the joint parameter as $(\theta^{\trans},\phi^{\trans})^{\trans}$. Let $\Lambda^F$ be the nuisance tangent space for the law of $(L,\pi_{\Pp})$ with $\theta$ fixed. Its orthogonal complement is taken in the Hilbert space of scalar mean-zero, square-integrable functions of $(L,\pi_{\Pp})$. Recall
\[
\Lambda_2=\{a\in\mathcal H_O:\E(a\mid L,\pi_{\Pp})=0\}.
\]
For a vector-valued full-data estimating function $D^*(L,\pi_{\Pp})$, let $u(D^*)(O)$ denote an observed-data representation satisfying
\begin{equation}
\E\{u(D^*)(O)\mid L,\pi_{\Pp}\}=D^*(L,\pi_{\Pp}).
\label{eq:s-general-unbiased-representation}
\end{equation}
Let $S_\phi^{\mathrm{obs}}$ be the observed-data likelihood score for $\phi$ with the law of $(L,\pi_{\Pp})$ held fixed, and write
\[
S_{\phi,\Lambda}=\Pi(S_\phi^{\mathrm{obs}}\mid\Lambda_2).
\]

\begin{lemma}[Morikawa--Kim observed-data representation]
\label{lem:s-mk-representation}
Under the conditions of Proposition A1 of \citet{rotnitzkyANALYSISSEMIPARAMETRICREGRESSION1997}, in the form stated in Lemma B.1 of \citet{morikawaSemiparametricOptimalEstimation2021a}, let $D_{\mathrm{eff}}^*$ be the full-data direction whose components belong to $(\Lambda^F)^\perp$. Then the untransformed joint efficient score can be written as
\begin{equation}
S_{\mathrm{eff}}^{\mathrm{raw}}
=u(D_{\mathrm{eff}}^*)
-\Pi\{u(D_{\mathrm{eff}}^*)\mid\Lambda_2\}
+\begin{pmatrix}0_p\\S_{\phi,\Lambda}\end{pmatrix},
\label{eq:s-general-efficient-representation}
\end{equation}
where $0_p$ is the $p$-dimensional zero vector.
\end{lemma}

\begin{proof}
This is Proposition A1 of \citet{rotnitzkyANALYSISSEMIPARAMETRICREGRESSION1997}, restated in the notation of Lemma B.1 of \citet{morikawaSemiparametricOptimalEstimation2021a} and with the joint parameter ordered as $(\theta^{\trans},\phi^{\trans})^{\trans}$.
\end{proof}

The observed-data representability needed in Lemma~\ref{lem:s-mk-representation} is explicit in the independent-survey experiment. When $\delta_{\Pp}=1$, the pair $(L,\pi_{\Pp})$ is observed. Hence, for any scalar or vector full-data function $h(L,\pi_{\Pp})$ with square-integrable components,
\[
\E\left\{\frac{\delta_{\Pp}}{\pi_{\Pp}}h(L,\pi_{\Pp})\,\middle|\,L,\pi_{\Pp}\right\}
=h(L,\pi_{\Pp}).
\]
If $\pi_{\Pp}\geq\epsilon_{\Pp}>0$ almost surely, then
\[
\E\left\|\frac{\delta_{\Pp}}{\pi_{\Pp}}h(L,\pi_{\Pp})\right\|^2
=\E\left\{\frac{\|h(L,\pi_{\Pp})\|^2}{\pi_{\Pp}}\right\}
\leq\epsilon_{\Pp}^{-1}\E\|h(L,\pi_{\Pp})\|^2.
\]
Thus the required observed-data representation is square integrable under the positivity condition used in the main paper. The remaining finite-dimensional conditions used in the application are the validation-information and nonsingular-Jacobian conditions already stated in the main paper and Section~\ref{supp:regularity}.

The proof of Lemma 4.1 of \citet{morikawaSemiparametricOptimalEstimation2021a} uses a further reduction specific to Z-estimation. The common target-direction component in the mechanism block is removed by a nonsingular constant transformation, leaving one projection onto $\Lambda_2^\perp$ and one projection onto $\Lambda_2$. The next theorem states this reduction explicitly.

\begin{theorem}[Projection characterization for Z-estimands]
\label{thm:s-z-projection}
Assume the setting of Lemma~\ref{lem:s-mk-representation}. Let
\[
U_0=U(\theta_0;L),\qquad \E(U_0)=0,
\]
where $U_0\in\mathbb R^p$ is square integrable and the law of
$(L,\pi_{\Pp})$ is otherwise unrestricted. Suppose that, for some
constant matrix $C$,
\[
D_{\mathrm{eff}}^*=C U_0,
\qquad
C=
\begin{pmatrix}
C_\theta\\
C_\phi
\end{pmatrix},
\qquad
C_\theta\in\mathbb R^{p\times p},\quad
C_\phi\in\mathbb R^{q\times p},\quad
\operatorname{rank}(C_\theta)=p.
\]
Let $H_U(O)$ be a $p$-dimensional square-integrable observed-data
function satisfying
\begin{equation}
\E(H_U\mid L,\pi_{\Pp})=U_0.
\label{eq:s-z-HU}
\end{equation}
Define
\begin{align}
S_\theta
&=\Pi(H_U\mid\Lambda_2^\perp)
=H_U-\Pi(H_U\mid\Lambda_2),
\label{eq:s-z-target-projection}\\
S_\phi
&=\Pi(S_\phi^{\mathrm{obs}}\mid\Lambda_2)
=S_{\phi,\Lambda}.
\label{eq:s-z-mechanism-projection}
\end{align}
After a nonsingular constant transformation of the joint efficient
score,
\[
S_{\mathrm{eff}}(O)
=
\begin{pmatrix}
S_\theta(O)\\
S_\phi(O)
\end{pmatrix}
\]
is an equivalent joint efficient estimating basis for
$\xi=(\theta^{\trans},\phi^{\trans})^{\trans}$. Writing the
corresponding estimating functions away from the truth as
$S_{\mathrm{eff}}(O;\xi)$, if
\[
J_0
=
\left.
\E\left\{
\frac{\partial S_{\mathrm{eff}}(O;\xi)}
{\partial\xi^{\trans}}
\right\}
\right|_{\xi=\xi_0}
\]
is nonsingular, then
\[
\varphi_{\xi,\mathrm{eff}}(O)
=
-J_0^{-1}S_{\mathrm{eff}}(O;\xi_0)
\]
is the joint efficient influence function; its $\theta$- and
$\phi$-blocks are the efficient influence functions for $\theta$ and
$\phi$, respectively. If the mechanism is known, the $\phi$-block is
omitted.

The function $S_\theta$ does not depend on the choice of $H_U$
satisfying~\eqref{eq:s-z-HU} and is the unique minimum-norm
observed-data representation of $U_0$. Moreover, for a fixed $H_U$ and
any closed linear subspace
$\widetilde\Lambda_2\subseteq\Lambda_2$,
\begin{equation}
S_{\theta,\mathrm{res}}
=
H_U-\Pi(H_U\mid\widetilde\Lambda_2)
\label{eq:s-z-restricted-projection}
\end{equation}
is the unique minimum-norm element among $H_U+a$ with every component
of $a$ in $\widetilde\Lambda_2$.
\end{theorem}

\begin{proof}
Since $\operatorname{rank}(C_\theta)=p$, the square matrix $C_\theta$
is nonsingular. By~\eqref{eq:s-z-HU}, $C H_U$ is an observed-data
representation of $C U_0=D_{\mathrm{eff}}^*$. Substitution into
Lemma~\ref{lem:s-mk-representation} and linearity of orthogonal
projection give
\begin{equation}
S_{\mathrm{eff}}^{\mathrm{raw}}
=
\begin{pmatrix}
C_\theta S_\theta\\
C_\phi S_\theta+S_\phi
\end{pmatrix}.
\label{eq:s-z-raw-phi-decomposition}
\end{equation}
Let $I_q$ denote the $q\times q$ identity matrix. The constant matrix
\[
M=
\begin{pmatrix}
C_\theta^{-1}&0\\
-C_\phi C_\theta^{-1}&I_q
\end{pmatrix}
\]
is nonsingular and satisfies
\begin{equation}
M S_{\mathrm{eff}}^{\mathrm{raw}}
=
\begin{pmatrix}
S_\theta\\
S_\phi
\end{pmatrix}.
\label{eq:s-z-block-transformation}
\end{equation}
Thus the common target-direction component $C_\phi S_\theta$ in the
untransformed mechanism block is removed, exactly as in the proof of
Lemma 4.1 of
\citet{morikawaSemiparametricOptimalEstimation2021a}. If
$J_{\mathrm{raw}}$ is the joint Jacobian of
$S_{\mathrm{eff}}^{\mathrm{raw}}$, then the Jacobian of the transformed
basis is $J_0=MJ_{\mathrm{raw}}$, and hence
\[
-J_0^{-1}M S_{\mathrm{eff}}^{\mathrm{raw}}
=
-J_{\mathrm{raw}}^{-1}S_{\mathrm{eff}}^{\mathrm{raw}}.
\]
The nonsingular transformation therefore leaves the normalized
influence function and the efficiency bound unchanged, proving the
efficient-basis and joint-EIF statements.

If two matrices represent the same full-data direction, their
difference multiplied by $H_U$ belongs componentwise to $\Lambda_2$;
after projection, their difference multiplied by $S_\theta$ belongs
to both $\Lambda_2$ and $\Lambda_2^\perp$ and is therefore zero. Thus
the projected basis is independent of the chosen representation
matrix. Likewise, any two functions satisfying~\eqref{eq:s-z-HU}
differ by a vector whose components belong to $\Lambda_2$, so their
projections onto $\Lambda_2^\perp$ agree.

Finally, the affine class of observed-data representations of $U_0$
is $H_U+\Lambda_2$ componentwise. The projection theorem shows that
$H_U-\Pi(H_U\mid\Lambda_2)$ is its unique minimum-norm element.
Replacing $\Lambda_2$ by the closed subspace
$\widetilde\Lambda_2$ gives
\eqref{eq:s-z-restricted-projection} and the restricted minimum-norm
claim.
\end{proof}

In both applications below, the target-specific step is to construct an observed-data function $H_U$ satisfying \eqref{eq:s-z-HU}. Section~\ref{supp:section4-projection} computes the full projection, and Section~\ref{supp:ra} computes its restricted analogue.

\subsection{Application to Section 4: independent surveys}
\label{supp:section4-projection}

For the model in Section 4 of the main paper, write
\[
\pi=\pi_{\NP}(L;\phi),
\qquad
\dot\pi=\frac{\partial\pi_{\NP}(L;\phi)}{\partial\phi}.
\]
The functions $q_\phi(L,\pi_{\Pp})$ and $O_{\Pp}$ were defined in Section~\ref{supp:regularity}. Within this subsection, their arguments are suppressed when no ambiguity arises.
The inverse-probability estimating function
\begin{equation}
H_{\mathrm{ind}}(O;\theta)
=\frac{\delta_{\Pp}}{\pi_{\Pp}}U(\theta;L)
\label{eq:s-Hind}
\end{equation}
satisfies $\E(H_{\mathrm{ind}}\mid L,\pi_{\Pp})=U(\theta;L)$ for every $\theta$. At $\theta_0$, the positivity condition in the main paper and square integrability of $U_0$ give
\[
\E\{\lVert H_{\mathrm{ind}}(O;\theta_0)\rVert^2\}
=\E\left(\frac{\lVert U_0\rVert^2}{\pi_{\Pp}}\right)<\infty.
\]
Thus $H_{\mathrm{ind}}(O;\theta_0)$ is the square-integrable observed-data representation required by Theorem~\ref{thm:s-z-projection}. The theorem gives the efficient target block by projecting~\eqref{eq:s-Hind} onto $\Lambda_2^\perp$ at $\theta_0$; a generic $\theta$ is retained below to display the resulting estimating function.

Define
\begin{align*}
A&=\delta_{\Pp}\left(1-\frac{\delta_{\NP}}{\pi}\right),
&
B&=\delta_{\NP}\left(1-\frac{\delta_{\Pp}}{\pi_{\Pp}}\right),\\
T&=1-\frac{\delta_{\NP}\delta_{\Pp}}{\pi\pi_{\Pp}}-A-B.
\end{align*}
The notation $A,B,T$ is used only in this subsection. For any square-integrable function $k_2(L)$, define
\begin{align}
R_\phi\{O;k_2\}
={}&1-\frac{\delta_{\NP}\delta_{\Pp}}{\pi\pi_{\Pp}}
-\delta_{\NP}\left(1-\frac{\delta_{\Pp}}{\pi_{\Pp}}\right)\{1+k_2(L)\}
\nonumber\\[-2pt]
&-\delta_{\Pp}\left(1-\frac{\delta_{\NP}}\pi\right)
\left[1+O_{\Pp}\{1-\pi-\pi k_2(L)\}\right].
\label{eq:s-Rphi}
\end{align}

\begin{lemma}[Pattern-space representation]
\label{lem:s-general-tangent}
For the pattern representation in~\eqref{eq:s-pattern},
\begin{equation}
\Lambda_2
=
\left\{
 h_2(L,\pi_{\Pp})A+h_3(L)B-h_4(X)T\;\middle|\;
 h_2,h_3,h_4\text{ square integrable}
\right\}.
\label{eq:s-general-tangent}
\end{equation}
\end{lemma}

\begin{proof}
The restriction $\E(h\mid L,\pi_{\Pp})=0$ is
\[
\pi\pi_{\Pp}h_{11}+(1-\pi)\pi_{\Pp}h_{01}
+\pi(1-\pi_{\Pp})h_{10}-(1-\pi)(1-\pi_{\Pp})h_{00}=0.
\]
Solving for $h_{11}$ and regrouping the four pattern indicators gives
\eqref{eq:s-general-tangent}. Conversely,
$\E(A\mid L,\pi_{\Pp})=\E(B\mid L,\pi_{\Pp})=\E(T\mid L,\pi_{\Pp})=0$.
\end{proof}

For a generic $h$ in~\eqref{eq:s-pattern}, write
\[
h^*=\Pi(h\mid\Lambda_2)
=h_2^*(L,\pi_{\Pp})A+h_3^*(L)B-h_4^*(X)T.
\]
The normal equations are
\begin{equation}
\E[\{h-h^*\}\{a(L,\pi_{\Pp})A+b(L)B-c(X)T\}]=0
\label{eq:s-general-normal}
\end{equation}
for every square-integrable $a,b,c$. The required conditional moments are
\begin{align}
\E(A^2\mid L,\pi_{\Pp})&=\pi_{\Pp}(\pi^{-1}-1),
&
\E(B^2\mid L,\pi_{\Pp})&=\pi(\pi_{\Pp}^{-1}-1),
\nonumber\\
\E(AB\mid L,\pi_{\Pp})&=(1-\pi)(1-\pi_{\Pp}),
&
\E(AT\mid L,\pi_{\Pp})&=(1-\pi)(1-\pi_{\Pp})(\pi^{-1}-1),
\nonumber\\
\E(BT\mid L,\pi_{\Pp})&=(1-\pi)(1-\pi_{\Pp})(\pi_{\Pp}^{-1}-1).
\label{eq:s-general-pattern-moments}
\end{align}

Define
\begin{align}
 k_{1,h}(L)
 &=\frac{
 \E[(1-\pi_{\Pp})\{(h_{11}-h_{10})
 -(1-\pi)(h_{11}-h_{01})\}\mid L]
 }{
 \E\{O_{\Pp}q_\phi(L,\pi_{\Pp})\mid L\}},
\label{eq:s-k1h}\\
 k_2^*(L)
 &=\frac{
 \E[O_{\Pp}\{1-q_\phi(L,\pi_{\Pp})\}\mid L]
 }{
 \E\{O_{\Pp}q_\phi(L,\pi_{\Pp})\mid L\}},
\label{eq:s-k2-general}\\
 \eta_h(X)
 &=\frac{
 \E[(1-\pi)(1-\pi_{\Pp})
 \{(h_{11}+h_{00})-(1-\pi)(h_{11}-h_{01})
 -\pi O_{\Pp}k_{1,h}(L)\}\mid X]
 }{
 \E[(1-\pi)(1-\pi_{\Pp})
 \{1+O_{\Pp}[1-\pi\{1+k_2^*(L)\}]\}\mid X]}.
\label{eq:s-etah}
\end{align}

\begin{lemma}[Projection coefficients under independent surveys]
\label{lem:s-general-projection}
The solution of~\eqref{eq:s-general-normal} is
\begin{align}
 h_3^*(L)
 &=-k_{1,h}(L)+k_2^*(L)\eta_h(X),
\label{eq:s-h3star}\\
 h_2^*(L,\pi_{\Pp})
 &=-\pi\{h_{11}-h_{01}-O_{\Pp}k_{1,h}(L)\}
 \nonumber\\[-2pt]
 &\quad+O_{\Pp}\{1-\pi-\pi k_2^*(L)\}\eta_h(X),
\label{eq:s-h2star}\\
 h_4^*(X)&=\eta_h(X).
\label{eq:s-h4star}
\end{align}
\end{lemma}

\begin{proof}
Because $a(L,\pi_{\Pp})$ is arbitrary, the $A$-normal equation holds conditionally on
$(L,\pi_{\Pp})$. Substitution of~\eqref{eq:s-general-pattern-moments} gives
\[
0=-\pi_{\Pp}(1-\pi)h_{11}+\pi_{\Pp}(1-\pi)h_{01}
-h_2^*\pi_{\Pp}(\pi^{-1}-1)
-h_3^*(1-\pi)(1-\pi_{\Pp})
+h_4^*(1-\pi)(1-\pi_{\Pp})(\pi^{-1}-1).
\]
The $B$-normal equation, conditional on $L$, then gives
\eqref{eq:s-h3star} and~\eqref{eq:s-h2star}. Substitution into the $T$-normal
equation conditional on $X$ gives~\eqref{eq:s-etah} and
\eqref{eq:s-h4star}.
\end{proof}

For the target component, take
\[
h_{11}=h_{01}=\frac{U(\theta;L)}{\pi_{\Pp}},
\qquad
h_{10}=h_{00}=0.
\]
Equations~\eqref{eq:s-k1h}--\eqref{eq:s-etah} then reduce to $k_1^*$,
$k_2^*$, and $\eta^*$ in Section 4 of the main paper. Theorem~\ref{thm:s-z-projection} gives
\[
S_{\theta}
=\frac{\delta_{\Pp}}{\pi_{\Pp}}U(\theta;L)
-
\Pi\left\{
\frac{\delta_{\Pp}}{\pi_{\Pp}}U(\theta;L)
\,\middle|\,\Lambda_2
\right\}.
\]
Substitution of~\eqref{eq:s-h3star}--\eqref{eq:s-h4star} yields
\begin{align}
S_{\theta}
={}&
\frac{\delta_{\Pp}}{\pi_{\Pp}}U(\theta;L)
+
\left\{
\delta_{\Pp}(\delta_{\NP}-\pi)O_{\Pp}
+\delta_{\NP}\left(1-\frac{\delta_{\Pp}}{\pi_{\Pp}}\right)
\right\}k_1^*(L)
\nonumber\\[-2pt]
&+\eta^*(X)R_\phi\{O;k_2^*\},
\label{eq:s-general-stheta}
\end{align}
where $R_\phi$ is defined in~\eqref{eq:s-Rphi}. In particular,
\begin{equation}
\E(S_{\theta}\mid L,\pi_{\Pp})=U(\theta;L).
\label{eq:s-general-representation}
\end{equation}
For a probability-sample record, setting $\delta_{\Pp}=1$ gives
\[
S_{\theta}\big|_{\delta_{\Pp}=1}
=\frac{U(\theta;L)}{\pi_{\Pp}}
-\pi O_{\Pp}k_1^*(L)
-\eta^*(X)O_{\Pp}\{1-\pi(1+k_2^*)\},
\]
which is independent of $\delta_{\NP}$. Thus probability-only and overlapping
records with the same $(L,\pi_{\Pp})$ contribute identically to the target equation
for fixed $\phi$ and augmentation functions. The validation equation and the
$\phi$-score still use $\delta_{\NP}$.

The observed-data likelihood score for $\phi$ is
\begin{align*}
S_{\phi}^{\mathrm{obs}}
={}&
\delta_{\NP}\delta_{\Pp}\frac{\dot\pi}{\pi}
-(1-\delta_{\NP})\delta_{\Pp}\frac{\dot\pi}{1-\pi}
\\[-2pt]
&+\delta_{\NP}(1-\delta_{\Pp})\frac{\dot\pi}{\pi}
-(1-\delta_{\NP})(1-\delta_{\Pp})m_{00}(X;\phi),
\end{align*}
where
\[
m_{00}(X;\phi)
=\E\left\{
\frac{\dot\pi}{1-\pi}
\,\middle|\,
\delta_{\NP}=\delta_{\Pp}=0,X
\right\}.
\]
By Theorem~\ref{thm:s-z-projection}, the untransformed mechanism score
generally contains a constant linear combination of $S_\theta$. We work with
the transformed efficient estimating basis in~\eqref{eq:s-z-block-transformation}
and, for notational simplicity, write its mechanism block as
$S_{\phi}=S_{\phi,\Lambda}=\Pi(S_{\phi}^{\mathrm{obs}}\mid\Lambda_2)$.
To evaluate this projection with Lemma~\ref{lem:s-general-projection}, set
\[
h_{11}=h_{10}=\frac{\dot\pi}{\pi},
\qquad
h_{01}=-\frac{\dot\pi}{1-\pi},
\qquad
h_{00}=m_{00}(X;\phi).
\]
Then $k_{1,h}=\kappa_1^*$ and $\eta_h=\kappa^*$, giving
\begin{align}
S_{\phi}
={}&
\delta_{\Pp}(\delta_{\NP}-\pi)
\left\{
\frac{\dot\pi}{\pi(1-\pi)}
-O_{\Pp}\kappa_1^*(L)
\right\}
\nonumber\\[-2pt]
&-\delta_{\NP}\left(1-\frac{\delta_{\Pp}}{\pi_{\Pp}}\right)\kappa_1^*(L)
-\kappa^*(X)R_\phi\{O;k_2^*\}.
\label{eq:s-general-sphi}
\end{align}
It satisfies
\begin{equation}
\E(S_{\phi}\mid L,\pi_{\Pp})=0.
\label{eq:s-general-phi-mean}
\end{equation}
Equations~\eqref{eq:s-general-representation} and
\eqref{eq:s-general-phi-mean} also imply orthogonality to tangent directions
of $f(\pi_{\Pp}\mid L)$.

The validation-information condition in the main paper also guarantees nondegeneracy of the transformed mechanism block. Let
\[
S_{\mathrm{val},0}
=\delta_{\Pp}\{\delta_{\NP}-\pi_0(L)\}V(L),
\qquad
I_{\mathrm{val}}
=\E\{\pi_{\Pp}\pi_0(L)[1-\pi_0(L)]V(L)V(L)^{\trans}\}.
\]
Then $S_{\mathrm{val},0}\in\Lambda_2$. Direct conditioning on $(L,\pi_{\Pp})$ and the four-pattern likelihood score gives
\[
\E\{S_{\mathrm{val},0}(S_{\phi}^{\mathrm{obs}})^{\trans}\}
=\E(S_{\mathrm{val},0}S_{\mathrm{val},0}^{\trans})
=I_{\mathrm{val}}.
\]
Because $S_{\phi}=\Pi(S_{\phi}^{\mathrm{obs}}\mid\Lambda_2)$ and $S_{\mathrm{val},0}\in\Lambda_2$,
\[
\E(S_{\mathrm{val},0}S_{\phi}^{\trans})=I_{\mathrm{val}},
\qquad
\Var(S_{\phi})-I_{\mathrm{val}}
=\E\{(S_{\phi}-S_{\mathrm{val},0})^{\otimes2}\}\succeq0.
\]
The weighted full-rank condition in Section 3 of the main paper makes $I_{\mathrm{val}}$ positive definite.

Let
\[
S_{\eff}
=\begin{pmatrix}S_{\theta}\\S_{\phi}\end{pmatrix},
\qquad
J_0
=\left.
\E\left\{\frac{\partial S_{\eff}(O;\xi)}{\partial\xi^{\trans}}\right\}
\right|_{\xi=\xi_0}.
\]
Theorem~\ref{thm:s-z-projection} implies that the two projected blocks form
an efficient estimating basis for $\xi$. Therefore, when $J_0$ is nonsingular,
the joint EIF is
\begin{equation}
\varphi_{\xi,\eff}(O)=-J_0^{-1}S_{\eff}(O;\xi_0).
\label{eq:s-joint-eif}
\end{equation}
Exact augmentation invariance is used for feasible nuisance estimation in
Section~\ref{supp:dml}; it is not needed to justify this oracle projection.

\subsection{Application to Section 5: restricted two-stage projection}
\label{supp:ra}

Under the conceptual two-stage model in Section 5 of the main paper, an
unbiased observed-data representation of $U(\theta;L)$ is
\begin{equation}
H_{\mathrm{2st}}(O;\theta)
=\delta_{\NP}U(\theta;L)
+(1-\delta_{\NP})\frac{\delta_{\Pp}}{\pi_{\Pp}}U(\theta;L).
\label{eq:s-h2st}
\end{equation}
Indeed, $\E(H_{\mathrm{2st}}\mid L,\pi_{\Pp})=U(\theta;L)$
for every $\theta$. At $\theta_0$,
\[
\E\{\lVert H_{\mathrm{2st}}(O;\theta_0)\rVert^2\}
=\E\left[
\left\{\pi_{\NP}(L)+\frac{1-\pi_{\NP}(L)}{\pi_{\Pp}}\right\}
\lVert U_0\rVert^2
\right]<\infty
\]
under the same positivity and moment conditions. Thus the required observed-data representation is again explicit. The projection statement is evaluated at $\theta_0$, while the generic notation defines the estimating equation. By contrast,
$\E(\delta_{\Pp}U/\pi_{\Pp}\mid L,\pi_{\Pp})=\{1-\pi_{\NP}(L)\}U$ because the
probability sample is drawn only from units outside the non-probability source.

Write $\pi_{\NP}=\pi_{\NP}(L)$ in this subsection. A generic
observed-data function under the two-stage model is
\[
h(O)
=\delta_{\NP}h_1(L)
+(1-\delta_{\NP})\delta_{\Pp}h_2(L,\pi_{\Pp})
+(1-\delta_{\NP})(1-\delta_{\Pp})h_3(X).
\]
Solving $\E(h\mid L,\pi_{\Pp})=0$ for $h_2$ gives
\begin{align}
\Lambda_{2,\mathrm{2st}}
=\Bigg\{&a(L)\left[
\left(1-\frac{\delta_{\NP}}{\pi_{\NP}}\right)
-(1-\delta_{\NP})
\left(1-\frac{\delta_{\Pp}}{\pi_{\Pp}}\right)
\right]
\nonumber\\
&+b(X)\left(1-\frac{\delta_{\NP}}{\pi_{\NP}}\right):
 a,b\text{ square integrable}\Bigg\}.
\label{eq:s-lambda-2st}
\end{align}
Treating the two-stage inclusion law as known or modeled at the oracle level,
Theorem~\ref{thm:s-z-projection} shows that the fully efficient target component is
\[
H_{\mathrm{2st}}
-\Pi(H_{\mathrm{2st}}\mid\Lambda_{2,\mathrm{2st}}).
\]
Because this projection generally depends on the unknown non-probability
inclusion mechanism, Section 5 restricts the augmentation space to
\[
\widetilde\Lambda_{2,\mathrm{2st}}
=
\left\{
(1-\delta_{\NP})
\left(1-\frac{\delta_{\Pp}}{\pi_{\Pp}}\right)c(X):
 c\text{ square integrable}
\right\}.
\]
Every element of the restricted affine class in Theorem~\ref{thm:s-z-projection} can be written as
\begin{align}
B_{\RA}(O;\theta,\gamma)
={}&\delta_{\NP}U(\theta;L)
+(1-\delta_{\NP})
\left[
\frac{\delta_{\Pp}}{\pi_{\Pp}}U(\theta;L)
+
\left(1-\frac{\delta_{\Pp}}{\pi_{\Pp}}\right)\gamma(X)
\right].
\label{eq:s-BRA}
\end{align}
The restricted projection is $B_{\RA}(O;\theta_0,\gamma^*)$, where $\gamma^*$ is the solution of
\[
\E\left[
B_{\RA}(O;\theta_0,\gamma^*)
(1-\delta_{\NP})
\left(1-\frac{\delta_{\Pp}}{\pi_{\Pp}}\right)h(X)
\right]=0
\]
for every square-integrable $h$. Conditioning on $X$ gives
\begin{equation}
\E\left[
\{1-\pi_{\NP}(L)\}O_{\Pp}
\{U(\theta_0;L)-\gamma^*(X)\}
\mid X
\right]=0,
\label{eq:s-gamma-normal}
\end{equation}
and hence
\begin{equation}
\gamma^*(X)
=
\frac{\E[\{1-\pi_{\NP}(L)\}O_{\Pp}U(\theta_0;L)\mid X]}
{\E[\{1-\pi_{\NP}(L)\}O_{\Pp}\mid X]}.
\label{eq:s-gamma-star}
\end{equation}
Thus $B_{\RA}(O;\theta_0,\gamma^*)$ is the restricted projection of the unbiased observed-data representation~\eqref{eq:s-h2st}.

The same estimating function remains centered under the independent-survey
model in Section 4. For every square-integrable $\gamma$,
\begin{equation}
\E\{B_{\RA}(O;\theta,\gamma)\mid L,\pi_{\Pp}\}
=U(\theta;L).
\label{eq:s-BRA-centering}
\end{equation}
Thus the two-stage model motivates the projection, whereas consistency of the
resulting estimating equation does not require the samples to have been
collected sequentially.

\subsection{Efficiency comparisons}
\label{supp:efficiency-comparisons}

Let $V_{\theta,\eff}$ and $V_{\theta,\Pp}$ denote the semiparametric efficiency bounds for $\theta$ in the full dual-frame experiment and the probability-sample-only experiment, respectively. A probability-sample-only procedure can be used in the full experiment by ignoring the non-probability records. Hence the full experiment contains every regular influence function available in the probability-sample-only experiment. Since the full EIF has minimum norm in the larger class,
\begin{equation}
V_{\theta,\eff}\preceq V_{\theta,\Pp}.
\label{eq:s-full-vs-P}
\end{equation}
This proves the efficiency-gain result in Section 4 of the main paper.

For the ordering in Section 5, write $U_0=U(\theta_0;L)$ and define, for a
common working augmentation $\gamma$,
\[
B_{\Pp}(O;\theta_0,\gamma)
=\gamma(X)+\frac{\delta_{\Pp}}{\pi_{\Pp}}\{U_0-\gamma(X)\}.
\]
Under the independent-survey model, both $B_{\Pp}$ and $B_{\RA}$ have
conditional mean $U_0$ given $(L,\pi_{\Pp})$, and
\begin{align}
\Var\{B_{\Pp}(O;\theta_0,\gamma)\mid L,\pi_{\Pp}\}
&=O_{\Pp}\{U_0-\gamma(X)\}^{\otimes2},
\label{eq:s-p-only-condvar}\\
\Var\{B_{\RA}(O;\theta_0,\gamma)\mid L,\pi_{\Pp}\}
&=\{1-\pi_{\NP}(L)\}O_{\Pp}
\{U_0-\gamma(X)\}^{\otimes2}.
\label{eq:s-ra-condvar-full}
\end{align}
The law of total covariance therefore yields
\begin{equation}
\Var\{B_{\Pp}(\gamma)\}-\Var\{B_{\RA}(\gamma)\}
=\E\left[
\pi_{\NP}(L)O_{\Pp}\{U_0-\gamma(X)\}^{\otimes2}
\right]\succeq0.
\label{eq:s-common-gamma-dominance}
\end{equation}
The variance-minimizing probability-sample augmentation is
\[
\gamma_{\Pp}^*(X)
=\frac{\E(O_{\Pp}U_0\mid X)}{\E(O_{\Pp}\mid X)},
\]
whereas the restricted projection gives $\gamma^*$ in
\eqref{eq:s-gamma-star}. Optimality within the two classes and
\eqref{eq:s-common-gamma-dominance} imply
\[
\Var\{B_{\RA}(\gamma^*)\}
\preceq\Var\{B_{\RA}(\gamma_{\Pp}^*)\}
\preceq\Var\{B_{\Pp}(\gamma_{\Pp}^*)\}
\preceq\Var\{B_{\Pp}(0)\}.
\]
Let $\widehat\theta_{\mathrm{sub},\gamma}$ and $\widehat\theta_{\Pp,\gamma}$ denote roots of the empirical estimating equations based on $B_{\RA}(\gamma)$ and $B_{\Pp}(\gamma)$, respectively. All four estimating functions have Jacobian
$A_0=\E\{\partial U(\theta_0;L)/\partial\theta^{\trans}\}$. Congruence by
$A_0^{-1}$ preserves the preceding ordering.

Under the regular asymptotic linearity conditions in Section~\ref{supp:effsub-feasible}, the oracle Eff-sub estimator is also a regular estimator in the full independent-survey model, with influence function
\[
\varphi_{\theta,\mathrm{sub}}
=-A_0^{-1}B_{\RA}(O;\theta_0,\gamma^*).
\]
Let $\varphi_{\theta,\eff}$ be the target block of the full-model EIF. By the defining projection property of the EIF,
\[
\AVar(\widehat\theta_{\mathrm{sub},\gamma^*})-V_{\theta,\eff}
=\E\left[
(\varphi_{\theta,\mathrm{sub}}-\varphi_{\theta,\eff})
(\varphi_{\theta,\mathrm{sub}}-\varphi_{\theta,\eff})^{\trans}
\right]\succeq0.
\]
Together with the preceding common-augmentation comparison, this gives
\[
V_{\theta,\eff}
\preceq\AVar(\widehat\theta_{\mathrm{sub},\gamma^*})
\preceq\AVar(\widehat\theta_{\Pp,\gamma_{\Pp}^*})
\preceq\AVar(\widehat\theta_{\Pp,0}),
\]
which proves the complete ordering in Section 5 of the main paper.

\section{Estimation, implementation, and asymptotic expansions}
\label{supp:dml}

Eff-DML is the pooled DML2 estimator in the double/debiased machine-learning (DML) terminology of \citet{chernozhukov2018double}; DML1, which averages fold-specific parameter estimates, is not used.

Recall that $\zeta=(k_1,k_2,\eta,\kappa_1,\kappa)$ and
\[
S(O;\xi,\zeta)
=
\begin{pmatrix}
S_{\theta}(O;\theta,\phi;k_1,k_2,\eta)\\
S_{\phi}(O;\phi;k_2,\kappa_1,\kappa)
\end{pmatrix}.
\]
This section gives exact augmentation invariance, the same-sample and cross-fitted expansions, and implementation details for the full efficient procedures and Eff-sub.

\subsection{Exact augmentation invariance}

Throughout this subsection, evaluate $\pi_{\NP}(L;\phi)$ at $\phi_0$ and condition on $(L,\pi_{\Pp})$. Conditional independence gives
\begin{align*}
\E(\delta_{\NP}\mid L,\pi_{\Pp})
&=\pi_{\NP}(L;\phi_0),
&
\E(\delta_{\Pp}\mid L,\pi_{\Pp})
&=\pi_{\Pp},
\nonumber\\
\E(\delta_{\NP}\delta_{\Pp}\mid L,\pi_{\Pp})
&=\pi_{\NP}(L;\phi_0)\pi_{\Pp}.
\end{align*}
Consequently,
\begin{align}
&\E\left[
\delta_{\Pp}
\{\delta_{\NP}-\pi_{\NP}(L;\phi_0)\}
\left(\frac1{\pi_{\Pp}}-1\right)
\,\middle|\,L,\pi_{\Pp}
\right]=0,
\label{eq:s-k1-first}\\
&\E\left[
\delta_{\NP}
\left(1-\frac{\delta_{\Pp}}{\pi_{\Pp}}\right)
\,\middle|\,L,\pi_{\Pp}
\right]=0,
\label{eq:s-k1-second}\\
&\E\left[
\delta_{\Pp}
\left\{1-
\frac{\delta_{\NP}}{\pi_{\NP}(L;\phi_0)}
\right\}
\,\middle|\,L,\pi_{\Pp}
\right]=0.
\label{eq:s-residual-third}
\end{align}
The first term of the residual has conditional mean
\[
\E\left[
1-
\frac{\delta_{\NP}\delta_{\Pp}}
{\pi_{\NP}(L;\phi_0)\pi_{\Pp}}
\,\middle|\,L,\pi_{\Pp}
\right]=0.
\]
Equations \eqref{eq:s-k1-second} and \eqref{eq:s-residual-third} show that the remaining two terms also have conditional mean zero for every working $k_2(L)$. Hence
\begin{equation}
\E\left[
R_{\phi_0}\{O;k_2\}
\,\middle|\,L,\pi_{\Pp}
\right]=0
\quad\text{for every }k_2.
\label{eq:s-general-residual-zero}
\end{equation}

Using \eqref{eq:s-k1-first}, \eqref{eq:s-k1-second}, and \eqref{eq:s-general-residual-zero} in the target component gives
\[
\E\{S_{\theta}\mid L,\pi_{\Pp}\}
=
\E\left(
\frac{\delta_{\Pp}}{\pi_{\Pp}}U(\theta_0;L)
\,\middle|\,L,\pi_{\Pp}
\right)
=U(\theta_0;L).
\]
The same identities applied to the $\phi$-component give
\[
\E\{S_{\phi}\mid L,\pi_{\Pp}\}=0.
\]
These equations hold for every square-integrable
$(k_1,k_2,\eta,\kappa_1,\kappa)$, proving the exact-invariance proposition.

Because the population moment is exactly constant over the admissible augmentation class, changing the working augmentation functions creates no population drift, whether or not they equal the optimal projection functions.

\subsection{Weighted sieve construction and tuning}
\label{supp:sieve-tuning}

Arguments of $\pi_{\NP}(L;\phi)$ and $q_\phi(L,\pi_{\Pp})$ are suppressed in this subsection when no ambiguity arises.

The two bounded $L$-level reparameterizations are fitted directly on probability-sample records. The definitions of $k_1^*$ and $k_2^*$ imply
\[
k_1^*(L)=U(\theta;L)\{1+k_2^*(L)\},
\]
because
$\E(O_{\Pp}\mid L)=\E\{O_{\Pp}q_\phi\mid L\}+\E[O_{\Pp}\{1-q_\phi\}\mid L]$.
Moreover,
\[
\frac{\pi_{\NP}}{1-\pi_{\NP}}\frac{1-q_\phi}{q_\phi}
=\frac{\pi_{\NP}(1-\pi_{\Pp})}{q_\phi}\in[0,1],
\]
and therefore
\[
0\leq\frac{\pi_{\NP}(L;\phi)}{1-\pi_{\NP}(L;\phi)}k_2^*(L)\leq1.
\]
Because $q_\phi\geq\pi_{\Pp}$,
\[
0\leq
\frac{\E(1-\pi_{\Pp}\mid L)}{\E\{O_{\Pp}q_\phi\mid L\}}
\leq1.
\]
Thus the bounded reparameterization is
\begin{align*}
0\leq \mu_2^*(L)
&=\frac{\pi_{\NP}(L;\phi)}{1-\pi_{\NP}(L;\phi)}k_2^*(L)\leq1,
&
0\leq \mu_\kappa^*(L)
&=\frac{\E(1-\pi_{\Pp}\mid L)}{\E\{O_{\Pp}q_\phi\mid L\}}\leq1.
\end{align*}
Write $\dot\pi_{\NP}(L;\phi)=\partial\pi_{\NP}(L;\phi)/\partial\phi$. Since $\dot\pi_{\NP}/\pi_{\NP}$ is a function of $L$, the two remaining $L$-level functions satisfy
\begin{equation}
k_1^*(L)=U(\theta;L)\{1+k_2^*(L)\},
\qquad
\kappa_1^*(L)
=-\frac{\dot\pi_{\NP}(L;\phi)}{\pi_{\NP}(L;\phi)}\mu_\kappa^*(L).
\label{eq:s-shape-identities}
\end{equation}

On probability-sample records, define
\[
W_L=\frac{O_{\Pp}q_\phi(L,\pi_{\Pp})}{\pi_{\Pp}},
\qquad
Z_2=\frac{\pi_{\NP}}{1-\pi_{\NP}}\frac{1-q_\phi}{q_\phi},
\qquad
Z_\kappa=\frac{\pi_{\Pp}}{q_\phi}.
\]
Both pseudo-outcomes lie in $[0,1]$. For $r\in\{2,\kappa\}$, let $Z_r$ denote the corresponding pseudo-outcome. Given a basis vector $b_L(L)$, write
\[
m_r(L;\beta_r)=\operatorname{expit}\{b_L(L)^{\trans}\beta_r\}.
\]
Direct conditioning on $(L,\pi_{\Pp})$ gives
\begin{align*}
\E\left[\delta_{\Pp}W_Lb_L(L)\{Z_2-m_2(L;\beta_2)\}\right]
&=\E\left[b_L(L)\E\{O_{\Pp}q_\phi\mid L\}
\{\mu_2^*(L)-m_2(L;\beta_2)\}\right],\\
\E\left[\delta_{\Pp}W_Lb_L(L)\{Z_\kappa-m_\kappa(L;\beta_\kappa)\}\right]
&=\E\left[b_L(L)\E\{O_{\Pp}q_\phi\mid L\}
\{\mu_\kappa^*(L)-m_\kappa(L;\beta_\kappa)\}\right].
\end{align*}
Thus the unpenalized population scores vanish when the fitted means equal the two bounded reparameterizations, up to sieve approximation.

The implementation uses the corresponding penalized sample equations. Let $\mathcal I_{\mathrm{tr}}$ denote the current full-sample or fold-specific training indices and set
\[
n_{\Pp,\mathrm{tr}}=\sum_{i\in\mathcal I_{\mathrm{tr}}}\delta_{\Pp,i}.
\]
Let $J_L$ be the dimension of $b_L(L)$ and $\lambda_{L,n}\geq0$ the ridge coefficient. If $W_{L,i}^{\mathrm c}$ denotes $W_{L,i}$ after the prespecified upper-tail cap, define
\[
\widetilde W_{L,i}
=\frac{W_{L,i}^{\mathrm c}}
{n_{\Pp,\mathrm{tr}}^{-1}
 \sum_{j\in\mathcal I_{\mathrm{tr}}}\delta_{\Pp,j}W_{L,j}^{\mathrm c}}.
\]
Let $B_L$ have rows $b_L(L_i)^{\trans}$ for $i\in\mathcal I_{\mathrm{tr}}$ with $\delta_{\Pp,i}=1$, and let $\widetilde{\mathcal W}_L$ be diagonal with entries $\widetilde W_{L,i}$ for these records. Define
\[
D_L=\operatorname{diag}(0,1,\ldots,1),
\qquad
a_L=\max\left\{J_L^{-1}\operatorname{tr}
\left(B_L^{\trans}\widetilde{\mathcal W}_L B_L\right),1\right\}.
\]
For each $r\in\{2,\kappa\}$, $\widehat\beta_r$ solves
\[
\sum_{i\in\mathcal I_{\mathrm{tr}}}
\delta_{\Pp,i}\widetilde W_{L,i}b_L(L_i)
\{Z_{r,i}-m_r(L_i;\beta_r)\}
-\lambda_{L,n}a_L D_L\beta_r=0.
\]
Equivalently, it minimizes the weighted fractional-logit objective
\[
\sum_{i\in\mathcal I_{\mathrm{tr}}}
\delta_{\Pp,i}\widetilde W_{L,i}
\left[\log\{1+\exp(b_L(L_i)^{\trans}\beta_r)\}
-Z_{r,i}b_L(L_i)^{\trans}\beta_r\right]
+\frac{\lambda_{L,n}a_L}{2}\beta_r^{\trans}D_L\beta_r.
\]
This is a fractional-response quasi-likelihood: $Z_r$ need not be Bernoulli, and the logistic link enforces $m_r(L;\beta_r)\in(0,1)$. The ridge does not penalize the intercept and uses $\lambda_{L,n}=n_{\Pp,\mathrm{tr}}^{-3/4}$ in the controlled simulations. We recover $k_2$ from $\widehat\mu_2=m_2(\cdot;\widehat\beta_2)$, reconstruct $k_1$ and $\kappa_1$ from~\eqref{eq:s-shape-identities}, and estimate the remaining real-valued $X$-conditional functions $\eta$ and $\kappa$ through weighted linear normal equations. Thus no separately fitted denominator is divided into a separately fitted numerator.

Let $\operatorname{vec}$ stack the columns of a matrix, and let $B_K(t)$ denote a $K$-column natural cubic spline basis constructed after $t$ is centered and scaled using the relevant training sample. The same training-sample center, scale, knots, and boundary knots are used for prediction. Let $K_L$ and $K_X$ denote the numbers of one-dimensional spline columns used in the $L$- and frame-covariate regressions. In the controlled simulations, $X$ denotes the scalar continuous covariate and $Z$ the binary covariate, so the general frame-covariate vector is $(X,Z)^{\trans}$ and the complete-data vector is $(X,Z,Y)^{\trans}$. The $L$-level basis is
\begin{equation}
b_{L,K_L}(L)
=\left[1,Z,B_{K_L}(X)^{\trans},B_{K_L}(Y)^{\trans},
\operatorname{vec}\{B_{K_L}(X)B_{K_L}(Y)^{\trans}\}^{\trans}\right]^{\trans},
\label{eq:s-L-spline-basis}
\end{equation}
with total dimension $J_L=2+2K_L+K_L^2$. The frame-covariate basis is
\begin{equation}
b_{X,K_X}(X,Z)
=\left[1,Z,B_{K_X}(X)^{\trans},ZB_{K_X}(X)^{\trans}\right]^{\trans},
\label{eq:s-X-spline-basis}
\end{equation}
with total dimension $J_X=2+2K_X$. Thus the $L$-level regression is nonparametric in the two continuous variables and includes their full tensor interaction, while the frame-level regression is a varying-coefficient spline across the two values of $Z$.

For smoothness index $s$ and continuous dimension $d$, the requested number of one-dimensional spline columns is
\begin{equation}
K_{\mathrm{req}}(n_{\Pp,\mathrm{tr}},d)
=\max\left\{K_{\min},
\left\lceil c\,n_{\Pp,\mathrm{tr}}^{1/(2s+d)}\right\rceil\right\}.
\label{eq:s-spline-K-rule}
\end{equation}
We set $K_L=K_{\mathrm{req}}(n_{\Pp,\mathrm{tr}},2)$ and $K_X=K_{\mathrm{req}}(n_{\Pp,\mathrm{tr}},1)$ before applying feasibility caps. The primary rule fixes $s=2$, $c=1$, and $K_{\min}=2$, with each total basis dimension $J\in\{J_L,J_X\}$ capped at $\lfloor n_{\Pp,\mathrm{tr}}/4\rfloor$. The rich sensitivity rule fixes $s=2$, $c=1.5$, and $K_{\min}=3$, with each $J\in\{J_L,J_X\}$ capped at $\lfloor n_{\Pp,\mathrm{tr}}/3\rfloor$. A requested dimension is reduced only when the total-term cap or the number of distinct training values makes it infeasible.

Realized dimensions may therefore vary with the training count and support, but neither rule is selected using residual-based fit criteria or Monte Carlo performance. The augmented probability-sample estimator (P-AUG) and Eff-sub use the frame-covariate basis; the same-sample efficient estimator (Eff-sieve) uses the full probability sample; Eff-DML selects dimensions separately within each training fold; and the oracle efficient benchmark (Oracle-Eff) applies the same rule to its larger independent training population. The primary and rich rules were designated, respectively, for the main analysis and sensitivity analysis before the new results were inspected.

The default ridge coefficient for the controlled-simulation spline fits is $n_{\Pp,\mathrm{tr}}^{-3/4}$. The upper weight-cap quantile and the interior margin used when recovering $k_2$ are $\max\{0.95,1-n_{\Pp,\mathrm{tr}}^{-3/4}\}$, so both safeguards vanish asymptotically.

For finite-sample numerical stability, the implementation also clips fitted probabilities to $[10^{-6},1-10^{-6}]$ and applies fixed absolute bounds of 50 to the recovered $k_2$, 10 to $\eta$, and 20 to $\kappa$ and the working $m_{00}$ prediction. These fixed safeguards are not part of the oracle projection. Exact centering at the true finite-dimensional parameters continues to hold for bounded working augmentation functions, but attainment of the efficiency bound requires the oracle augmentation functions to lie inside the fixed bounds, or a theoretical sequence of bounds that diverges while the probability-clipping threshold tends to zero. Under the theoretical positivity condition, probability clipping is asymptotically inactive.

For a sieve of dimension $J_N$, the usual $L^2$ error for an $s$-smooth function of $d$ continuous variables is
\[
O_{\Prob}\left\{\left(\frac{J_N}{n_{\Pp,\mathrm{tr}}}\right)^{1/2}+J_N^{-s/d}\right\},
\]
and balancing the two terms gives $J_N\asymp n_{\Pp,\mathrm{tr}}^{d/(2s+d)}$ \citep{chenLargeSampleSieve2007}. Equations~\eqref{eq:s-L-spline-basis}--\eqref{eq:s-spline-K-rule} implement this balance because $J_L\asymp K_L^2$ for the two-dimensional continuous component and $J_X\asymp K_X$ for the one-dimensional component. At minimum, $J_N\to\infty$ and $J_N/n_{\Pp,\mathrm{tr}}\to0$ are needed for $L^2$ consistency.

The same-sample estimator additionally requires the empirical-process condition in Assumption~\ref{ass:s-sieve}. A convenient primitive route is that the local score-difference class have entropy of order $J_N\log(A/\epsilon)$, for a fixed $A>1$, and $L^2(P_0)$ radius $r_N\to0$. Under uniformly bounded envelopes, the local maximal inequality of \citet{vanDerVaartWellnerLocalMaximal2011} gives the sufficient condition
\[
r_N\sqrt{J_N\log(A/r_N)}
+\frac{J_N\log(A/r_N)}{\sqrt N}
\longrightarrow0.
\]
For an envelope $F_N$ satisfying a uniform sub-Weibull bound
\[
\sup_N \E\left[\exp\left\{(F_N/C_F)^\nu\right\}\right]<\infty
\]
for fixed $C_F,\nu>0$, truncation gives the conservative sufficient condition
\[
r_N\sqrt{J_N\log(A/r_N)}
+\frac{J_N\log(A/r_N)(\log N)^{1/\nu}}{\sqrt N}
\longrightarrow0.
\]
With
\[
r_N\asymp
\left(\frac{J_N}{n_{\Pp,\mathrm{tr}}}\right)^{1/2}+J_N^{-s/d},
\]
$n_{\Pp,\mathrm{tr}}\asymp N$, and the balanced choice of $J_N$, these conditions hold up to logarithmic factors when $s>d/2$. Under the corresponding envelope condition for the squared score-difference class, the same entropy calculation also gives the empirical second-moment requirement in Assumption~\ref{ass:s-sieve}. The reported rules use $s=2$ and $d\in\{1,2\}$. Exact augmentation invariance removes the conventional $N^{-1/4}$-type product-rate condition, but the same-sample estimator must still satisfy the displayed empirical-process condition.

For the cross-fitted estimator, any rate yielding convergence of the fitted joint estimating function in $L^2(P_0)$ is sufficient for the stated first-order expansion; convergence to the optimal projection is additionally needed to attain the efficiency bound. Weighted cross-validation could instead choose the dimension within each training sample, but it was not used in the reported experiments.

\subsection{Same-sample profiled-root expansion}

Let $\zeta^\dagger$ be the deterministic working limit in Assumption~\ref{ass:s-sieve} and abbreviate
\[
S^\dagger(O)=S(O;\xi_0,\zeta^\dagger),
\qquad
J^\dagger
=P_0\{\partial_{\xi^{\trans}}S(O;\xi_0,\zeta^\dagger)\}.
\]
For the full-sample augmentation map $\widehat\zeta_{\mathrm{sieve}}(\xi)$, define
\[
\widehat\Psi_{\mathrm{sieve}}(\xi)
=\mathbb P_N S
\{O;\xi,\widehat\zeta_{\mathrm{sieve}}(\xi)\}.
\]
For every realized full-sample fit $\widehat\zeta_{\mathrm{sieve}}(\xi_0)$, exact invariance implies
\begin{equation*}
P_0S
\{O;\xi_0,\widehat\zeta_{\mathrm{sieve}}(\xi_0)\}=0,
\qquad
P_0S^\dagger=0.
\end{equation*}
Therefore,
\begin{align}
\sqrt N\,\widehat\Psi_{\mathrm{sieve}}(\xi_0)
={}&\mathbb G_N S^\dagger
\nonumber\\
&+\mathbb G_N\left[
S\{O;\xi_0,\widehat\zeta_{\mathrm{sieve}}(\xi_0)\}
-S^\dagger(O)
\right].
\label{eq:s-same-sample-decomp}
\end{align}
There is no population drift term. Under Assumption~\ref{ass:s-sieve}, equivalently the same-sample stochastic-equicontinuity condition in Theorem~3 of the main paper, the second line of~\eqref{eq:s-same-sample-decomp} is $o_{\Prob}(1)$.

Exact invariance also removes the chain-rule contribution from the fitted augmentation map. For any differentiable map $\xi\mapsto\zeta(\xi)$, the derivative at the truth of
\[
\xi\longmapsto P_0S\{O;\xi,\zeta(\xi)\}
\]
is therefore the partial derivative with respect to $\xi$ holding $\zeta$ fixed. Hence the partial Jacobian is the correct first-order Jacobian for the expansion and sandwich variance estimation. This is a population statement and does not require the numerical root finder to use the partial Jacobian at every intermediate iterate. Under the stated uniform convergence condition, both the local profiled Jacobian and the partial Jacobian converge to $J^\dagger$.

Let $\widehat\xi_{\mathrm{sieve}}$ be a consistent local approximate root with
$\|\widehat\Psi_{\mathrm{sieve}}(\widehat\xi_{\mathrm{sieve}})\|=o_{\Prob}(N^{-1/2})$, and define the integrated local Jacobian
\[
\widehat J_{N,\mathrm{sieve}}^{\mathrm{int}}
=\int_0^1
\frac{\partial\widehat\Psi_{\mathrm{sieve}}}
{\partial\xi^{\trans}}
\{\xi_0+t(\widehat\xi_{\mathrm{sieve}}-\xi_0)\}\,dt.
\]
The mean-value identity is
\[
\widehat\Psi_{\mathrm{sieve}}(\widehat\xi_{\mathrm{sieve}})
=\widehat\Psi_{\mathrm{sieve}}(\xi_0)
+\widehat J_{N,\mathrm{sieve}}^{\mathrm{int}}
(\widehat\xi_{\mathrm{sieve}}-\xi_0).
\]
Because $\sqrt N\widehat\Psi_{\mathrm{sieve}}(\xi_0)=O_{\Prob}(1)$ and
$\widehat J_{N,\mathrm{sieve}}^{\mathrm{int}}\to J^\dagger$, nonsingularity first gives
$\widehat\xi_{\mathrm{sieve}}-\xi_0=O_{\Prob}(N^{-1/2})$. Combining the same identity with~\eqref{eq:s-same-sample-decomp} then yields
\[
\sqrt N(\widehat\xi_{\mathrm{sieve}}-\xi_0)
=-(J^\dagger)^{-1}\mathbb G_N S^\dagger+o_{\Prob}(1).
\]
No approximation-bias term appears even when the sieve converges to a nonoptimal $\zeta^\dagger$. If the fitted joint estimating function converges in $L^2(P_0)$ to the joint function obtained from the optimal projection, Slutsky's theorem gives the efficiency bound.

\subsection{First-order expansion for Eff-DML}

This subsection proves the Eff-DML part of Theorem~3 in the main paper. Cross-fitting makes the fitted augmentation functions fixed on each evaluation fold; exact augmentation invariance removes population drift, and Assumption~\ref{ass:s-crossfit} controls the remaining held-out empirical term.

Let $I_1,\ldots,I_K$ be fixed folds, let $\widehat\zeta^{[-k]}(\xi)$ be fitted without fold $I_k$, and define
\[
\widehat\Psi_{\mathrm{DML2}}(\xi)
=\frac1N\sum_{k=1}^K\sum_{i\in I_k}
S\{O_i;\xi,\widehat\zeta^{[-k]}(\xi)\}.
\]
At $\xi_0$, set
\[
\Delta_k(O)
=S\{O;\xi_0,\widehat\zeta^{[-k]}(\xi_0)\}
-S(O;\xi_0,\zeta^\dagger).
\]
Fold assignment is stratified by the finite membership pattern $R=(\delta_{\NP},\delta_{\Pp})$. Let $p_r=\Pr(R=r)>0$, let $n_r$ be the number of observations in pattern $r$, and let $n_{kr}$ be the number assigned to fold $k$. Random allocation within each pattern gives $n_{kr}=n_r/K+O(1)$. Conditional on the training data $\mathcal D_{-k}$, write $\mu_{kr}=\E\{\Delta_k(O)\mid R=r,\mathcal D_{-k}\}$. Exact invariance gives $\sum_r p_r\mu_{kr}=0$, while Assumption~\ref{ass:s-crossfit} and the finite number of positive-probability patterns imply that the pattern-specific means and conditional second moments of $\Delta_k$ converge to zero.

Let $I_{kr}=\{i\in I_k:R_i=r\}$. Conditional on the training data and the held-out pattern labels,
\begin{align*}
\frac1{\sqrt N}\sum_{i\in I_k}\Delta_k(O_i)
={}&\frac1{\sqrt N}\sum_r\sum_{i\in I_{kr}}
\{\Delta_k(O_i)-\mu_{kr}\}\\
&+\sum_r\frac{n_{kr}-Np_r/K}{\sqrt N}\mu_{kr}.
\end{align*}
The first term is $o_{\Prob}(1)$ by the conditional variance bound. The second is also $o_{\Prob}(1)$ because $n_r-Np_r=O_{\Prob}(\sqrt N)$, $n_{kr}-n_r/K=O(1)$, and $\mu_{kr}=o_{\Prob}(1)$. Since $K$ is fixed,
\[
\frac1{\sqrt N}\sum_{k=1}^K\sum_{i\in I_k}\Delta_k(O_i)
=o_{\Prob}(1),
\]
and therefore
\[
\sqrt N\widehat\Psi_{\mathrm{DML2}}(\xi_0)
=\mathbb G_N S^\dagger+o_{\Prob}(1).
\]

Let $\widehat\xi_{\mathrm{DML2}}$ be a consistent local approximate root satisfying
$\|\widehat\Psi_{\mathrm{DML2}}(\widehat\xi_{\mathrm{DML2}})\|=o_{\Prob}(N^{-1/2})$. The integrated pooled Jacobian along the segment from $\xi_0$ to $\widehat\xi_{\mathrm{DML2}}$ converges to $J^\dagger$. The mean-value identity first gives $\widehat\xi_{\mathrm{DML2}}-\xi_0=O_{\Prob}(N^{-1/2})$ and then
\[
\sqrt N(\widehat\xi_{\mathrm{DML2}}-\xi_0)
=-(J^\dagger)^{-1}\mathbb G_N S^\dagger+o_{\Prob}(1).
\]
This proves the Eff-DML statement in Theorem~3. The EIF derivation in Section~\ref{supp:efficiency} and the Eff-sieve expansion above do not use this cross-fitting argument.

\subsection{Numerical profiling}

The implementation computes the local approximate roots by repeatedly refitting the augmentation functions. For the same-sample estimator, a damped partial-Jacobian Newton iteration is retained because it is stable in the controlled-complexity sieve setting. For Eff-DML, the folds are fixed once and the fold-specific augmentation functions are refitted on their training observations whenever the profiled estimating function is evaluated. A short partial-Jacobian iteration supplies a warm start when available, but it does not define the final estimator.

The DML root search uses a centered finite-difference Jacobian of the pooled profiled estimating function, with the augmentation functions refitted at every perturbed parameter value. The direction is computed by a parameter- and estimating-function-scaled Levenberg--Marquardt system. A trust region bounds the standardized parameter displacement, and step halving continues until the standardized estimating-function objective decreases.

If these steps do not produce a decrease, a local Nelder--Mead minimization of the same objective is used as a fallback. This regularization stabilizes the search direction without altering the defining estimating equation. After a local approximate root is found, the sandwich variance estimator uses the partial Jacobian with the final augmentation functions held fixed.

Only prespecified, data-driven starts are used. Each implementation first attempts its preferred preliminary start. If it fails, at most ten additional values are tried: the last conditional-inclusion coefficient is shifted by $\pm0.5$ and $\pm1$ preliminary standard errors; $\theta$ is shifted by $\pm0.5$ standard errors; the conditional-inclusion coefficients are shrunk halfway toward zero or set to zero; and two opposite alternating-sign joint half-standard-error perturbations are used. For Eff-DML, the ordered candidate pool begins with the Eff-sieve estimate, the validation estimate, and fixed convex combinations, subject to the same ten-retry cap.

Every candidate solves the same profiled estimating equation. The first candidate meeting the approximate-root criteria is retained; if all attempts fail, the replication is recorded as a numerical failure with nonfinite estimates. The starts depend only on the observed preliminary estimate and its standard errors, never on a true parameter or oracle root.

The default rule requires at least two iterations and stops when both (i) the accepted parameter update, standardized by the preliminary standard errors, has Euclidean norm at most $2\times10^{-3}$ and (ii) the largest componentwise studentized mean estimating function is at most $2\times10^{-2}$. The finite-difference relative step is $2\times10^{-4}$, the initial Levenberg--Marquardt penalty is $10^{-3}$, the standardized trust radius is five, and 50 iterations is a safety limit.

The implementation records the profiled- and partial-Jacobian singular values, line-search and trust-region factors, and a weak-identification flag. A fit is unsuccessful if no finite local step reduces the standardized objective or the numerical stopping criteria are not met. The stated tolerances are finite-sample numerical stopping criteria. The asymptotic results separately assume the approximate-root condition, which is not implied solely by the numerical convergence flag.

\subsection{Sandwich variance estimation}

For either implementation, let $\widehat\xi$ be the final local approximate root and let $\widehat S_i(\xi)$ be the joint estimating-function contribution for unit $i$, with the final augmentation functions held fixed. Write $\widehat S_i=\widehat S_i(\widehat\xi)$ and define
\[
\widehat J
=\left.\mathbb P_N\left\{\frac{\partial\widehat S_i(\xi)}{\partial\xi^{\trans}}\right\}\right|_{\xi=\widehat\xi},
\qquad
\widehat\Omega
=\mathbb P_N(\widehat S_i\widehat S_i^{\trans}),
\qquad
\widehat V
=\widehat J^{-1}\widehat\Omega\widehat J^{-\trans}.
\]
For the same-sample estimator, the empirical second-moment condition in Assumption~\ref{ass:s-sieve}, together with local continuity and the Jacobian condition, gives
\[
\widehat J\longrightarrow J^\dagger,
\qquad
\widehat\Omega
\longrightarrow
\Omega^\dagger
=\E\{S^\dagger(O)S^\dagger(O)^{\trans}\}.
\]
For the cross-fitted estimator, the same conclusion follows from the held-out conditional law of large numbers, the pattern-stratified decomposition above, and the uniform $(2+c)$ moment bound in Assumption~\ref{ass:s-crossfit}.
Thus
\[
\widehat V
\longrightarrow
(J^\dagger)^{-1}\Omega^\dagger(J^\dagger)^{-\trans}.
\]
Here $\widehat V$ estimates the asymptotic variance of $\sqrt N(\widehat\xi-\xi_0)$; the estimated variance of $\widehat\xi$ is $\widehat V/N$. When the fitted joint estimating function converges to the joint function obtained from the optimal projection, this limit is the general-model efficiency bound. In all cases, inference for $\theta$ must use the upper-left block of the joint sandwich variance estimator rather than the empirical variance of the target efficient estimating function alone.

\subsection{Feasible Eff-sub implementation and cross-fitted expansion}
\label{supp:effsub-feasible}

The restricted normal equation can be estimated directly from probability-sample records outside the non-probability source, without first forming a conditional-ratio representation. For a working model $\gamma_\alpha(X)$ with basis $b_X(X)$, define $\alpha^*$ by
\[
\E\left[
\delta_{\Pp}(1-\delta_{\NP})
\frac{O_{\Pp}}{\pi_{\Pp}}b_X(X)
\{U(\theta_0;L)-\gamma_\alpha(X)\}
\right]=0.
\]
Conditioning on $(L,\pi_{\Pp})$ transforms this equation into
\[
\E\left[
\{1-\pi_{\NP}(L)\}O_{\Pp}b_X(X)
\{U(\theta_0;L)-\gamma_\alpha(X)\}
\right]=0,
\]
the basis-restricted version of~\eqref{eq:s-gamma-normal}. If the working model contains $\gamma^*$, then $\gamma_{\alpha^*}=\gamma^*$; otherwise $\gamma_{\alpha^*}$ is its weighted projection onto the working model.

For the mean, write $U(\theta;L)=\theta-Y$ and parameterize $\gamma_{\theta,\alpha}(X)=\theta-m_\alpha(X)$. The working outcome regression is fitted among records with $\delta_{\Pp}=1$ and $\delta_{\NP}=0$, using raw weight
\[
W_{\gamma}=\frac{O_{\Pp}}{\pi_{\Pp}}
=\frac{1-\pi_{\Pp}}{\pi_{\Pp}^2}.
\]
Let $W_{\gamma,i}^{\mathrm c}$ be the raw weight after the same prespecified upper-tail cap used for the other working regressions, and normalize it to mean one over the eligible training records, yielding $\widetilde W_{\gamma,i}$. Let $J_X$ be the dimension of $b_X(X)$, let $B_X$ have rows $b_X(X_i)^{\trans}$ for the eligible training records, and let $\widetilde{\mathcal W}_\gamma$ be diagonal with entries $\widetilde W_{\gamma,i}$. Define
\[
D_X=\operatorname{diag}(0,1,\ldots,1),
\qquad
a_\gamma=\max\left\{J_X^{-1}\operatorname{tr}
\left(B_X^{\trans}\widetilde{\mathcal W}_\gamma B_X\right),1\right\},
\]
and let $\lambda_{\gamma,n}\geq0$ be the ridge coefficient. With $m_\alpha(X)=b_X(X)^{\trans}\alpha$ in the controlled simulations, the fold-specific estimate solves
\[
\sum_{i\in\mathcal I_{\mathrm{tr}}}
\delta_{\Pp,i}(1-\delta_{\NP,i})
\widetilde W_{\gamma,i}b_X(X_i)
\{Y_i-b_X(X_i)^{\trans}\alpha\}
-\lambda_{\gamma,n}a_\gamma D_X\alpha=0.
\]
The controlled simulations use the $X$-spline basis in~\eqref{eq:s-X-spline-basis} and $\lambda_{\gamma,n}=n_{\Pp,\mathrm{tr}}^{-3/4}$. The Culture and Community in a Time of Crisis (CCTC) analysis uses the same eligible records and weights with the fixed-ridge linear learner described in Section~\ref{supp:cctc}. Thus the feasible estimator is obtained by a single weighted regression; no separately fitted numerator and denominator are formed.

Let $k(i)$ denote the fold containing unit $i$, and let $\widehat m^{[-k(i)]}$ be trained without that fold. The resulting cross-fitted estimator is
\begin{equation}
\widehat\theta_{\RA}
=\frac1N\sum_{i=1}^N\left[
\delta_{\NP,i}Y_i
+(1-\delta_{\NP,i})\left\{
\frac{\delta_{\Pp,i}}{\pi_{\Pp,i}}Y_i
+\left(1-\frac{\delta_{\Pp,i}}{\pi_{\Pp,i}}\right)
\widehat m^{[-k(i)]}(X_i)
\right\}
\right].
\label{eq:s-effsub-mean}
\end{equation}
If both inclusion mechanisms are noninformative given $X$, the population working regression reduces to $m(X)=\E(Y\mid X)$.

Let $\widehat\gamma^{[-k]}$ converge in $L^2$ to a finite working limit $\bar\gamma$, and define
\[
J_{\mathrm{sub}}
=\E\left\{\frac{\partial U(\theta_0;L)}{\partial\theta^{\trans}}\right\},
\]
which is assumed nonsingular. The population moment is constant in $\gamma$:
\[
P_0B_{\RA}(\theta_0,\gamma)=0
\quad\text{for all }\gamma.
\]
Consequently,
\[
P_0\{B_{\RA}(\theta_0,\widehat\gamma)-B_{\RA}(\theta_0,\bar\gamma)\}=0
\]
exactly, not merely to first order. Conditional on the training folds,
\[
\sqrt N(\mathbb P_N-P_0)
\{B_{\RA}(\theta_0,\widehat\gamma)-B_{\RA}(\theta_0,\bar\gamma)\}
=o_{\Prob}(1)
\]
by cross-fitting and $L^2$ convergence. Taylor expansion in $\theta$ then yields
\[
\sqrt N(\widehat\theta_{\RA}-\theta_0)
=-J_{\mathrm{sub}}^{-1}
\frac1{\sqrt N}\sum_{i=1}^N
B_{\RA}(O_i;\theta_0,\bar\gamma)
+o_{\Prob}(1).
\]
The empirical sandwich variance estimator is consistent by the same law-of-large-numbers argument used for the efficient procedure. If $\bar\gamma=\gamma^*$, Eff-sub attains the restricted-class variance bound.

\section{Empirical-study specifications and complete results}
\label{supp:simulation}

\subsection{Controlled-simulation design}

Each Monte Carlo replication begins with a newly generated finite population of size $N=10{,}000$. The labels S1--S4 denote the four scenarios defined in the main paper, O1--O3 denote the three outcome mechanisms, and M1--M2 denote the two non-probability inclusion mechanisms. Generate
\[
X\sim N(0,1),\qquad Z\sim\operatorname{Bernoulli}(0.5),
\]
and use the outcome and non-probability inclusion mechanisms stated in the simulation section of the main paper. The fitted non-probability inclusion model is logistic with regressors $(1,X,Y)$ in S1, S2, and S4. In S3 the same fitted family is deliberately misspecified.

The conditional mean of the design inclusion probability is
\[
\bar\pi_{\Pp}(X)=0.005+\expit\{-3-0.25(X-2)^2\}.
\]
With $a_{\Pp}=0.002$, $b_{\Pp}=0.08$, and
\[
\mu(X)=\frac{\bar\pi_{\Pp}(X)-a_{\Pp}}{b_{\Pp}-a_{\Pp}},
\]
generate
\[
B\mid L\sim\operatorname{Beta}\{50\mu(X),50[1-\mu(X)]\},
\qquad
\pi_{\Pp}=a_{\Pp}+(b_{\Pp}-a_{\Pp})B.
\]
The two sample inclusion indicators are independent conditional on $(L,\pi_{\Pp})$. Because the Gaussian covariates and outcomes are unbounded, the logistic non-probability inclusion probabilities are not uniformly bounded away from zero and one; the simulations therefore extend beyond the uniform-positivity sufficient condition used in the theory.

In S4, every original population unit retains a single record and an original-unit identifier. For a random half of the true overlaps, the observed non-probability inclusion indicator is set to zero while the probability-sample inclusion indicator and the corresponding $\pi_{\Pp}$ are retained. No record is duplicated. Because the non-probability inclusion indicator is then misclassified for some units, S4 is a sensitivity experiment outside the identification theorem; coverage is reported with this qualification.

The complete study contains 500 Monte Carlo replications, and the number of cross-fitting folds is $K=2$. Fold labels are randomly assigned within the four observed sample-membership patterns and then held fixed throughout each profiled fit. The primary target used for bias, standard-error, and coverage calculations is the superpopulation mean $\theta_0=0$. For replication $r$, we also compute
\[
\theta_N^{(r)}=N^{-1}\sum_{i=1}^N Y_i^{(r)}
\]
and report secondary bias and root mean squared error (RMSE) relative to this realized finite-population mean. Table~\ref{tab:s-method-map} summarizes the methods used in the study.

\begin{longtable}{>{\raggedright\arraybackslash}p{0.18\textwidth}>{\raggedright\arraybackslash}p{0.72\textwidth}}
\caption{Methods used in the simulation study.}\label{tab:s-method-map}\\
\toprule
Method & Description \\
\midrule
\endfirsthead
\multicolumn{2}{c}{\tablename\ \thetable\ (continued)}\\
\toprule
Method & Description \\
\midrule
\endhead
\midrule\multicolumn{2}{r}{Continued on next page}\\
\endfoot
\bottomrule
\endlastfoot
P & Probability-sample Horvitz--Thompson estimator \\
P-AUG & Augmented probability-sample estimator \\
NP & Non-probability inverse-probability-weighted estimator using the validation-sample estimate of $\phi$ \\
CLW & Pseudo-likelihood weighting benchmark of Chen, Li, and Wu \\
CLW+P & Minimum-variance combination of CLW and P \\
Eff-sub & Cross-fitted sub-efficient estimator \\
Eff-0 & Joint estimating procedure with the working augmentation collection fixed at the zero-function collection \\
Eff-sieve & Same-sample sieve estimator based on the joint efficient estimating function \\
Eff-DML & Pooled cross-fitted DML2 estimator based on the joint efficient estimating function \\
Oracle-Eff & Joint efficient estimating function with the true finite-dimensional parameters and augmentation functions learned from an independent population; reported only under correct specification of the non-probability inclusion model \\
\end{longtable}

In the method label Eff-0, ``0'' denotes zero augmentation; subscripts $0$ on parameter-dependent quantities elsewhere denote evaluation at the true parameter values.

The validation-sample logistic estimator initializes the efficient procedures. Eff-sieve uses full-sample augmentation fits, whereas Eff-DML uses fixed folds and the pooled profiled equation. Both use the weighted spline constructions and local solvers in Section~\ref{supp:dml}. Oracle-Eff learns the five augmentation functions from a large independent population at the generating finite-dimensional parameters; it is reported only under correct specification and serves as an efficiency-bound diagnostic.

The experiment on heterogeneity in the probability sampling design uses S1 and retains the same conditional mean $\bar\pi_{\Pp}(X)$ while replacing the concentration 50 by
\[
c\in\{100,20,5\}.
\]
All other elements of the data-generating mechanism are held fixed. The comparison includes P, P-AUG, NP, Eff-sub, Eff-0, Eff-sieve, Eff-DML, and Oracle-Eff. The reported summaries include the realized average conditional variance of $\pi_{\Pp}\mid L$ as well as performance of the target and conditional inclusion parameters.

The weak-identification experiment uses
\[
Y=0.8X+\epsilon,
\qquad
\epsilon\sim N(0,\sigma^2),
\]
with a prespecified grid of residual standard deviations. The non-probability inclusion mechanism is M1 and the probability sampling design is the one described above. As $\sigma$ decreases, $X$ and $Y$ become increasingly collinear in the validation regression. Writing $\phi_Y$ for the coefficient of $Y$ in the validation logistic regression, the reported diagnostics include its RMSE, the smallest singular value and condition number of each method-specific mechanism-equation Jacobian, target RMSE and coverage, and estimator failure rates. These method-specific diagnostics are distinct from the validation-Jacobian diagnostics reported for the preliminary estimator.

\subsection{Controlled-simulation results and spline sensitivity}

The complete method-by-method results underlying the selected tables in the main paper are reported below. In the tables, RMSE denotes root mean squared error, SD denotes Monte Carlo standard deviation, SE denotes the mean estimated standard error, and MCSE denotes Monte Carlo standard error. The main experiment and weak-identification experiment have no recorded numerical failures. Under the primary spline rule in the design-heterogeneity experiment, Eff-sieve has one diagnosed local-profile failure among 500 replications at $c=5$ after the default start and all ten prespecified reinitializations; the corresponding failure rate is $0.002$ and the summary uses the 499 finite fits. All other primary entries use 500 replications. The rich-rule sensitivity analysis has no numerical failures.

\begin{longtable}{@{}llrrrrr@{}}
\caption{Complete Monte Carlo performance in Scenarios S1--S4 under the primary spline rule. Target-estimation summaries.}\label{tab:supp-main-performance-full}\\
\toprule
Scenario & Method & \shortstack{Bias\\(MCSE)} & SD & \shortstack{RMSE\\(MCSE)} & \shortstack{Mean\\SE} & \shortstack{Coverage\\(MCSE)} \\
\midrule
\endfirsthead
\multicolumn{7}{c}{\tablename\ \thetable\ (continued)}\\
\toprule
Scenario & Method & \shortstack{Bias\\(MCSE)} & SD & \shortstack{RMSE\\(MCSE)} & \shortstack{Mean\\SE} & \shortstack{Coverage\\(MCSE)} \\
\midrule
\endhead
\midrule\multicolumn{7}{r}{Continued on next page}\\
\endfoot
\bottomrule
\endlastfoot
S1 & P & 0.006 (0.004) & 0.096 & 0.097 (0.003) & 0.096 & 0.920 (0.012) \\
S1 & P-AUG & 0.006 (0.003) & 0.073 & 0.073 (0.002) & 0.068 & 0.906 (0.013) \\
S1 & Eff-sub & 0.003 (0.003) & 0.057 & 0.057 (0.002) & 0.054 & 0.936 (0.011) \\
S1 & Eff-0 & 0.009 (0.004) & 0.096 & 0.097 (0.003) & 0.095 & 0.918 (0.012) \\
S1 & Eff-sieve & 0.001 (0.001) & 0.029 & 0.029 (0.001) & 0.026 & 0.932 (0.011) \\
S1 & Eff-DML & 0.005 (0.002) & 0.035 & 0.035 (0.002) & 0.042 & 0.966 (0.008) \\
S1 & Oracle-Eff & 0.006 (0.001) & 0.025 & 0.026 (0.001) & 0.021 & 0.924 (0.012) \\
S1 & NP & 0.041 (0.013) & 0.302 & 0.304 (0.025) & 0.276 & 0.992 (0.004) \\
S1 & CLW & -4.314 (3.480) & 77.807 & 77.849 (30.061) & 37.376 & 0.220 (0.019) \\
S1 & CLW+P & -0.163 (0.008) & 0.180 & 0.243 (0.006) & 0.079 & 0.430 (0.022) \\
S2 & P & -0.003 (0.003) & 0.076 & 0.076 (0.003) & 0.075 & 0.946 (0.010) \\
S2 & P-AUG & 0.000 (0.001) & 0.027 & 0.027 (0.001) & 0.025 & 0.916 (0.012) \\
S2 & Eff-sub & 0.000 (0.001) & 0.023 & 0.023 (0.001) & 0.021 & 0.926 (0.012) \\
S2 & Eff-0 & 0.001 (0.003) & 0.076 & 0.076 (0.002) & 0.074 & 0.946 (0.010) \\
S2 & Eff-sieve & 0.000 (0.000) & 0.011 & 0.011 (0.000) & 0.011 & 0.938 (0.011) \\
S2 & Eff-DML & 0.001 (0.001) & 0.013 & 0.013 (0.000) & 0.018 & 0.962 (0.009) \\
S2 & Oracle-Eff & 0.000 (0.001) & 0.011 & 0.011 (0.001) & 0.010 & 0.942 (0.010) \\
S2 & NP & 0.066 (0.013) & 0.294 & 0.301 (0.025) & 0.274 & 0.992 (0.004) \\
S2 & CLW & 9.570 (5.181) & 115.855 & 116.134 (45.442) & 37.186 & 0.944 (0.010) \\
S2 & CLW+P & 0.005 (0.003) & 0.065 & 0.065 (0.002) & 0.060 & 0.910 (0.013) \\
S3 & P & 0.000 (0.004) & 0.086 & 0.086 (0.003) & 0.085 & 0.940 (0.011) \\
S3 & P-AUG & 0.003 (0.002) & 0.047 & 0.048 (0.002) & 0.048 & 0.936 (0.011) \\
S3 & Eff-sub & 0.003 (0.002) & 0.046 & 0.046 (0.001) & 0.045 & 0.932 (0.011) \\
S3 & Eff-0 & 0.003 (0.004) & 0.086 & 0.086 (0.003) & 0.085 & 0.934 (0.011) \\
S3 & Eff-sieve & -0.006 (0.001) & 0.026 & 0.026 (0.001) & 0.024 & 0.930 (0.011) \\
S3 & Eff-DML & -0.006 (0.001) & 0.026 & 0.027 (0.001) & 0.029 & 0.956 (0.009) \\
S3 & NP & -0.035 (0.009) & 0.207 & 0.209 (0.012) & 0.190 & 0.970 (0.008) \\
S3 & CLW & 0.044 (0.003) & 0.072 & 0.084 (0.003) & 0.070 & 0.898 (0.014) \\
S3 & CLW+P & 0.052 (0.003) & 0.065 & 0.083 (0.002) & 0.067 & 0.882 (0.014) \\
S4 & P & 0.006 (0.004) & 0.096 & 0.097 (0.003) & 0.096 & 0.920 (0.012) \\
S4 & P-AUG & 0.006 (0.003) & 0.074 & 0.074 (0.003) & 0.069 & 0.918 (0.012) \\
S4 & Eff-sub & -0.010 (0.003) & 0.060 & 0.061 (0.003) & 0.062 & 0.934 (0.011) \\
S4 & Eff-0 & 0.009 (0.004) & 0.096 & 0.097 (0.003) & 0.095 & 0.918 (0.012) \\
S4 & Eff-sieve & -0.004 (0.001) & 0.028 & 0.028 (0.001) & 0.027 & 0.936 (0.011) \\
S4 & Eff-DML & 0.000 (0.002) & 0.035 & 0.035 (0.001) & 0.048 & 0.944 (0.010) \\
S4 & NP & 0.089 (0.086) & 1.933 & 1.933 (0.284) & 1.297 & 1.000 (0.000) \\
S4 & CLW & -5.739 (4.420) & 98.839 & 98.907 (38.346) & 36.424 & 0.232 (0.019) \\
S4 & CLW+P & -0.160 (0.008) & 0.179 & 0.240 (0.006) & 0.079 & 0.438 (0.022) \\
\end{longtable}

\begin{longtable}{llrrr}
\caption{Complete Monte Carlo performance in Scenarios S1--S4 under the primary spline rule. Numerical and finite-population diagnostics.}\label{tab:supp-main-performance-full-diagnostics}\\
\toprule
Scenario & Method & Failure & \shortstack{Finite\\RMSE} & $\phi_Y$ RMSE \\
\midrule
\endfirsthead
\multicolumn{5}{c}{\tablename\ \thetable\ (continued)}\\
\toprule
Scenario & Method & Failure & \shortstack{Finite\\RMSE} & $\phi_Y$ RMSE \\
\midrule
\endhead
\midrule\multicolumn{5}{r}{Continued on next page}\\
\endfoot
\bottomrule
\endlastfoot
S1 & P & 0.000 & 0.096 & -- \\
S1 & P-AUG & 0.000 & 0.073 & -- \\
S1 & Eff-sub & 0.000 & 0.057 & -- \\
S1 & Eff-0 & 0.000 & 0.096 & 0.312 \\
S1 & Eff-sieve & 0.000 & 0.027 & 0.068 \\
S1 & Eff-DML & 0.000 & 0.035 & 0.078 \\
S1 & Oracle-Eff & 0.000 & 0.023 & 0.095 \\
S1 & NP & 0.000 & 0.303 & 0.312 \\
S1 & CLW & 0.000 & 77.849 & -- \\
S1 & CLW+P & 0.000 & 0.243 & -- \\
S2 & P & 0.000 & 0.076 & -- \\
S2 & P-AUG & 0.000 & 0.026 & -- \\
S2 & Eff-sub & 0.000 & 0.021 & -- \\
S2 & Eff-0 & 0.000 & 0.075 & 2.433 \\
S2 & Eff-sieve & 0.000 & 0.009 & 0.981 \\
S2 & Eff-DML & 0.000 & 0.012 & 1.100 \\
S2 & Oracle-Eff & 0.000 & 0.008 & 0.972 \\
S2 & NP & 0.000 & 0.301 & 2.384 \\
S2 & CLW & 0.000 & 116.135 & -- \\
S2 & CLW+P & 0.000 & 0.065 & -- \\
S3 & P & 0.000 & 0.085 & -- \\
S3 & P-AUG & 0.000 & 0.047 & -- \\
S3 & Eff-sub & 0.000 & 0.046 & -- \\
S3 & Eff-0 & 0.000 & 0.086 & -- \\
S3 & Eff-sieve & 0.000 & 0.025 & -- \\
S3 & Eff-DML & 0.000 & 0.025 & -- \\
S3 & NP & 0.000 & 0.210 & -- \\
S3 & CLW & 0.000 & 0.083 & -- \\
S3 & CLW+P & 0.000 & 0.083 & -- \\
S4 & P & 0.000 & 0.096 & -- \\
S4 & P-AUG & 0.000 & 0.073 & -- \\
S4 & Eff-sub & 0.000 & 0.061 & -- \\
S4 & Eff-0 & 0.000 & 0.096 & 0.450 \\
S4 & Eff-sieve & 0.000 & 0.027 & 0.067 \\
S4 & Eff-DML & 0.000 & 0.034 & 0.074 \\
S4 & NP & 0.000 & 1.934 & 0.450 \\
S4 & CLW & 0.000 & 98.907 & -- \\
S4 & CLW+P & 0.000 & 0.241 & -- \\
\end{longtable}

\clearpage
\begin{longtable}{rlrrrrrr}
\caption{Complete results for heterogeneity in the probability sampling design under the primary spline rule.}\label{tab:supp-design-heterogeneity-full}\\
\toprule
$c$ & Method & Bias & SD & RMSE & \shortstack{Mean\\SE} & Coverage & Failure \\
\midrule
\endfirsthead
\multicolumn{8}{c}{\tablename\ \thetable\ (continued)}\\
\toprule
$c$ & Method & Bias & SD & RMSE & \shortstack{Mean\\SE} & Coverage & Failure \\
\midrule
\endhead
\midrule\multicolumn{8}{r}{Continued on next page}\\
\endfoot
\bottomrule
\endlastfoot
100 & P & 0.005 & 0.099 & 0.099 & 0.093 & 0.920 & 0.000 \\
100 & P-AUG & 0.003 & 0.072 & 0.072 & 0.065 & 0.912 & 0.000 \\
100 & Eff-sub & 0.000 & 0.055 & 0.055 & 0.051 & 0.924 & 0.000 \\
100 & Eff-0 & 0.008 & 0.098 & 0.098 & 0.093 & 0.916 & 0.000 \\
100 & Eff-sieve & 0.000 & 0.028 & 0.028 & 0.026 & 0.922 & 0.000 \\
100 & Eff-DML & 0.003 & 0.035 & 0.035 & 0.042 & 0.946 & 0.000 \\
100 & Oracle-Eff & 0.006 & 0.024 & 0.025 & 0.021 & 0.906 & 0.000 \\
100 & NP & 0.071 & 0.336 & 0.343 & 0.290 & 0.986 & 0.000 \\
20 & P & 0.004 & 0.103 & 0.103 & 0.102 & 0.928 & 0.000 \\
20 & P-AUG & 0.000 & 0.077 & 0.077 & 0.071 & 0.918 & 0.000 \\
20 & Eff-sub & 0.002 & 0.061 & 0.061 & 0.057 & 0.930 & 0.000 \\
20 & Eff-0 & 0.008 & 0.103 & 0.103 & 0.100 & 0.926 & 0.000 \\
20 & Eff-sieve & -0.001 & 0.036 & 0.036 & 0.026 & 0.922 & 0.000 \\
20 & Eff-DML & 0.004 & 0.037 & 0.037 & 0.046 & 0.934 & 0.000 \\
20 & Oracle-Eff & 0.007 & 0.026 & 0.027 & 0.022 & 0.922 & 0.000 \\
20 & NP & 0.044 & 0.329 & 0.332 & 0.283 & 0.994 & 0.000 \\
5 & P & 0.004 & 0.133 & 0.133 & 0.123 & 0.892 & 0.000 \\
5 & P-AUG & 0.007 & 0.098 & 0.098 & 0.088 & 0.912 & 0.000 \\
5 & Eff-sub & 0.002 & 0.074 & 0.074 & 0.069 & 0.948 & 0.000 \\
5 & Eff-0 & 0.010 & 0.131 & 0.131 & 0.120 & 0.892 & 0.000 \\
5 & Eff-sieve & 0.001 & 0.039 & 0.039 & 0.028 & 0.932 & 0.002 \\
5 & Eff-DML & 0.007 & 0.041 & 0.042 & 0.062 & 0.954 & 0.000 \\
5 & Oracle-Eff & 0.006 & 0.025 & 0.026 & 0.023 & 0.930 & 0.000 \\
5 & NP & 0.057 & 0.281 & 0.286 & 0.271 & 0.996 & 0.000 \\
\end{longtable}

\clearpage
\begin{longtable}{rlrrrr}
\caption{Complete weak-identification results under the primary spline rule. Target-estimation summaries.}\label{tab:supp-weak-identification-full}\\
\toprule
$\sigma$ & Method & RMSE & \shortstack{Mean\\SE} & Coverage & Failure \\
\midrule
\endfirsthead
\multicolumn{6}{c}{\tablename\ \thetable\ (continued)}\\
\toprule
$\sigma$ & Method & RMSE & \shortstack{Mean\\SE} & Coverage & Failure \\
\midrule
\endhead
\midrule\multicolumn{6}{r}{Continued on next page}\\
\endfoot
\bottomrule
\endlastfoot
0.50 & P & 0.085 & 0.084 & 0.938 & 0.000 \\
0.50 & P-AUG & 0.049 & 0.047 & 0.936 & 0.000 \\
0.50 & Eff-sub & 0.040 & 0.041 & 0.946 & 0.000 \\
0.50 & Eff-0 & 0.085 & 0.083 & 0.936 & 0.000 \\
0.50 & Eff-sieve & 0.037 & 0.031 & 0.898 & 0.000 \\
0.50 & Eff-DML & 0.042 & 0.077 & 0.966 & 0.000 \\
0.50 & Oracle-Eff & 0.031 & 0.031 & 0.946 & 0.000 \\
0.50 & NP & 0.419 & 0.337 & 0.982 & 0.000 \\
0.25 & P & 0.078 & 0.077 & 0.940 & 0.000 \\
0.25 & P-AUG & 0.033 & 0.031 & 0.920 & 0.000 \\
0.25 & Eff-sub & 0.027 & 0.027 & 0.932 & 0.000 \\
0.25 & Eff-0 & 0.078 & 0.076 & 0.940 & 0.000 \\
0.25 & Eff-sieve & 0.019 & 0.018 & 0.922 & 0.000 \\
0.25 & Eff-DML & 0.023 & 0.039 & 0.956 & 0.000 \\
0.25 & Oracle-Eff & 0.016 & 0.017 & 0.960 & 0.000 \\
0.25 & NP & 0.343 & 0.292 & 0.990 & 0.000 \\
0.10 & P & 0.076 & 0.075 & 0.946 & 0.000 \\
0.10 & P-AUG & 0.027 & 0.025 & 0.916 & 0.000 \\
0.10 & Eff-sub & 0.023 & 0.021 & 0.926 & 0.000 \\
0.10 & Eff-0 & 0.076 & 0.074 & 0.946 & 0.000 \\
0.10 & Eff-sieve & 0.011 & 0.011 & 0.938 & 0.000 \\
0.10 & Eff-DML & 0.013 & 0.018 & 0.962 & 0.000 \\
0.10 & Oracle-Eff & 0.011 & 0.010 & 0.942 & 0.000 \\
0.10 & NP & 0.301 & 0.274 & 0.992 & 0.000 \\
0.05 & P & 0.076 & 0.075 & 0.944 & 0.000 \\
0.05 & P-AUG & 0.025 & 0.023 & 0.910 & 0.000 \\
0.05 & Eff-sub & 0.022 & 0.021 & 0.932 & 0.000 \\
0.05 & Eff-0 & 0.076 & 0.074 & 0.952 & 0.000 \\
0.05 & Eff-sieve & 0.010 & 0.010 & 0.954 & 0.000 \\
0.05 & Eff-DML & 0.012 & 0.017 & 0.960 & 0.000 \\
0.05 & Oracle-Eff & 0.010 & 0.009 & 0.952 & 0.000 \\
0.05 & NP & 0.279 & 0.260 & 0.988 & 0.000 \\
0.02 & P & 0.076 & 0.075 & 0.946 & 0.000 \\
0.02 & P-AUG & 0.024 & 0.023 & 0.914 & 0.000 \\
0.02 & Eff-sub & 0.021 & 0.020 & 0.928 & 0.000 \\
0.02 & Eff-0 & 0.075 & 0.074 & 0.952 & 0.000 \\
0.02 & Eff-sieve & 0.009 & 0.010 & 0.962 & 0.000 \\
0.02 & Eff-DML & 0.011 & 0.013 & 0.954 & 0.000 \\
0.02 & Oracle-Eff & 0.010 & 0.008 & 0.956 & 0.000 \\
0.02 & NP & 0.267 & 0.249 & 0.986 & 0.000 \\
\end{longtable}

\begin{longtable}{rlrrr}
\caption{Complete weak-identification results under the primary spline rule. Method-specific mechanism-equation Jacobian conditioning diagnostics.}\label{tab:supp-weak-identification-full-diagnostics}\\
\toprule
$\sigma$ & Method & $\phi_Y$ RMSE & \shortstack{Minimum\\singular value} & \shortstack{Condition\\number} \\
\midrule
\endfirsthead
\multicolumn{5}{c}{\tablename\ \thetable\ (continued)}\\
\toprule
$\sigma$ & Method & $\phi_Y$ RMSE & \shortstack{Minimum\\singular value} & \shortstack{Condition\\number} \\
\midrule
\endhead
\midrule\multicolumn{5}{r}{Continued on next page}\\
\endfoot
\bottomrule
\endlastfoot
0.50 & P & -- & -- & -- \\
0.50 & P-AUG & -- & -- & -- \\
0.50 & Eff-sub & -- & -- & -- \\
0.50 & Eff-0 & 0.480 & $2.51\times10^{-4}$ & 10.5 \\
0.50 & Eff-sieve & 0.207 & 0.002 & 85.7 \\
0.50 & Eff-DML & 0.238 & 0.002 & 171.5 \\
0.50 & Oracle-Eff & 0.193 & 0.002 & 95.9 \\
0.50 & NP & 0.480 & $2.51\times10^{-4}$ & 10.5 \\
0.25 & P & -- & -- & -- \\
0.25 & P-AUG & -- & -- & -- \\
0.25 & Eff-sub & -- & -- & -- \\
0.25 & Eff-0 & 0.965 & $6.62\times10^{-5}$ & 37.6 \\
0.25 & Eff-sieve & 0.394 & $5.52\times10^{-4}$ & 347.7 \\
0.25 & Eff-DML & 0.457 & $4.05\times10^{-4}$ & 635.6 \\
0.25 & Oracle-Eff & 0.384 & $5.30\times10^{-4}$ & 400.8 \\
0.25 & NP & 0.965 & $6.62\times10^{-5}$ & 37.6 \\
0.10 & P & -- & -- & -- \\
0.10 & P-AUG & -- & -- & -- \\
0.10 & Eff-sub & -- & -- & -- \\
0.10 & Eff-0 & 2.433 & $1.07\times10^{-5}$ & 229.6 \\
0.10 & Eff-sieve & 0.981 & $8.93\times10^{-5}$ & 2175.4 \\
0.10 & Eff-DML & 1.100 & $6.42\times10^{-5}$ & 3550.0 \\
0.10 & Oracle-Eff & 0.970 & $8.54\times10^{-5}$ & 2504.0 \\
0.10 & NP & 2.384 & $1.07\times10^{-5}$ & 229.5 \\
0.05 & P & -- & -- & -- \\
0.05 & P-AUG & -- & -- & -- \\
0.05 & Eff-sub & -- & -- & -- \\
0.05 & Eff-0 & 4.799 & $2.68\times10^{-6}$ & 913.3 \\
0.05 & Eff-sieve & 1.964 & $2.24\times10^{-5}$ & 8703.4 \\
0.05 & Eff-DML & 2.193 & $1.57\times10^{-5}$ & 20572.9 \\
0.05 & Oracle-Eff & 1.886 & $2.14\times10^{-5}$ & 10117.3 \\
0.05 & NP & 3.853 & $2.69\times10^{-6}$ & 909.8 \\
0.02 & P & -- & -- & -- \\
0.02 & P-AUG & -- & -- & -- \\
0.02 & Eff-sub & -- & -- & -- \\
0.02 & Eff-0 & 11.459 & $4.31\times10^{-7}$ & 5703.3 \\
0.02 & Eff-sieve & 4.862 & $3.58\times10^{-6}$ & 54414.6 \\
0.02 & Eff-DML & 5.536 & $2.57\times10^{-6}$ & 92251.1 \\
0.02 & Oracle-Eff & 4.607 & $3.39\times10^{-6}$ & 63920.4 \\
0.02 & NP & 3.968 & $4.34\times10^{-7}$ & 5665.8 \\
\end{longtable}

\subsection{Prespecified rich-spline sensitivity analysis}
\label{supp:rich-spline}

The rich rule in~\eqref{eq:s-spline-K-rule} was fixed before the new Monte Carlo results were examined and uses the same populations, samples, and seeds as the primary rule. It is therefore a prespecified sensitivity analysis. In the nonlinear O1 settings, the Eff-sieve RMSE changes from $0.029$ to $0.024$ in S1 and from $0.028$ to $0.024$ in S4; the Eff-DML RMSE changes from $0.035$ to $0.027$ in both scenarios.

\begin{figure}[p]
\centering
\includegraphics[width=0.92\linewidth]{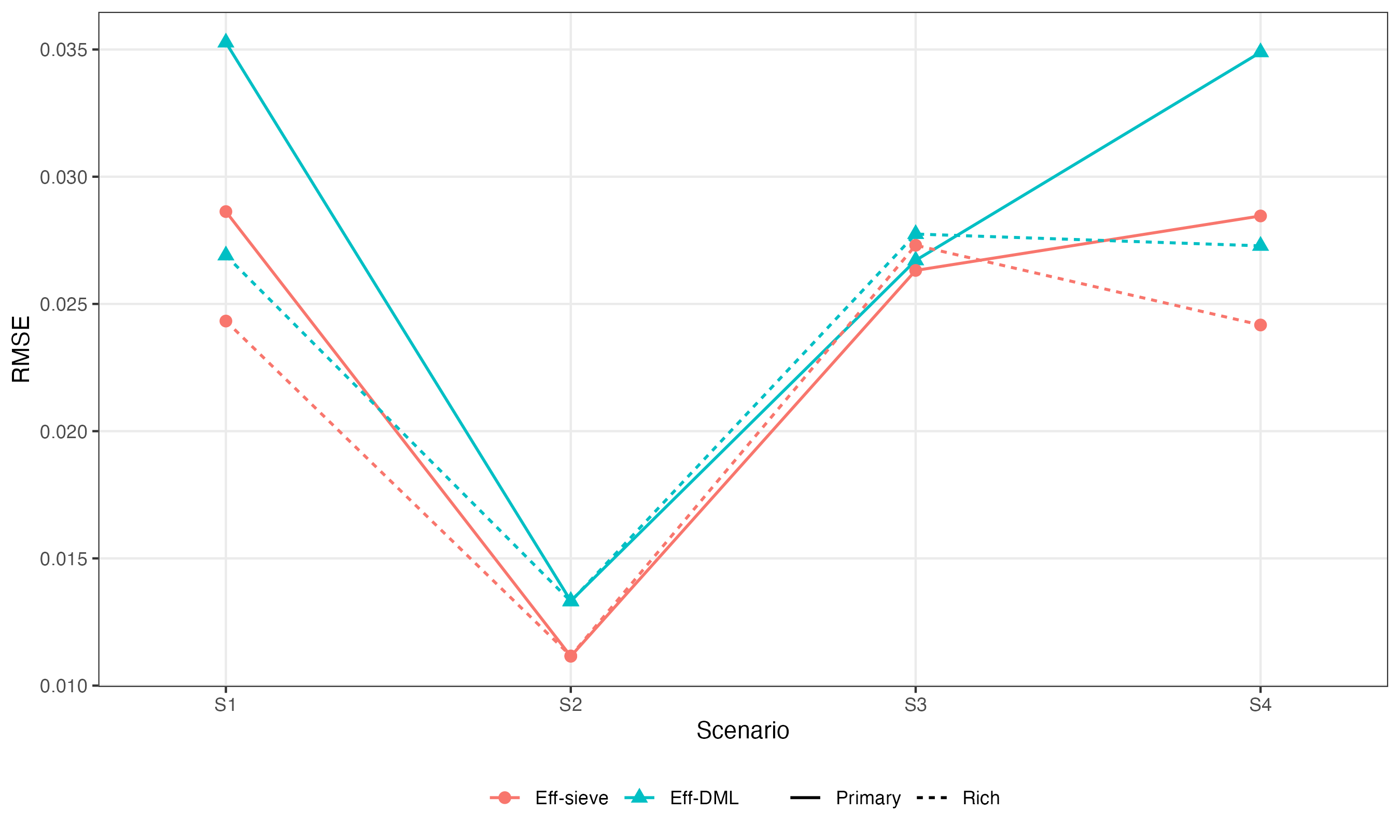}
\caption{RMSE sensitivity to the prespecified primary and rich spline rules. The two rules use the same 500 Monte Carlo seeds.}
\label{fig:supp-primary-rich-sensitivity}
\end{figure}

In S2 and S3, the Eff-sieve and Eff-DML changes are at most $0.001$ after rounding, and the P-AUG and Eff-sub changes are at most $0.002$. Under the most heterogeneous probability design, the rich rule has no failures and gives RMSEs of $0.026$ for Eff-sieve and $0.032$ for Eff-DML\@. Thus the substantive conclusions are stable across the two rules. Tables~\ref{tab:supp-spline-dimensions}--\ref{tab:supp-primary-rich-rmse} and Figure~\ref{fig:supp-primary-rich-sensitivity} summarize the comparison; complete rich-rule results follow.

\begin{longtable}{lllrrrr}
\caption{Primary and rich spline dimensions. Each entry is the minimum/median/maximum realized value across the corresponding replications and, for cross-fitted estimators, training folds.}\label{tab:supp-spline-dimensions}\\
\toprule
Rule & Experiment & Method & $K_L$ & $J_L$ & $K_X$ & $J_X$ \\
\midrule
\endfirsthead
\multicolumn{7}{c}{\tablename\ \thetable\ (continued)}\\
\toprule
Rule & Experiment & Method & $K_L$ & $J_L$ & $K_X$ & $J_X$ \\
\midrule
\endhead
\midrule\multicolumn{7}{r}{Continued on next page}\\
\endfoot
\bottomrule
\endlastfoot
Primary & S1--S4 & P-AUG & -- & -- & 3/3/3 & 8/8/8 \\
Primary & S1--S4 & Eff-sub & -- & -- & 3/3/3 & 8/8/8 \\
Primary & S1--S4 & Eff-sieve & 3/3/3 & 17/17/17 & 3/4/4 & 8/10/10 \\
Primary & S1--S4 & Eff-DML & 3/3/3 & 17/17/17 & 3/3/3 & 8/8/8 \\
Primary & S1--S4 & Oracle-Eff & 5/5/5 & 37/37/37 & 6/6/6 & 14/14/14 \\
Primary & Design & P-AUG & -- & -- & 3/3/3 & 8/8/8 \\
Primary & Design & Eff-sub & -- & -- & 3/3/3 & 8/8/8 \\
Primary & Design & Eff-sieve & 3/3/3 & 17/17/17 & 3/4/4 & 8/10/10 \\
Primary & Design & Eff-DML & 3/3/3 & 17/17/17 & 3/3/3 & 8/8/8 \\
Primary & Design & Oracle-Eff & 5/5/5 & 37/37/37 & 6/6/6 & 14/14/14 \\
Primary & Weak ID & P-AUG & -- & -- & 3/3/3 & 8/8/8 \\
Primary & Weak ID & Eff-sub & -- & -- & 3/3/3 & 8/8/8 \\
Primary & Weak ID & Eff-sieve & 3/3/3 & 17/17/17 & 3/4/4 & 8/10/10 \\
Primary & Weak ID & Eff-DML & 3/3/3 & 17/17/17 & 3/3/3 & 8/8/8 \\
Primary & Weak ID & Oracle-Eff & 5/5/5 & 37/37/37 & 6/6/6 & 14/14/14 \\
Rich & S1--S4 & P-AUG & -- & -- & 4/4/5 & 10/10/12 \\
Rich & S1--S4 & Eff-sub & -- & -- & 4/4/5 & 10/10/12 \\
Rich & S1--S4 & Eff-sieve & 4/4/4 & 26/26/26 & 5/5/5 & 12/12/12 \\
Rich & S1--S4 & Eff-DML & 4/4/4 & 26/26/26 & 4/4/5 & 10/10/12 \\
Rich & S1--S4 & Oracle-Eff & 7/7/7 & 65/65/65 & 9/9/9 & 20/20/20 \\
Rich & Design & P-AUG & -- & -- & 4/4/5 & 10/10/12 \\
Rich & Design & Eff-sub & -- & -- & 4/4/5 & 10/10/12 \\
Rich & Design & Eff-sieve & 4/4/4 & 26/26/26 & 5/5/5 & 12/12/12 \\
Rich & Design & Eff-DML & 4/4/4 & 26/26/26 & 4/4/5 & 10/10/12 \\
Rich & Design & Oracle-Eff & 7/7/7 & 65/65/65 & 9/9/9 & 20/20/20 \\
Rich & Weak ID & P-AUG & -- & -- & 4/4/5 & 10/10/12 \\
Rich & Weak ID & Eff-sub & -- & -- & 4/4/5 & 10/10/12 \\
Rich & Weak ID & Eff-sieve & 4/4/4 & 26/26/26 & 5/5/5 & 12/12/12 \\
Rich & Weak ID & Eff-DML & 4/4/4 & 26/26/26 & 4/4/5 & 10/10/12 \\
Rich & Weak ID & Oracle-Eff & 7/7/7 & 65/65/65 & 9/9/9 & 20/20/20 \\
\end{longtable}

\begin{longtable}{llrrr}
\caption{RMSE comparison between the prespecified primary and rich spline rules for the augmentation-based methods in Scenarios S1--S4.}\label{tab:supp-primary-rich-rmse}\\
\toprule
Scenario & Method & \shortstack{Primary\\RMSE} & \shortstack{Rich\\RMSE} & \shortstack{Rich $-$\\Primary} \\
\midrule
\endfirsthead
\multicolumn{5}{c}{\tablename\ \thetable\ (continued)}\\
\toprule
Scenario & Method & \shortstack{Primary\\RMSE} & \shortstack{Rich\\RMSE} & \shortstack{Rich $-$\\Primary} \\
\midrule
\endhead
\midrule\multicolumn{5}{r}{Continued on next page}\\
\endfoot
\bottomrule
\endlastfoot
S1 & P-AUG & 0.073 & 0.063 & -0.010 \\
S1 & Eff-sub & 0.057 & 0.053 & -0.004 \\
S1 & Eff-sieve & 0.029 & 0.024 & -0.004 \\
S1 & Eff-DML & 0.035 & 0.027 & -0.008 \\
S2 & P-AUG & 0.027 & 0.029 & 0.002 \\
S2 & Eff-sub & 0.023 & 0.024 & 0.002 \\
S2 & Eff-sieve & 0.011 & 0.011 & 0.000 \\
S2 & Eff-DML & 0.013 & 0.013 & 0.000 \\
S3 & P-AUG & 0.048 & 0.049 & 0.002 \\
S3 & Eff-sub & 0.046 & 0.048 & 0.002 \\
S3 & Eff-sieve & 0.026 & 0.027 & 0.001 \\
S3 & Eff-DML & 0.027 & 0.028 & 0.001 \\
S4 & P-AUG & 0.074 & 0.065 & -0.008 \\
S4 & Eff-sub & 0.061 & 0.058 & -0.003 \\
S4 & Eff-sieve & 0.028 & 0.024 & -0.004 \\
S4 & Eff-DML & 0.035 & 0.027 & -0.008 \\
\end{longtable}

\clearpage
\begin{longtable}{@{}llrrrrr@{}}
\caption{Complete Monte Carlo performance in Scenarios S1--S4 under the rich spline rule. Target-estimation summaries.}\label{tab:supp-rich-main-performance-full}\\
\toprule
Scenario & Method & \shortstack{Bias\\(MCSE)} & SD & \shortstack{RMSE\\(MCSE)} & \shortstack{Mean\\SE} & \shortstack{Coverage\\(MCSE)} \\
\midrule
\endfirsthead
\multicolumn{7}{c}{\tablename\ \thetable\ (continued)}\\
\toprule
Scenario & Method & \shortstack{Bias\\(MCSE)} & SD & \shortstack{RMSE\\(MCSE)} & \shortstack{Mean\\SE} & \shortstack{Coverage\\(MCSE)} \\
\midrule
\endhead
\midrule\multicolumn{7}{r}{Continued on next page}\\
\endfoot
\bottomrule
\endlastfoot
S1 & P & 0.006 (0.004) & 0.096 & 0.097 (0.003) & 0.096 & 0.920 (0.012) \\
S1 & P-AUG & 0.005 (0.003) & 0.063 & 0.063 (0.002) & 0.061 & 0.916 (0.012) \\
S1 & Eff-sub & 0.003 (0.002) & 0.053 & 0.053 (0.002) & 0.051 & 0.942 (0.010) \\
S1 & Eff-0 & 0.009 (0.004) & 0.096 & 0.097 (0.003) & 0.095 & 0.918 (0.012) \\
S1 & Eff-sieve & 0.001 (0.001) & 0.024 & 0.024 (0.001) & 0.023 & 0.936 (0.011) \\
S1 & Eff-DML & 0.002 (0.001) & 0.027 & 0.027 (0.001) & 0.031 & 0.962 (0.009) \\
S1 & Oracle-Eff & 0.006 (0.001) & 0.022 & 0.023 (0.001) & 0.020 & 0.926 (0.012) \\
S1 & NP & 0.041 (0.013) & 0.302 & 0.304 (0.025) & 0.276 & 0.992 (0.004) \\
S1 & CLW & -4.314 (3.480) & 77.807 & 77.849 (30.061) & 37.376 & 0.220 (0.019) \\
S1 & CLW+P & -0.163 (0.008) & 0.180 & 0.243 (0.006) & 0.079 & 0.430 (0.022) \\
S2 & P & -0.003 (0.003) & 0.076 & 0.076 (0.003) & 0.075 & 0.946 (0.010) \\
S2 & P-AUG & 0.000 (0.001) & 0.029 & 0.029 (0.001) & 0.026 & 0.908 (0.013) \\
S2 & Eff-sub & 0.000 (0.001) & 0.024 & 0.024 (0.001) & 0.022 & 0.906 (0.013) \\
S2 & Eff-0 & 0.001 (0.003) & 0.076 & 0.076 (0.002) & 0.074 & 0.946 (0.010) \\
S2 & Eff-sieve & 0.000 (0.000) & 0.011 & 0.011 (0.000) & 0.011 & 0.934 (0.011) \\
S2 & Eff-DML & 0.001 (0.001) & 0.013 & 0.013 (0.000) & 0.024 & 0.968 (0.008) \\
S2 & Oracle-Eff & 0.000 (0.001) & 0.011 & 0.011 (0.001) & 0.010 & 0.948 (0.010) \\
S2 & NP & 0.066 (0.013) & 0.294 & 0.301 (0.025) & 0.274 & 0.992 (0.004) \\
S2 & CLW & 9.570 (5.181) & 115.855 & 116.134 (45.442) & 37.186 & 0.944 (0.010) \\
S2 & CLW+P & 0.005 (0.003) & 0.065 & 0.065 (0.002) & 0.060 & 0.910 (0.013) \\
S3 & P & 0.000 (0.004) & 0.086 & 0.086 (0.003) & 0.085 & 0.940 (0.011) \\
S3 & P-AUG & 0.003 (0.002) & 0.049 & 0.049 (0.002) & 0.049 & 0.930 (0.011) \\
S3 & Eff-sub & 0.003 (0.002) & 0.048 & 0.048 (0.002) & 0.046 & 0.928 (0.012) \\
S3 & Eff-0 & 0.003 (0.004) & 0.086 & 0.086 (0.003) & 0.085 & 0.934 (0.011) \\
S3 & Eff-sieve & -0.006 (0.001) & 0.027 & 0.027 (0.001) & 0.024 & 0.918 (0.012) \\
S3 & Eff-DML & -0.005 (0.001) & 0.027 & 0.028 (0.001) & 0.030 & 0.956 (0.009) \\
S3 & NP & -0.035 (0.009) & 0.207 & 0.209 (0.012) & 0.190 & 0.970 (0.008) \\
S3 & CLW & 0.044 (0.003) & 0.072 & 0.084 (0.003) & 0.070 & 0.898 (0.014) \\
S3 & CLW+P & 0.052 (0.003) & 0.065 & 0.083 (0.002) & 0.067 & 0.882 (0.014) \\
S4 & P & 0.006 (0.004) & 0.096 & 0.097 (0.003) & 0.096 & 0.920 (0.012) \\
S4 & P-AUG & 0.005 (0.003) & 0.065 & 0.065 (0.003) & 0.061 & 0.930 (0.011) \\
S4 & Eff-sub & -0.010 (0.003) & 0.057 & 0.058 (0.003) & 0.057 & 0.938 (0.011) \\
S4 & Eff-0 & 0.009 (0.004) & 0.096 & 0.097 (0.003) & 0.095 & 0.918 (0.012) \\
S4 & Eff-sieve & -0.003 (0.001) & 0.024 & 0.024 (0.001) & 0.023 & 0.942 (0.010) \\
S4 & Eff-DML & 0.000 (0.001) & 0.027 & 0.027 (0.001) & 0.030 & 0.962 (0.009) \\
S4 & NP & 0.089 (0.086) & 1.933 & 1.933 (0.284) & 1.297 & 1.000 (0.000) \\
S4 & CLW & -5.739 (4.420) & 98.839 & 98.907 (38.346) & 36.424 & 0.232 (0.019) \\
S4 & CLW+P & -0.160 (0.008) & 0.179 & 0.240 (0.006) & 0.079 & 0.438 (0.022) \\
\end{longtable}

\begin{longtable}{llrrr}
\caption{Complete Monte Carlo performance in Scenarios S1--S4 under the rich spline rule. Numerical and finite-population diagnostics.}\label{tab:supp-rich-main-performance-full-diagnostics}\\
\toprule
Scenario & Method & Failure & \shortstack{Finite\\RMSE} & $\phi_Y$ RMSE \\
\midrule
\endfirsthead
\multicolumn{5}{c}{\tablename\ \thetable\ (continued)}\\
\toprule
Scenario & Method & Failure & \shortstack{Finite\\RMSE} & $\phi_Y$ RMSE \\
\midrule
\endhead
\midrule\multicolumn{5}{r}{Continued on next page}\\
\endfoot
\bottomrule
\endlastfoot
S1 & P & 0.000 & 0.096 & -- \\
S1 & P-AUG & 0.000 & 0.063 & -- \\
S1 & Eff-sub & 0.000 & 0.052 & -- \\
S1 & Eff-0 & 0.000 & 0.096 & 0.312 \\
S1 & Eff-sieve & 0.000 & 0.023 & 0.063 \\
S1 & Eff-DML & 0.000 & 0.026 & 0.070 \\
S1 & Oracle-Eff & 0.000 & 0.021 & 0.089 \\
S1 & NP & 0.000 & 0.303 & 0.312 \\
S1 & CLW & 0.000 & 77.849 & -- \\
S1 & CLW+P & 0.000 & 0.243 & -- \\
S2 & P & 0.000 & 0.076 & -- \\
S2 & P-AUG & 0.000 & 0.028 & -- \\
S2 & Eff-sub & 0.000 & 0.023 & -- \\
S2 & Eff-0 & 0.000 & 0.075 & 2.433 \\
S2 & Eff-sieve & 0.000 & 0.009 & 0.981 \\
S2 & Eff-DML & 0.000 & 0.012 & 1.135 \\
S2 & Oracle-Eff & 0.000 & 0.008 & 0.968 \\
S2 & NP & 0.000 & 0.301 & 2.384 \\
S2 & CLW & 0.000 & 116.135 & -- \\
S2 & CLW+P & 0.000 & 0.065 & -- \\
S3 & P & 0.000 & 0.085 & -- \\
S3 & P-AUG & 0.000 & 0.049 & -- \\
S3 & Eff-sub & 0.000 & 0.048 & -- \\
S3 & Eff-0 & 0.000 & 0.086 & -- \\
S3 & Eff-sieve & 0.000 & 0.026 & -- \\
S3 & Eff-DML & 0.000 & 0.026 & -- \\
S3 & NP & 0.000 & 0.210 & -- \\
S3 & CLW & 0.000 & 0.083 & -- \\
S3 & CLW+P & 0.000 & 0.083 & -- \\
S4 & P & 0.000 & 0.096 & -- \\
S4 & P-AUG & 0.000 & 0.065 & -- \\
S4 & Eff-sub & 0.000 & 0.057 & -- \\
S4 & Eff-0 & 0.000 & 0.096 & 0.450 \\
S4 & Eff-sieve & 0.000 & 0.023 & 0.063 \\
S4 & Eff-DML & 0.000 & 0.026 & 0.070 \\
S4 & NP & 0.000 & 1.934 & 0.450 \\
S4 & CLW & 0.000 & 98.907 & -- \\
S4 & CLW+P & 0.000 & 0.241 & -- \\
\end{longtable}

\clearpage
\begin{longtable}{rlrrrrrr}
\caption{Complete probability-design heterogeneity results under the rich spline rule.}\label{tab:supp-rich-design-heterogeneity-full}\\
\toprule
$c$ & Method & Bias & SD & RMSE & \shortstack{Mean\\SE} & Coverage & Failure \\
\midrule
\endfirsthead
\multicolumn{8}{c}{\tablename\ \thetable\ (continued)}\\
\toprule
$c$ & Method & Bias & SD & RMSE & \shortstack{Mean\\SE} & Coverage & Failure \\
\midrule
\endhead
\midrule\multicolumn{8}{r}{Continued on next page}\\
\endfoot
\bottomrule
\endlastfoot
100 & P & 0.005 & 0.099 & 0.099 & 0.093 & 0.920 & 0.000 \\
100 & P-AUG & 0.003 & 0.065 & 0.065 & 0.058 & 0.918 & 0.000 \\
100 & Eff-sub & 0.000 & 0.052 & 0.052 & 0.048 & 0.932 & 0.000 \\
100 & Eff-0 & 0.008 & 0.098 & 0.098 & 0.093 & 0.916 & 0.000 \\
100 & Eff-sieve & 0.000 & 0.024 & 0.024 & 0.023 & 0.934 & 0.000 \\
100 & Eff-DML & 0.002 & 0.028 & 0.028 & 0.029 & 0.942 & 0.000 \\
100 & Oracle-Eff & 0.006 & 0.023 & 0.024 & 0.019 & 0.906 & 0.000 \\
100 & NP & 0.071 & 0.336 & 0.343 & 0.290 & 0.986 & 0.000 \\
20 & P & 0.004 & 0.103 & 0.103 & 0.102 & 0.928 & 0.000 \\
20 & P-AUG & 0.001 & 0.069 & 0.069 & 0.064 & 0.936 & 0.000 \\
20 & Eff-sub & 0.002 & 0.057 & 0.057 & 0.054 & 0.930 & 0.000 \\
20 & Eff-0 & 0.008 & 0.103 & 0.103 & 0.100 & 0.926 & 0.000 \\
20 & Eff-sieve & 0.000 & 0.025 & 0.025 & 0.023 & 0.926 & 0.000 \\
20 & Eff-DML & 0.002 & 0.029 & 0.029 & 0.032 & 0.944 & 0.000 \\
20 & Oracle-Eff & 0.007 & 0.024 & 0.025 & 0.020 & 0.906 & 0.000 \\
20 & NP & 0.044 & 0.329 & 0.332 & 0.283 & 0.994 & 0.000 \\
5 & P & 0.004 & 0.133 & 0.133 & 0.123 & 0.892 & 0.000 \\
5 & P-AUG & 0.005 & 0.084 & 0.084 & 0.079 & 0.926 & 0.000 \\
5 & Eff-sub & 0.002 & 0.068 & 0.068 & 0.065 & 0.950 & 0.000 \\
5 & Eff-0 & 0.010 & 0.131 & 0.131 & 0.120 & 0.892 & 0.000 \\
5 & Eff-sieve & 0.002 & 0.026 & 0.026 & 0.024 & 0.926 & 0.000 \\
5 & Eff-DML & 0.006 & 0.032 & 0.032 & 0.115 & 0.944 & 0.000 \\
5 & Oracle-Eff & 0.006 & 0.023 & 0.024 & 0.021 & 0.936 & 0.000 \\
5 & NP & 0.057 & 0.281 & 0.286 & 0.271 & 0.996 & 0.000 \\
\end{longtable}

\clearpage
\begin{longtable}{rlrrrr}
\caption{Complete weak-identification results under the rich spline rule. Target-estimation summaries.}\label{tab:supp-rich-weak-identification-full}\\
\toprule
$\sigma$ & Method & RMSE & \shortstack{Mean\\SE} & Coverage & Failure \\
\midrule
\endfirsthead
\multicolumn{6}{c}{\tablename\ \thetable\ (continued)}\\
\toprule
$\sigma$ & Method & RMSE & \shortstack{Mean\\SE} & Coverage & Failure \\
\midrule
\endhead
\midrule\multicolumn{6}{r}{Continued on next page}\\
\endfoot
\bottomrule
\endlastfoot
0.50 & P & 0.085 & 0.084 & 0.938 & 0.000 \\
0.50 & P-AUG & 0.051 & 0.049 & 0.934 & 0.000 \\
0.50 & Eff-sub & 0.041 & 0.041 & 0.934 & 0.000 \\
0.50 & Eff-0 & 0.085 & 0.083 & 0.936 & 0.000 \\
0.50 & Eff-sieve & 0.036 & 0.030 & 0.896 & 0.000 \\
0.50 & Eff-DML & 0.043 & 0.096 & 0.964 & 0.000 \\
0.50 & Oracle-Eff & 0.031 & 0.032 & 0.952 & 0.000 \\
0.50 & NP & 0.419 & 0.337 & 0.982 & 0.000 \\
0.25 & P & 0.078 & 0.077 & 0.940 & 0.000 \\
0.25 & P-AUG & 0.034 & 0.032 & 0.924 & 0.000 \\
0.25 & Eff-sub & 0.028 & 0.028 & 0.920 & 0.000 \\
0.25 & Eff-0 & 0.078 & 0.076 & 0.940 & 0.000 \\
0.25 & Eff-sieve & 0.019 & 0.017 & 0.910 & 0.000 \\
0.25 & Eff-DML & 0.024 & 0.050 & 0.946 & 0.000 \\
0.25 & Oracle-Eff & 0.016 & 0.017 & 0.958 & 0.000 \\
0.25 & NP & 0.343 & 0.292 & 0.990 & 0.000 \\
0.10 & P & 0.076 & 0.075 & 0.946 & 0.000 \\
0.10 & P-AUG & 0.029 & 0.026 & 0.908 & 0.000 \\
0.10 & Eff-sub & 0.024 & 0.022 & 0.906 & 0.000 \\
0.10 & Eff-0 & 0.076 & 0.074 & 0.946 & 0.000 \\
0.10 & Eff-sieve & 0.011 & 0.011 & 0.934 & 0.000 \\
0.10 & Eff-DML & 0.013 & 0.024 & 0.968 & 0.000 \\
0.10 & Oracle-Eff & 0.011 & 0.010 & 0.948 & 0.000 \\
0.10 & NP & 0.301 & 0.274 & 0.992 & 0.000 \\
0.05 & P & 0.076 & 0.075 & 0.944 & 0.000 \\
0.05 & P-AUG & 0.028 & 0.025 & 0.898 & 0.000 \\
0.05 & Eff-sub & 0.023 & 0.022 & 0.914 & 0.000 \\
0.05 & Eff-0 & 0.076 & 0.074 & 0.952 & 0.000 \\
0.05 & Eff-sieve & 0.010 & 0.010 & 0.954 & 0.000 \\
0.05 & Eff-DML & 0.012 & 0.016 & 0.964 & 0.000 \\
0.05 & Oracle-Eff & 0.010 & 0.009 & 0.948 & 0.000 \\
0.05 & NP & 0.279 & 0.260 & 0.988 & 0.000 \\
0.02 & P & 0.076 & 0.075 & 0.946 & 0.000 \\
0.02 & P-AUG & 0.026 & 0.024 & 0.916 & 0.000 \\
0.02 & Eff-sub & 0.022 & 0.021 & 0.920 & 0.000 \\
0.02 & Eff-0 & 0.075 & 0.074 & 0.952 & 0.000 \\
0.02 & Eff-sieve & 0.009 & 0.009 & 0.958 & 0.000 \\
0.02 & Eff-DML & 0.011 & 0.015 & 0.964 & 0.000 \\
0.02 & Oracle-Eff & 0.009 & 0.008 & 0.952 & 0.000 \\
0.02 & NP & 0.267 & 0.249 & 0.986 & 0.000 \\
\end{longtable}

\begin{longtable}{rlrrr}
\caption{Complete weak-identification results under the rich spline rule. Method-specific mechanism-equation Jacobian conditioning diagnostics.}\label{tab:supp-rich-weak-identification-full-diagnostics}\\
\toprule
$\sigma$ & Method & $\phi_Y$ RMSE & \shortstack{Minimum\\singular value} & \shortstack{Condition\\number} \\
\midrule
\endfirsthead
\multicolumn{5}{c}{\tablename\ \thetable\ (continued)}\\
\toprule
$\sigma$ & Method & $\phi_Y$ RMSE & \shortstack{Minimum\\singular value} & \shortstack{Condition\\number} \\
\midrule
\endhead
\midrule\multicolumn{5}{r}{Continued on next page}\\
\endfoot
\bottomrule
\endlastfoot
0.50 & P & -- & -- & -- \\
0.50 & P-AUG & -- & -- & -- \\
0.50 & Eff-sub & -- & -- & -- \\
0.50 & Eff-0 & 0.480 & $2.51\times10^{-4}$ & 10.5 \\
0.50 & Eff-sieve & 0.207 & 0.002 & 80.1 \\
0.50 & Eff-DML & 0.242 & 0.002 & 200.7 \\
0.50 & Oracle-Eff & 0.195 & 0.002 & 99.4 \\
0.50 & NP & 0.480 & $2.51\times10^{-4}$ & 10.5 \\
0.25 & P & -- & -- & -- \\
0.25 & P-AUG & -- & -- & -- \\
0.25 & Eff-sub & -- & -- & -- \\
0.25 & Eff-0 & 0.965 & $6.62\times10^{-5}$ & 37.6 \\
0.25 & Eff-sieve & 0.395 & $5.96\times10^{-4}$ & 326.9 \\
0.25 & Eff-DML & 0.471 & $4.26\times10^{-4}$ & 657.5 \\
0.25 & Oracle-Eff & 0.387 & $5.31\times10^{-4}$ & 406.8 \\
0.25 & NP & 0.965 & $6.62\times10^{-5}$ & 37.6 \\
0.10 & P & -- & -- & -- \\
0.10 & P-AUG & -- & -- & -- \\
0.10 & Eff-sub & -- & -- & -- \\
0.10 & Eff-0 & 2.433 & $1.07\times10^{-5}$ & 229.6 \\
0.10 & Eff-sieve & 0.981 & $9.66\times10^{-5}$ & 2047.2 \\
0.10 & Eff-DML & 1.135 & $6.78\times10^{-5}$ & 5050.2 \\
0.10 & Oracle-Eff & 0.969 & $8.53\times10^{-5}$ & 2549.7 \\
0.10 & NP & 2.384 & $1.07\times10^{-5}$ & 229.5 \\
0.05 & P & -- & -- & -- \\
0.05 & P-AUG & -- & -- & -- \\
0.05 & Eff-sub & -- & -- & -- \\
0.05 & Eff-0 & 4.799 & $2.68\times10^{-6}$ & 913.3 \\
0.05 & Eff-sieve & 1.957 & $2.42\times10^{-5}$ & 8189.0 \\
0.05 & Eff-DML & 2.232 & $1.67\times10^{-5}$ & 16373.3 \\
0.05 & Oracle-Eff & 1.882 & $2.14\times10^{-5}$ & 10473.0 \\
0.05 & NP & 3.853 & $2.69\times10^{-6}$ & 909.8 \\
0.02 & P & -- & -- & -- \\
0.02 & P-AUG & -- & -- & -- \\
0.02 & Eff-sub & -- & -- & -- \\
0.02 & Eff-0 & 11.459 & $4.31\times10^{-7}$ & 5703.3 \\
0.02 & Eff-sieve & 4.849 & $3.88\times10^{-6}$ & 51191.6 \\
0.02 & Eff-DML & 5.644 & $2.73\times10^{-6}$ & 123935.5 \\
0.02 & Oracle-Eff & 4.612 & $3.39\times10^{-6}$ & 65192.0 \\
0.02 & NP & 3.968 & $4.34\times10^{-7}$ & 5665.8 \\
\end{longtable}

\subsection{CCTC data processing and model details}
\label{supp:cctc}

The public CCTC data were developed for comparing methods that combine probability and non-probability samples \citep{benoitbryanMulrow2020}. The cleaned Frame~1 file is treated as a finite population of $N=113{,}549$. The archived samples may reflect fixed-size or other complex-design dependence, so this study is not used to verify the independent-Poisson efficiency bound directly.

The outcome is the binary indicator \texttt{q7\_22} for attendance at a classical music event. Each replication uses a probability sample of size $1{,}000$ with observed design inclusion probabilities and a non-probability source of size $4{,}000$ drawn from an undercoverage frame.

The non-probability inclusion model used by NP, Eff-sieve, and Eff-DML is
\[
\pi_{\NP}(L;\phi)
=\expit\{\phi^{\trans}(1,Y,L_{\mathrm{sel}}^{\trans})^{\trans}\},
\]
where $L_{\mathrm{sel}}$ contains age, metropolitan status, nine-category region, and race indicators. The probability sample observes $Y$, the non-probability inclusion indicator, and the design inclusion probability, so the validation estimator is ordinary logistic regression within the probability sample.

The source population has 293 variables. After identifiers, the outcome, variables with population-level missingness, and redundant recodes are removed, the default frame construction retains 112 fully observed predictors; categorical factors are expanded into indicator variables. The five general augmentation functions are fitted on the full sample for Eff-sieve and with $K=2$ training/evaluation folds for Eff-DML. Because tensor-product splines are impractical in this high-dimensional frame, the analysis uses an additive linear basis with fixed ridge coefficient $10^{-2}$.

The two bounded $L$-level reparameterizations are fitted on probability-sample records by ridge-regularized fractional-logit regression. The identities in Section~\ref{supp:sieve-tuning} recover $k_2$, $k_1$, and $\kappa_1$, while $\eta$ and $\kappa$ are fitted by ridge-regularized linear regression on probability-sample records and evaluated on the complete frame. P-AUG uses a weighted penalized linear regression on all probability-sample records, and Eff-sub uses the weighted regression in Section~\ref{supp:effsub-feasible} on probability-sample records outside the non-probability source. This application-specific tuning is separate from the growing-spline rules used in the controlled simulations.

The final method list is P, P-AUG, NP, CLW, CLW+P, Eff-sub, Eff-sieve, Eff-DML, and NP-mean. Eff-sieve and Eff-DML use the local profiling, prespecified starts, failure rule, and joint sandwich variance estimation in Section~\ref{supp:dml}; no oracle start is used. NP-mean is included only as a descriptive benchmark.

For each replication, $Y$ and $\pi_{\Pp}$ are masked according to the observed-data pattern before the estimators are fitted, while the fixed-population target is retained separately for evaluation. The reported diagnostics include estimator failures, validation-Jacobian behavior, and stabilized CLW weight behavior.

\subsection{Complete CCTC results and diagnostics}

\begin{longtable}{lrrrrr}
\caption{Complete CCTC repeated-sampling results. Monte Carlo standard errors (MCSEs) are based on 500 replications and are shown in parentheses.}\label{tab:supp-cctc-full}\\
\toprule
Method & \shortstack{Bias\\(MCSE)} & SD & \shortstack{RMSE\\(MCSE)} & \shortstack{Mean\\SE} & \shortstack{Coverage\\(MCSE)} \\
\midrule
\endfirsthead
\multicolumn{6}{c}{\tablename\ \thetable\ (continued)}\\
\toprule
Method & \shortstack{Bias\\(MCSE)} & SD & \shortstack{RMSE\\(MCSE)} & \shortstack{Mean\\SE} & \shortstack{Coverage\\(MCSE)} \\
\midrule
\endhead
\midrule\multicolumn{6}{r}{Continued on next page}\\
\endfoot
\bottomrule
\endlastfoot
P & -0.001 (0.001) & 0.018 & 0.018 (0.001) & 0.025 & 0.986 (0.005) \\
P-AUG & -0.001 (0.001) & 0.015 & 0.015 (0.000) & 0.016 & 0.974 (0.007) \\
Eff-sub & -0.001 (0.001) & 0.015 & 0.015 (0.000) & 0.016 & 0.964 (0.008) \\
Eff-sieve & -0.045 (0.001) & 0.023 & 0.051 (0.001) & 0.009 & 0.116 (0.014) \\
Eff-DML & -0.048 (0.001) & 0.028 & 0.055 (0.001) & 0.014 & 0.252 (0.019) \\
CLW & 0.021 (0.000) & 0.008 & 0.022 (0.000) & 0.008 & 0.284 (0.020) \\
CLW+P & 0.019 (0.000) & 0.008 & 0.020 (0.000) & 0.008 & 0.342 (0.021) \\
NP & 2.187 (0.492) & 11.005 & 11.209 (2.966) & 0.779 & 0.718 (0.020) \\
NP-mean & 0.017 (0.000) & 0.008 & 0.018 (0.000) & 0.008 & 0.466 (0.022) \\
\end{longtable}

\begin{longtable}{lrr}
\caption{Complete CCTC repeated-sampling diagnostics.}\label{tab:supp-cctc-full-diagnostics}\\
\toprule
Method & Failure & \shortstack{Outside\\$[0,1]$} \\
\midrule
\endfirsthead
\multicolumn{3}{c}{\tablename\ \thetable\ (continued)}\\
\toprule
Method & Failure & \shortstack{Outside\\$[0,1]$} \\
\midrule
\endhead
\midrule\multicolumn{3}{r}{Continued on next page}\\
\endfoot
\bottomrule
\endlastfoot
P & 0.000 & 0 \\
P-AUG & 0.000 & 0 \\
Eff-sub & 0.000 & 0 \\
Eff-sieve & 0.000 & 0 \\
Eff-DML & 0.000 & 0 \\
CLW & 0.000 & 0 \\
CLW+P & 0.000 & 0 \\
NP & 0.000 & 150 \\
NP-mean & 0.000 & 0 \\
\end{longtable}

All nine methods return finite estimates in all 500 replications under the stabilized CLW implementation. For CLW, the fitted propensity model has dimension 10, the median minimum propensity is $0.0214$, the mean maximum inverse weight is $47.0$, and the mean effective sample size is $3{,}918$ out of $4{,}000$ non-probability units. These diagnostics confirm that the revised CLW implementation is numerically stable, although its residual bias shows that covariate-only ignorability is not adequate for the CCTC mechanism.

NP remains an intentionally unnormalized Horvitz--Thompson benchmark. Its estimates lie outside $[0,1]$ in 150 of the 500 replications, which explains its very large RMSE\@.

Eff-sieve and Eff-DML are numerically stable, but both are centered below the finite-population target. Their similar biases indicate that this result is not driven by foldwise sample splitting. The pattern is instead consistent with misspecification or positivity failure in the low-dimensional working non-probability inclusion model and with departures of the archived probability samples from the independent-Poisson experiment used in the efficiency theory.

\end{document}